\documentclass[a4paper,11pt]{article}
\usepackage[dvipsnames]{xcolor}
\usepackage[utf8]{inputenc}
\usepackage[bbgreekl]{mathbbol}
\usepackage{geometry}
\usepackage{multicol}

\usepackage{extarrows}
\usepackage{xcolor}
\usepackage{amsmath, array, amssymb, amsfonts,amsthm}
\usepackage{upgreek}
\usepackage{xfrac}
\usepackage[inline]{enumitem}
\setlist{nolistsep}
\usepackage[french, italian, english]{babel}
\usepackage[all]{xy}
\usepackage{txfonts}  
\usepackage{sectsty}
\usepackage{booktabs}
\usepackage{caption}
\usepackage{dsfont}
\usepackage{mathtools}
\usepackage{slashed}
\usepackage[makeroom]{cancel}
\usepackage[hidelinks]{hyperref}
\usepackage{textcomp}
\usepackage{multicol}
\usepackage[title]{appendix}
\usepackage[square,numbers,sort,compress,merge]{natbib}
\let\cite\citep 
\usepackage{hyperref}
\hypersetup{colorlinks=true, urlcolor=blue, citecolor=blue, linktoc=page}

\usepackage{tikz}
\usepackage{tikz-cd}
\usetikzlibrary{cd}
\tikzcdset{
arrow style=tikz,
diagrams={>={Straight Barb[scale=0.8]}}
}
\usetikzlibrary{matrix,arrows,decorations.pathmorphing}

\DeclareMathAlphabet{\mathpzc}{OT1}{pzc}{m}{it}

\usepackage{fancybox}

\usepackage{mathrsfs}
\usepackage{scrextend}

\makeatletter
\renewcommand*\env@matrix[1][\arraystretch]{%
  \edef\arraystretch{#1}%
  \hskip -\arraycolsep
  \let\@ifnextchar\new@ifnextchar
  \array{*\c@MaxMatrixCols c}}
\makeatother

\newcommand{\defeq}{\vcentcolon=}
\newcommand{\rdefeq}{=\vcentcolon}
\renewcommand\P{\mathcal{P}}

\newcommand\RR{\mathbb{R}}
\newcommand\CC{\mathbb{C}}
\newcommand\EE{\mathbb{E}}
\newcommand\C{\mathcal{C}}

\newcommand\id{\textit{id}}
\newcommand\T{\mathcal{T}}
\newcommand\G{\mathcal{G}}
\renewcommand\H{\mathcal{H}}
\newcommand\Hl{\H_\text{\tiny loc}}

\renewcommand\S{\mathcal{S}}

\newcommand\SU{\mathcal{SU}}
\newcommand\U{\mathcal{U}}

\newcommand\K{\mathcal{K}}
\newcommand\J{\mathcal{J}}
\renewcommand\O{\mathcal{O}}

\newcommand\D{\mathcal{D}}

\renewcommand\u{\text{\bf u}}

\newcommand\mC{\mathscr{C}}
\newcommand\mH{\mathscr{H}}
\newcommand\mK{\mathscr{K}}

\renewcommand\epsilon{\varepsilon}

\newcommand\rarrow{\rightarrow}

\newcommand\aut{\mathfrak{aut}}

\newcommand\LieG{\mathfrak{g}}
\newcommand\LieH{\mathfrak{h}}

\renewcommand\t{\tilde}

\renewcommand\b{\bar }

\renewcommand\d{\partial}
\newcommand\s{\sigma}
\newcommand\bs{\boldsymbol}

\renewcommand\-{^{-1}}
\newcommand\Ad{\text{Ad}}

\renewcommand\id{\text{id}}

\makeatletter

\newcommand{\Rmnum}[1]{\expandafter\@slowromancap\romannumeral #1@}
\makeatother

\makeatletter
\newcommand{\leqnomode}{\tagsleft@true\let\veqno\@@leqno}
\newcommand{\reqnomode}{\tagsleft@false\let\veqno\@@eqno}
\makeatother

\DeclareMathOperator{\Diff}{Diff}
\DeclareMathOperator{\Aut}{Aut}
\DeclareMathOperator{\Der}{Der}

\DeclareMathOperator{\vol}{vol}

\theoremstyle{definition}

\makeatother

\usepackage{float}

\begin{document}


\title{Relational bundle geometric formulation \\
of non-relativistic quantum mechanics}

\author{J. T. \textsc{François} $\,{}^{a,\,b,\,c,\,*}$ \and L. \textsc{Ravera} $\,{}^{d,\,e,\,f,\,\star}$ }

\date{}

\maketitle
\begin{center}
\vskip -0.6cm
\noindent
${}^a$ Department of Mathematics \& Statistics, Masaryk University -- MUNI.\\
Kotlářská 267/2, Veveří, Brno, Czech Republic.\\[2mm]
 
${}^b$  Department of Philosophy -- University of Graz. \\
 Heinrichstraße 26/5, 8010 Graz, Austria.\\[2mm]
 
${}^c$ Department of Physics, Mons University -- UMONS.\\
20 Place du Parc, 7000 Mons, Belgium.
\\[2mm]

${}^d$ DISAT, Politecnico di Torino -- PoliTo. \\
Corso Duca degli Abruzzi 24, 10129 Torino, Italy. \\[2mm]

${}^e$ Istituto Nazionale di Fisica Nucleare, Section of Torino -- INFN. \\
Via P. Giuria 1, 10125 Torino, Italy. \\[2mm]

${}^f$ \emph{Grupo de Investigación en Física Teórica} -- GIFT. \\
Universidad Cat\'{o}lica De La Sant\'{i}sima Concepci\'{o}n, Concepción, Chile. \\[2mm]

\vspace{1mm}

${}^*$ {\small{jordan.francois@uni-graz.at}} \qquad \quad ${}^\star$ {\small{lucrezia.ravera@polito.it}}
\end{center}


\vspace{-3mm}

\begin{abstract}
We present a  bundle geometric formulation of non-relativistic many-particles Quantum Mechanics. 
A wave function is seen to be a $\CC$-valued \emph{cocyclic} tensorial 0-form on 
configuration space-time seen as a principal bundle, while
the Schrödinger equation flows from its covariant derivative, with the action functional supplying a (flat) \emph{cocyclic connection} 1-form on the configuration bundle. 
In line with the historical motivations of Dirac and Feynman, 
ours is thus a \emph{Lagrangian} geometric formulation of   QM, in which the Dirac-Feynman path integral arises in a geometrically natural way.

Applying the \emph{dressing field method}, we obtain a  relational reformulation of this geometric non-relativistic QM: 
a relational wave function is realised as a basic cocyclic 0-form on the configuration bundle. 
In this relational QM, any particle position can be used as a dressing field, i.e. as a ``physical reference frame". 
The dressing field method naturally accounts for the freedom in choosing the dressing field, 
which is readily understood as a covariance of the relational formulation under changes of physical reference frame.  
\end{abstract}


\vspace{-3mm}

\tableofcontents

\bigskip


\section{Introduction}  
\label{Introduction}  

Contrary to the other fundamental pillars of modern physics, General Relativity (GR) and Gauge Field Theory (GFT) -- or broadly, general-relativistic Gauge Field Theory (gRGFT) --
 which are fundamentally understood in geometric terms,  
Quantum Mechanics (QM) is usually still primarily conceived algebraically. And indeed many of its key puzzling features, from an interpretive standpoint, stems from its algebraic linearity. 
Yet, the geometric structure of QM has been a topic of interest for many. 
Notably, the field of geometric quantization \cite{Woodhouse1980-1998} pioneered by Souriau, Kostant and Kirillov  has been fruitful, mainly in mathematics and group representation theory. 
Also, isolated works throughout the years have made interesting contributions, e.g. 
\cite{Kibble1979, Schilling1996, Ashtekar-Schilling1999}.\footnote{Recently was published the result \cite{Janiska-Modugno2022} of a years-long concerted effort to couch QM in differential geometric terms. Our approach, to be described momentarily, is entirely independent.}

As part of our program of a relational (re)formulation of fundamental physics, started in \cite{JTF-Ravera2024c, JTF-Ravera2024gRGFT} (see also \cite{JTF-Ravera2024-SUSY,Francois:2024xqi}),
in this work we propose an original contribution to this tradition. 
We give a bundle geometric formulation of non-relativistic QM which relies on a \emph{generalisation} of standard bundle geometric notions, where group representations are replaced by 1-cocycles for the action of the group: \emph{cocyclic bundle geometry}  \cite{Francois2019_II} --  originally called \emph{twisted bundle geometry}.
A key notion of which is that of \emph{cocyclic connection}, extending the usual Ehresmann (and/or Cartan) connections.\footnote{The notion of cocyclic (or twisted) connection features prominently in twistor geometry \cite{Attard-Francois2016_II}, as well as in its real counterpart, conformal tractor geometry \cite{Attard-Francois2016_I} -- and in projective geometry. Tractor and twistors are cocyclic gauge fields regarding Weyl rescaling symmetry, and the conformal tractor and twistor connections are cocyclic Cartan connections.}
We further stress that ours is naturally a \emph{Lagrangian} approach to QM, which 
soundly resonates with the historical
motivations of Dirac \cite{Dirac1933} and Feynman \cite{Feynman1948, Feynman1965},\footnote{Dirac sought a Lagrangian formulation of QM so that it could be made compatible with Special Relativity from first principles. Feynman was motivated to find a Lagrangian approach to QM by the particular approach he and Wheeler took at the time to tackle QED, which rested on a classical version of electrodynamics that could not be cast in an Hamiltonian form.} leading them to the path integral (PI) formulation. 
As a matter of fact, the  Dirac-Feynman  PI fits well with our geometric approach.

Our bundle geometric version of QM makes it possible to apply the \emph{dressing field method} (DFM) of symmetry reduction \cite{Francois2014, JTF-Ravera2024gRGFT, Francois2023-a}, thereby obtaining a geometric \emph{relational} formulation of QM.
It is distinct from  relational QM à la Rovelli \cite{Rovelli1996}, even though the interpretation of our results is broadly in line with it. \emph{In spirit}, it is also broadly in line with the heuristic idea behind the recent burgeoning literature on ``quantum reference frames" \cite{Aharonov-Kaufherr1984, Spekkens-et-al2007,Giacomini-et-al2019}, though we approach the subject in a  mathematically more sophisticated way,  akin to geometric quantization. 

\medskip

The paper is organised as follows:
For the benefit of the reader, in section \ref{Bundle geometry and the dressing field method} we provide the essential background material. 
In section \ref{Cocyclic bundle geometry in a nutshell}
we remind the basics of bundle geometry. 
This is necessary for two reasons: 
First to introduce \emph{cocyclic} bundle geometry, in terms of which we shall formulate QM, and 
secondly because bundle geometry is the proper framework to express the DFM.
Which we do in section \ref{The dressing field method: basics}. 

With this preparatory work done, in section \ref{Bundle geometric non-relativistic Mechanics and Quantum Mechanics}
we give our bundle geometric formulation of non-relativistic~QM. 
The configuration space $\mC_{|t}$ of a set of $N$  particles at a time $t$ is seen as the fiber of a principal bundle $\mC$, which we call the configuration space-time bundle, over the Newtonian time line $\mathcal T$, its base space.
The gauge group of $\mC$ is shown to exactly capture the so-called \emph{extended Galilean transformations}  \cite{Greenberger1979}. 
The action functional is shown to induce
a \emph{flat cocyclic connection} 1-form.
It  induces a \emph{cocyclic covariant derivative} on $\CC$-valued \emph{cocyclic tensorial forms} on $\mC$.
Quantum mechanical wave functions $\psi$ are the subspace of form degree 0 of its kernel. 
The cocyclic covariant derivative of $\psi$
splits -- in  bundle coordinates -- 
as two \mbox{equations}: 
One~gives the quantum mechanical prescription for the conjugate momentum operator $\hat{\mathrm{\uppi}} =-i\hbar\tfrac{\d}{\d x}$, 
while the other gives the Schrödinger equation. 
We~then show how the
Dirac-Feynman PI formulation may arise  naturally from this geometric framework.

In section \ref{Relational bundle geometric classical  and quantum Mechanics}
we  describe a bundle geometric relational reformulation of both classical and quantum mechanics, achieved via the~DFM.
Relational wave functions are realized as  \emph{dressed}, or \emph{basic}, 0-forms on  $\mC$.
Correspondingly,  a \emph{relational Schrödinger equation}  gives the 
quantum dynamics.
The DFM further naturally accounts for the freedom in choosing the dressing field: in this case  any particle position can be used as a dressing field,  i.e. as a \emph{physical reference frame}. 
The dressed, relational formulation of QM via the DFM thus automatically implements a ``covariance under change of  physical reference frame". 

In the conclusion \ref{Conclusion}, we discuss the salient points our results and we indicate 
the next phase of our program. 
The~main text being kept general, appendix \ref{Free particles} illustrates  our formalism for  free particles.

\section{Bundle geometry and the dressing field method}
\label{Bundle geometry and the dressing field method}

\subsection{Cocyclic bundle geometry in a nutshell}
\label{Cocyclic bundle geometry in a nutshell}

The differential geometry of principal bundles is the  underpinning of our formulation of non-relativistic
classical and quantum mechanics, and of the Dressing Field Method.

The central object is a principal bundle $P$ over a base manifold $M$ with structure group $H$ and projection $\pi :P \rarrow M$, $p \mapsto \pi(p)=x$. 
A fiber over $x\in M$, $\pi\-(x)=P_{|x} \subset P$, is an orbit of the right action of the structure group, $P\times H \rarrow P$, $(p, h) \mapsto ph =: R_h p$, which is free and transitive on fibers, and s.t. $\pi \circ R_h =\pi$. 
This means the base is isomorphic as a manifold to the space of fibers $P/H \simeq M$.
The linearisation of the action of $H$ induces vectors tangent to the fibers:  $\forall X \in \LieH$ corresponds a \emph{fundamental} vertical vector $X^v_{|p}$ at $p$. 
At $p \in P$, the span of these vectors  is a subvector space $V_p P$ of the tangent space $T_p P$. 
The canonical vertical subbundle is $VP = \cup_p V_pP \subset T P$.  
We denote by $\Gamma(VP)$ the space of sections of $VP$, i.e. vertical vector fields $X^v: P \rarrow VP$.

The natural maximal group of transformation of $P$ is its group of automorphisms
\begin{align}
    \Aut(P):=\big\{ \Xi \in \Diff(P)\ |\ \Xi(ph)=\Xi(p)h \big\}.
\end{align}
It is the subgroup of $\Diff(P)$ that preserves the fibration structure by sending fibers to fibers, and thus induces diffeomorphisms of the base space $M$, i.e. projects onto $\Diff(M)$. 
Its maximal normal subgroup is the group of  vertical automorphisms
\begin{align}
    \Aut_v(P):=\big\{ \Xi \in \Aut(P)\ |\ \pi \circ \Xi=\pi\big\}.
\end{align}
It is the subgroup of those automorphisms that induce the identity transformation on $M$. 
It is isomorphic to the \emph{gauge group} of $P$, 
\begin{align}
    \H:= \big\{ \gamma: P \rarrow H\ |\ R^*_h \gamma =h\- \gamma h \big\},
\end{align}
the isomorphism being given by $\Xi(p)=p\gamma(p)$. 
The action of $\Aut_v(P)\simeq \H$ on differential forms of $P$ \emph{defines} their \emph{active gauge transformations}.
We have thus the characteristic short exact sequence (SES) of groups of $P$,
\begin{align}
\label{SES-P}
\id_P\rarrow \Aut_v(P) \simeq \H \xrightarrow{\triangleleft} \Aut(P)  \xrightarrow{\t \pi}  \Diff(M) \rarrow  \id_M,
\end{align}
whose linear version, the SES of Lie algebras, characterizes the Atiyah Lie algebroid canonically associated to $P$. 

\paragraph{Spaces of remarkable differential forms}

As a smooth manifold $P$ has a graded ring of differential forms $\Omega^\bullet(P)=\bigoplus_{k\geq 0} \Omega^k(P)$, with graded Lie algebra of derivations $\Der^\bullet\big( \Omega^\bullet(P)\big)=\bigoplus_{k\geq 0} \Der^k\big( \Omega^\bullet(P)\big)$ s.t. for $D_1 \in \Der^p$ and $D_2 \in \Der^q$ we have $[D_1, D_2]\defeq D_1 \circ D_2 -(-)^{pq} D_2 \circ D_1 \in \Der^{p+q}$.
In particular we have the exterior de Rham derivative $d \in \Der^1$, the inner derivation $\iota_X \in \Der^{-1}$ for any $X\in \Gamma(TP)$, and the Lie derivative $\mathfrak L_X \defeq [d, \iota_X]$.
The de Rham complex of $P$ is $\big(\Omega^\bullet(P), d \big)$. 

The action of the structure group on $P$ induces a right action on forms 
$\Omega^\bullet(P) \times H \rarrow  \Omega^\bullet(P)$, $(\alpha, h) \mapsto R^*_h \alpha$, which defines their \emph{equivariance}. 
The action by pullback of the group of automorphisms is also a right action:  
$\Omega^\bullet(P) \times \Aut(P) \rarrow  \Omega^\bullet(P)$, 
$(\alpha, \Xi) \mapsto\Xi^*\alpha$. 
In particular, gauge transformations are defined by the right action 
$\Omega^\bullet(P) \times \Aut_v(P)\simeq \H \rarrow  \Omega^\bullet(P)$, 
$(\alpha, \Xi \sim \gamma) \mapsto\Xi^*\alpha \rdefeq \alpha^\gamma$.

There are a number of remarkable spaces of forms worth describing. 
The first, which can be defined canonically, 
are \emph{horizontal} forms, i.e. those that vanish when evaluated on vertical vector fields
\begin{align}
\label{horiz-forms}
\Omega^\bullet_{\text{hor}}(P):=\big\{ \alpha \in \Omega^\bullet(P)\ |\  \iota_{X^v}\alpha=0  \text{ for } X^v \in \Gamma(VP) \big\}.
\end{align}
Then, there is the space of
 \emph{equivariant} forms, whose equivariance is ``simple". 
In standard bundle geometry, equivariant forms are valued in representations $(\rho, V)$ of $H$ which control their equivariance:
\begin{align}
\label{standard-equiv-forms}
\Omega^\bullet_{\text{eq}}(P, \rho):=\big\{ \alpha \in \Omega^\bullet(P, V)\ |\  R_h^*\alpha= \rho(h)\- \alpha \big\}.
\end{align}
Their linear equivariance is then given by the Lie derivative along fundamental vector fields: 
$\mathfrak L_{X^v} \alpha= -\rho_*(X)\, \alpha$. 

This is generalised in \emph{cocyclic} bundle geometry (formerly introduced as
 \emph{twisted} bundle geometry \cite{Francois2019_II}), where one defines \emph{cocyclic} equivariant forms whose equivariance is controlled by a 1-cocycle for the action of the structure group:
Given a Lie group $G$ we define
\begin{equation}
\label{1-cocycle}
\begin{aligned}
C: P \times H &\rarrow G, \\
    (p, h)    &\mapsto C(p, h), 
    \quad \text{s.t. }\ \  
    C(p, hh')=C(p, h)\cdot C(ph, h'). 
\end{aligned}
\end{equation}
We see that a 1-cocycle generalises representations. 
From this defining property follows that
\begin{align}
 \label{prop-cocycle}
 C(p, e_H)=e_G =C(ph, e_H), \quad 
 C(p, h)\- = C(ph, h\-), \quad
  \text{and} \quad
  C(p, h\-) = C(ph\-, h)\-.
\end{align}
The linearisation of such a group cocycle gives a 1-cocycle for the action of $\LieH$ on $P$:
\begin{equation}
\label{inf-1-cocycle}
\begin{aligned}
a: P \times \LieH &\rarrow \LieG, \\
    (p, X)    &\mapsto a(X, p), 
    \quad \text{s.t. }\ \  
    X^v \cdot a(Y, p) - {Y}^v \cdot a(X, p) + [a(X, p),\  a(Y, p)]_\LieG= a\big([X, Y ], p\big). 
\end{aligned}
\end{equation}
This can indeed be deduced from \eqref{1-cocycle} via 
$a(X, p)\defeq \tfrac{d}{d\tau}\, C(p, h_\tau)\, \big|_{\tau=0}$ and $X\defeq \tfrac{d}{d\tau}\, h_\tau\, \big|_{\tau=0} \, \in \LieH$. 
One may observe that, in the context of the field space of a gauge theory, $P\rarrow \Phi$, this generalises the Wess-Zumino consistency condition for (classical and/or quantum) gauge anomalies. 
See \cite{Francois2019_II, JTF-Ravera2024gRGFT}. 

Now, for $V$ a $G$-space, we define \emph{cocyclic equivariant} forms as
\begin{align}
\label{twisted-equiv-forms}
\Omega^\bullet_{\text{eq}}(P, C):=\big\{ \alpha \in \Omega^\bullet(P, V)\ |\  R_h^*\alpha= C(\ \, , h)\- \alpha \big\}.
\end{align}
One may check that the 1-cocycle defining property ensures compatibility with the right action of $H$, $R^*_{hh'}=R^*_{h}  R^*_{h'}$.  The linear equivariance is then given by 
$\mathfrak L_{X^v} \alpha = -a(X,\_\, )\, \alpha$. 
The 1-cocycle property \eqref{inf-1-cocycle} ensures compatibility with $[\mathfrak L_{X^v}, \mathfrak L_{Y^v}]= \mathfrak L_{[X^v, Y^v]} = \mathfrak L_{[X, Y]^v}$ -- which uses the  fact that the map $\LieH \rarrow \Gamma(VP)$, $X \mapsto X^v$, is a Lie algebra morphism.  
\medskip

Forms that are both equivariant and horizontal are said \emph{tensorial}.
As above, \emph{cocyclic tensorial} forms generalise   standard tensorial forms:
\begin{align}
\label{stand-twisted-tens-forms}
\Omega^\bullet_{\text{tens}}(P, \rho):= \Omega^\bullet_{\text{hor}}(P) \cap \Omega^\bullet_{\text{eq}}(P, \rho) \quad
 \subset \quad
\Omega^\bullet_{\text{tens}}(P, C):= \Omega^\bullet_{\text{hor}}(P) \cap \Omega^\bullet_{\text{eq}}(P, C).
\end{align}
Observe that $\Omega^0_{\text{eq}}(P, \rho)=\Omega^0_{\text{tens}}(P, \rho)$ and $\Omega^0_{\text{eq}}(P, C)=\Omega^0_{\text{tens}}(P, C)$ since 0-forms are horizontal.

It is an elementary result of bundle geometry that there is a bijection 
$\Omega^k_{\text{tens}}(P, \rho) \simeq \Omega^k(M) \otimes E$ between the space of standard tensorial forms and the space of forms on $M$ valued in associated vector bundles
$E= P \times_\rho V \defeq P\times V /\sim$, where the equivalence relation is $[p,v]=(p, v)\sim (ph, \rho(h)\-v)$: the bijection is given by $\t\alpha_{|x}=[p, \alpha_{|p}]=[ph, R^*_h\alpha_{|ph}]=[ph, \rho(h)\-\alpha_{|p}]$.
In degree $0$, this is the well-known isomorphism
$\Omega^0_{\text{tens}}(P, \rho) \simeq \Omega^0(M) \otimes E \simeq \Gamma(E)$ between representation-valued equivariant maps on $P$ and sections of associated bundles $\Gamma(E)=\{s: M \rightarrow E \}$.
This~in particular gives another characterisation of the gauge group $\H$, whose elements are $H$-valued equivariant 0-forms, as the space of sections of the associated bundle $\text{Conj}(P) = P\times_{\text{Conj}} H=P\times H/\sim$, with equivalence relation given by $[p,h']=(p,h')\sim (ph, h\- h' h)$: so we have the isomorphisms 
$\Aut_v(P)\simeq \H \simeq \Gamma\big(\text{Conj}(P) \big)$.

The result extends to the bijection $\Omega^k_{\text{tens}}(P, C) \simeq \Omega^k(M) \otimes E^C$ between the space of cocyclic tensorial forms and the space of forms on $M$ valued in \emph{cocyclic associated bundles} $E^C= P \times_C V \defeq P\times V /\sim$, where the equivalence relation is $[p,v]=(p, v)\sim (ph, C(p, h)\-v)$: the bijection being  $\t\alpha_{|x}=[p, \alpha_{|p}]=[ph, R^*_h\alpha_{|ph}]=[ph, C(p, h)\-\alpha_{|p}]$. In degree $0$, this gives the isomorphism $\Omega^0_{\text{tens}}(P, C) \simeq \Gamma(E^C)$, with $\Gamma(E^C)=\{\t s: M \rightarrow E^C \}$. 
\medskip

Finally, one may define the important space of \emph{basic} forms
\begin{align}
\label{basic-forms}
\Omega^\bullet_{\text{basic}}(P):=\big\{ \alpha \in \Omega^\bullet(P)\ |\  \exists \, \beta \in \Omega^\bullet (M) \text{ s.t. } \alpha=\pi^*\beta \big\}.   
\end{align}
These are horizontal: $\alpha (X^v) = \pi^*(\beta) (X^v)=\beta (\pi_* X^v)=0$. 
And they are \emph{invariant}, i.e. have trivial equivariance: 
$R^*_h \alpha = R^*_h \pi^* \beta = (\pi \circ R_h)^*\beta = \pi^*\beta \rdefeq \alpha$. 
This makes for an alternative definition of basic forms, as follows: $\Omega^\bullet_{\text{basic}}(P):=\big\{ \alpha \in \Omega^\bullet(P)\ |\ \iota_{X^v} \alpha =0 \text{ and } R^*_h\alpha=\alpha \big\}= \Omega^\bullet_{\text{hor}}(P) \cap \Omega^\bullet_{\text{inv}}(P)$.

\paragraph{Connections and covariant derivatives}

The exterior derivative is not a derivation of the spaces of tensorial forms. To define  first-order differential operators preserving those spaces, the \emph{covariant derivatives}, we need non-canonical structures on $P$: the appropriate notions of connection. 

In standard bundle geometry, the solution is well-known and given by the celebrated Ehresmann (or principal) connection 1-form $\omega \in \Omega^1(P, \LieH)$ defined as
\begin{align}
\label{Ehresmann-connection}
R^*_h \omega = \Ad_{h\-} \omega, \ \ \text{i.e.} \ \omega \in  \Omega_\text{eq}^1(P, \Ad), \quad
\text{and} \quad 
\omega_p(X^v_{|p})=X \in \LieH, \ \ \forall X^v_{|p} \in V_p P,
\end{align}
whose associated curvature can be found via the Cartan structure equation $\Omega =d\omega + \tfrac{1}{2}[\omega, \omega]\ \in \Omega^2_\text{tens}(P, \Ad)$. 
The~connection $\omega$ allows to define the standard covariant derivative 
$D = d\ + \rho_*(\omega): \Omega^\bullet_\text{tens}(P, \rho) \rarrow \Omega^{\bullet+1}_\text{tens}(P, \rho)$, which is s.t. $D\circ D = \rho_*(\Omega)$. 
Furthermore, we have the Bianchi identity: $D \Omega=0$.

By $\Omega^k_{\text{tens}}(P, \rho) \simeq \Omega^k(M) \otimes E$, a connection $\omega$ on $P$  induces a connection on the space $\Omega^\bullet(M)\otimes E$ by

\begin{minipage}[c]{0.45\linewidth}
 \begin{align*}
 D: \Omega^k_{\text{tens}}(P, \rho) &\rightarrow \Omega^{k+1}_{\text{tens}}(P, \rho) , \\
        \alpha &\mapsto D \alpha,
\end{align*}
\end{minipage} 
$\Rightarrow$
\begin{minipage}[c]{0.45\linewidth}
\begin{align*}
 \nabla: \Omega^k(M) \otimes E &\rightarrow \Omega^{k+1}(M) \otimes E , \\
        \t \alpha &\mapsto \nabla \t \alpha =[ \ \textunderscore \ , D\alpha].
\end{align*}
\end{minipage}

\medskip
\noindent In particular, for $k=0$ we have $\varphi \in \Omega^0_\text{tens}(P, \rho) \sim s \in \Gamma(E)$, and the above restricts to a connection on the space of sections of associated bundles:
\begin{equation}
\begin{aligned}
\nabla: \Gamma(E) &\rarrow \Omega^1(M)\otimes E , \\
s &\mapsto \nabla s =[\ \textunderscore \ , D \varphi].
\end{aligned}
\end{equation}
Finally, we observe that the space of  connections $\C$ is an affine space modelled on the vector space $\Omega^1_\text{tens}(P, \LieH)$, meaning that for $\omega \in \C$ and $\alpha \in \Omega^1_\text{tens}(P, \LieH)$, $\omega':=\omega+\alpha \in \C$.\footnote{Or, as is clear from the above defining properties, for $\omega, \omega' \in \C$, $\omega+\omega' \notin \C$  and $\omega'-\omega \in \Omega^1_\text{tens}(P, \LieH)$.}

This generalises to cocyclic bundle geometry.
One first defines the notion of \emph{cocyclic connection} 1-form $\varpi \in  \Omega^1(P, \LieG)$ by
\begin{equation}
\label{twisted-connection}
\begin{aligned}
R^*_h \varpi_{|ph} &= \Ad_{C(p,h)\-} \varpi_{|p} + C(p, h)\-d C(\ \, , h)_{|p},  \\[.5mm]
\varpi_p(X^v_{|p})&=\tfrac{d}{d\tau} C(p, h_\tau )\big|_{\tau=0}\defeq a(X, p) \in \LieG,
\end{aligned}
\end{equation}
where $X\defeq \tfrac{d}{d\tau}\, h_\tau \big|_{\tau=0} \in \LieH$.
The  \emph{cocyclic curvature} is a cocyclic tensorial 2-form defined by the Cartan structure equation $\b\Omega \defeq d\varpi + \tfrac{1}{2}[\varpi, \varpi]\ \in \Omega^2_\text{tens}(P, C)$. 
A cocyclic connection  allows to define  a \emph{cocyclic  covariant derivative}
$\b D = d\ + \varpi: \Omega^\bullet_\text{tens}(P, C) \rarrow \Omega^{\bullet+1}_\text{tens}(P, C)$, which is s.t. $\b D\circ \b D = \b\Omega$. 
The cocyclic curvature satisfies a Bianchi identity, $\b D \b\Omega=0$.

By the isomorphism $\Omega^k_{\text{tens}}(P, C) \simeq \Omega^k(M) \otimes E^C$, a cocyclic connection $\varpi$ on $P$  induces a cocyclic connection on the space $\Omega^\bullet(M)\otimes E^C$ by

\begin{minipage}[c]{0.45\linewidth}
 \begin{align*}
\b D: \Omega^k_{\text{tens}}(P, C) &\rightarrow \Omega^{k+1}_{\text{tens}}(P, C) , \\
        \alpha &\mapsto \b D \alpha,
\end{align*}
\end{minipage} 
$\Rightarrow$
\begin{minipage}[c]{0.45\linewidth}
\begin{align*}
 \b\nabla: \Omega^k(M) \otimes E^C &\rightarrow \Omega^{k+1}(M) \otimes E^C , \\
        \t \alpha &\mapsto \b \nabla \t \alpha =[ \ \_ \ , \b D\alpha].
\end{align*}
\end{minipage}

\medskip
\noindent For $k=0$ we have $\varphi \in \Omega^0_\text{tens}(P, C) \sim \t s \in \Gamma(E^C)$ and the above restricts to a connection on the space of sections of cocyclic associated bundles:
\begin{equation}
\begin{aligned}
\b \nabla: \Gamma(E^C) &\rarrow \Omega^1(M)\otimes E^C , \\
\t s &\mapsto \b \nabla \t s =[\ \_ \ , \b D  \varphi].
\end{aligned} 
\end{equation}
The space $\bar{\mathcal{C}}$ of cocyclic connections  is an affine space modeled on cocyclic tensorial 1-forms $\Omega^1_\text{tens}(P, C)$, so that for $\varpi \in \bar{\mathcal{C}}$ and $ \alpha \in \Omega^1_\text{tens}(P, C)$, 
$\varpi'=\varpi + \alpha$. 
\medskip

We observe that the exterior derivative $d$ is a covariant derivative for basic forms, $d: \Omega^\bullet_\text{basic}(P) \rarrow \Omega^{\bullet+1}_\text{basic}(P)$. 
This means that $\big( \Omega^\bullet_\text{basic}(P), d \big)$ is a subcomplex of the de Rham complex of $P$, called the \emph{basic subcomplex}, defining the \emph{equivariant cohomology} of $P$ \cite{Guillemin-Sternberg1999}. 
It is of special importance, notably in Physics, since it is isomorphic to the de Rham complex of the base $M$:  $\big(\Omega^\bullet_\text{basic}(P), d \big)\simeq \big(\Omega^\bullet(M), d \big)$.
As we shall see shortly, the dressing field method (DFM) is a way to associate to a form on $P$ a corresponding basic form.

\paragraph{Gauge transformations}

As observed above, the group of automorphisms $\Aut(P)$ acts on $\Omega^\bullet(P)$ by pullback: it~preserves the various spaces defined above, in particular tensorial forms $\Omega^\bullet_\text{tens}(P)$ and the space of connections $\mathcal C$. 
This stems from the functoriality of the definition of those spaces, automorphisms being special cases of morphisms in the bundle category. 
It follows that \emph{gauge transformations}, defined by the action of vertical automorphisms $\Aut_v(P)$, preserve these spaces too. 
Since $\Aut_v(P)\simeq \H$, gauge transformations may be expressed via elements of $\H$: 
We define the gauge transformation of $\alpha \in \Omega^\bullet(P)$ by $\alpha^\gamma(\mathfrak X) \defeq \Xi^*\alpha(\mathfrak X)$, for $\gamma\in \H \sim \Xi \in \Aut_v(P)$ and $\mathfrak X \in \Gamma(TP)$.
The explicit result is found by $\Xi^*\alpha(\mathfrak X) =\alpha(\Xi_*\mathfrak X)$ and via the well-known result
\begin{equation}
\begin{aligned}
  \label{pushforward_X}
  \Xi_* \mathfrak X_{|p} 
  = R_{\gamma(p)*} \mathfrak X_{|p} + \{ \gamma\- d\gamma_{|p}(\mathfrak X_{|p})\}^v_{|\Xi(p)}
  =R_{\gamma(p)*} \big( \mathfrak X_{|p} + \{  d\gamma \gamma\-_{|p}(\mathfrak X_{|p})\}^v_{|p} \big).
\end{aligned}
\end{equation}
From this follows that standard and cocyclic tensorial forms have homogeneous gauge transformations, controlled by their equivariance: they are ``gauge tensorial", hence their name,
\begin{equation}
\begin{aligned}
\label{GT-tens-forms}
& \alpha^\gamma = \rho(\gamma)\-\alpha, \quad \text{for } \ \ \alpha \in \Omega^\bullet_\text{tens}(P, \rho), \\[.2mm]
& \alpha^\gamma = C(\gamma)\-\alpha, \quad \text{for } \ \ \alpha \in \Omega^\bullet_\text{tens}(P, C),
\end{aligned}
\end{equation}
where we introduce the shorthand notation $C\big(\gamma(p)\big)\defeq C\big(p , \gamma(p)\big)$. 
The $\H$-transformations of standard and cocyclic connections are
\begin{equation}
\label{GT-connections}
\begin{aligned}
    & \omega^\gamma\defeq \Xi^* \omega =\gamma\- \omega\, \gamma + \gamma\- d \gamma, \quad \text{for } \ \ \omega \in \C, \\[.2mm]
    & \varpi^\gamma \defeq \Xi^* \varpi= C(\gamma)\-\varpi \, C(\gamma)+ C(\gamma)\- d C(\gamma), \quad \text{for } \ \ \varpi \in \bar{\mathcal{C}}.
\end{aligned}
\end{equation}
The first result is well-known, and easily read from  \eqref{pushforward_X}. For the proof of the second one see section 4.2 of \cite{Francois2019_II}.
The result \eqref{GT-tens-forms} gives, in particular, the $\H$-transformations of the curvatures $\Omega$ and $\b\Omega$ of the Ehresmann and cocyclic connections, as well as that of the (cocyclic) covariant derivative $\b D\alpha$ of a (cocyclic) form $\alpha$. 
It also gives the action of the gauge group $\H$ on its own elements: for $\eta, \gamma \in \H$ we have $\eta^\gamma =\gamma\- \eta \gamma$. 

The  action of $\Aut_v(P)\simeq \H$, giving \eqref{GT-tens-forms}-\eqref{GT-connections}, are called \emph{active} gauge transformations.
These are to be distinguished from \emph{passive} gauge transformations, the name given to the \emph{gluings} of local representatives on the base $M$, stemming from the local structure of $P$.

\paragraph{Local structure}

A principal bundle $P$ is  locally trivial,  i.e. for any open sets $U \subset M$ we have $P_{|U}\simeq U \times H$: the isomorphism being realised by a \emph{trivialisation} $\phi =(\pi, t) : P_{|U} \rarrow U \times H$, $p \mapsto \phi(p)=\big(\pi(p)\!=\!x,\ t(p)\big)$, which provides a ``bundle coordinate chart" on $P_{|U}$.
A trivialisation is compatible with
the structure group action, $\phi \circ R_h = R_h \circ \phi$, meaning that the fiber coordinate map $t$ is s.t. $R^*_h t =t h$, or $t(ph)=t(p)h$. Often $t$ itself is called a trivialisation.
A local section of $P$ is a map $\s : U \rarrow P_{|U}$, $x \mapsto \s(x)$ -- 
which allows to define a trivialisation of $P_{|U}$.
Any two local sections $\s$ and $\s'$ over $U$ -- or over overlapping  sets $U \cap U'$ -- are related via $\s'=\s g$, where $g$ is a transition function of $P$, that is a map $g: U \rarrow H$, $x \mapsto g(x)$. 

Given a local section $\s$, one may define the \emph{local representative} of  a form $\beta \in \Omega^\bullet(P)$ as   $b \defeq \s^* \beta \in \Omega^\bullet(U)$. 
In~particular, the local representative of a standard or cocyclic tensorial form is $a \defeq \s^*\alpha \in \Omega^\bullet(U, V)$, that of Ehresmann and cocyclic connections are respectively $A\defeq \s^* \omega \in \Omega^\bullet(U, \LieH)$ and $\b A \defeq \s^* \varpi \in \Omega^\bullet(U, \LieG)$.
One may also define the pullback of a 1-cocycle $\mathsf C\defeq \s^* C : U \times H \rarrow G$, $(x, h) \mapsto \mathsf C(x, h) =C\big(\s(x), h\big)$. It  allows to define the map $\mathsf C(g)\defeq  C(\s, g) : U \rarrow G$, which are \emph{cocyclic transition function} -- i.e. none other than the transition functions of cocyclic associated bundles $E^C$. 
Under change of local section $\s'=\s g$, i.e. of bundle coordinates, the local representatives $a'\defeq {\s'}^* a$, $A'\defeq {\s'}^* \omega$ and $\b A' \defeq {\s'}^* \varpi$  satisfy \emph{gluing relations}
\begin{equation}
\begin{aligned}
\label{PassiveGT}
a'&=\rho(g)\- a, \qquad  A' = g\- A g+ g\-dg, \\
a'&=\mathsf C(g)\- a, \qquad  \b A' = \mathsf C(g)\- \b A \mathsf C(g)+ \mathsf C(g)\-d\mathsf C(g).
\end{aligned}
\end{equation}
These are also known as \emph{passive gauge transformations}. 
These must be conceptually distinguished from \emph{local active gauge transformations}: Let us define the local gauge group $\Hl \defeq \{ \upgamma \defeq \s^*\gamma: U \rarrow H\, |\, \gamma \in \H, \text{ and } \upeta^\upgamma =\upgamma\- \upeta \upgamma 
\}$, and consider the map
$\mathsf C(\upgamma)\defeq C(\s, \upgamma): U\rarrow G$. 
Now, through a given section $\s$, the local representatives of the active gauge transformations \eqref{GT-tens-forms}-\eqref{GT-connections} are
\begin{equation}
\begin{aligned}
\label{ActiveGT}
a'&=\rho(\upgamma)\- a, \qquad  A' = \upgamma\- A \upgamma+ \upgamma\-d\upgamma, \\
a'&=\mathsf C(\upgamma)\- a, \qquad  \b A' = \mathsf C(\upgamma)\- \b A \mathsf C(\upgamma)+ \mathsf C(\upgamma)\-d\mathsf C(\upgamma).
\end{aligned}
\end{equation}
These are the local active gauge transformations, formally indistinguishable from the passive  gauge transformations \eqref{PassiveGT}, but conceptually as distinct as coordinate changes and diffeomorphisms in GR. 

\subsection{The dressing field method: basics}
\label{The dressing field method: basics}

The dressing field method (DFM) has been conceived as a tool of gauge symmetry reduction in Gauge Field Theory \cite{GaugeInvCompFields, Francois2014, Francois2021} (see also \cite{Zajac2023, Riello2024}) whose foundational implications were first explored in \cite{Francois2018, Berghofer-et-al2023} (see also \cite{Teh-et-al2021}). 
It~is~extended to general-relativistic theories, whose covariance group is the  group of diffeomorphisms, in \cite{Francois2023-a}. 
The~most general and synthetic mathematical and conceptual presentation, for general-relativistic Gauge Field Theory, is to be found in \cite{JTF-Ravera2024gRGFT}. 

Mathematically, it is a result about realisation of subbundles and of basic, or partially basic, forms on a bundle. 
In that respect it is close to the well-known  bundle reduction theorem, a standard result  of bundle geometry \cite{Trautman, Westenholz, Sternberg}.
A key difference is that, in the DFM the  dressing map, or \emph{dressing field}, is assumed to have a functional dependence on one or several forms of $P$. In physical terms, the dressing field is a functional of the degrees of freedom (d.o.f.) of the theory under consideration. 
The DFM thus has a straightforward \emph{relational} interpretation:\footnote{Which sets it apart from gauge-fixing, a notion, in Gauge Field Theory, with which it is often  conflated. See \cite{Francois-Berghofer2024}.}
the basic forms produced, or dressed fields, are naturally seen as relational variables as articulated precisely in \cite{JTF-Ravera2024c, Francois2023-a, JTF-Ravera2024gRGFT} -- and close to the views expressed e.g. in \cite{Rovelli1991, Rovelli2002, Rovelli2004, Rovelli2014, Tamborino2012}.

We shall now give a synthetic review of the  fundamentals of the DFM, as it is essential to our subsequent relational reformulation of non-relativistic classical and quantum mechanics.

\paragraph{Dressing field and basic dressed fields}

Suppose there exists a subgroup $K \subseteq H$ of the structure group, to which corresponds a subgroup $\K \subset \H$ of the gauge group, which is isomorphic to the subgroup $\Aut_v(P)^K \subset \Aut_v(P)$ of the group of vertical automorphisms.
Consider further a Lie group $G$  s.t.   $K\subseteq G \subseteq H$. 
A~$K$-\emph{dressing field} is  defined as
\begin{equation}
\begin{aligned}
\label{dressing-field-def}
u: P \rarrow G, \quad 
&\text{s.t.} \quad 
u^\kappa \defeq\, \Xi^* u= \kappa\- u, \quad \Xi \in \Aut_v(P)^K\sim \kappa \in \K, \\
&\text{or s.t.} \quad
R^*_ku \defeq\,k\- u,  \quad k\in K.
\end{aligned}
\end{equation}
We denote the space of $G$-valued $K$-dressing fields on $P$ by $\D r[K, G]$: $K$ is called the equivariance group of the dressing field and $G$ its target group. 
In the DFM, one  considers more generally dressing fields having a functional dependence on objects of $P$ (typically forms):
\begin{equation}
\begin{aligned}
\label{field-dep-dressing-field}
\u: \Omega^\bullet(P) &\rarrow \D r[K,G], &&\\
         \alpha &\mapsto 
\Big\{\ \u[\alpha]: P \rarrow G,  &&\text{s.t.} \quad
\u[\alpha]^\kappa\defeq \Xi^*\u[\alpha] =\kappa\- \u[\alpha]\
\big.,  \\
&\phantom{bb} &&\text{or s.t.} \quad
R^*_k\, \u[\alpha] \defeq\,k\- \u[\alpha],  \quad k\in K\ \Big\}.
\end{aligned}
\end{equation}
Alternatively, the property can be defined as  -- or required compatible with -- an equivariance w.r.t. transformation of the functional argument:
$\u[\alpha]^\kappa\defeq \u[\alpha^\kappa] = \u[\Xi^* \alpha] = \kappa\- \u[\alpha]$.
In physical applications, dynamical d.o.f. are represented by (or embedded into) objects on $P$, so one may understand dressing fields \eqref{field-dep-dressing-field} as being  determined by physical (sub)systems.
We may also observe that for $K=G=H$ a dressing field \eqref{dressing-field-def} is equivalent to a trivialisation $t\defeq u\-$, i.e. provides a bundle coordinate chart. 
Therefore, dressing fields \eqref{field-dep-dressing-field} provide  bundle coordinatisations via physical degrees of freedom.

Given such a $K$-dressing field, one may define the dressing map
\begin{equation}
\begin{aligned}
 \label{dressing-map}
 f_\u: P &\rarrow P'\subset P,  \qquad \qquad \qquad \text{s.t.} \quad f_\u \circ R_k = f_\u,
 \quad \text{and}\quad 
 f_\u \circ \Xi = f_\u \quad \text{for}\quad \Xi \in \Aut_v(P)^K\sim \K,\\
      p &\mapsto f_\u(p)\defeq p\u(p),
\end{aligned}
\end{equation}
where $P'=P/K$ is a quotient space since $f_\u$ is constant along the action of $K$.
It allows to define, for any form $\beta \in \Omega^\bullet(P)$, its \emph{dressed counterpart} 
\begin{align}
\label{dressed-forms-1}
\beta^\u\defeq  f_\u^* \beta\quad  \in \Omega^\bullet_\text{$K$-basic}(P).
\end{align}
It is $K$-basic since it has both trivial $K$-equivariance 
and is $K$-horizontal, so that it is $\K\simeq \Aut_v(P)^K$-invariant,  
$(\beta^\u)^\kappa \defeq \Xi^* (\beta^\u)=\Xi^* f^*_\u\beta = f_\u^*\beta \rdefeq \beta^\u$.
A dressed form $\beta^\u$ therefore naturally ``lives" on the quotient space $P/K$. 
In~particular, the dressings of an Ehresmann connection $\omega$ and standard tensorial form $\alpha$ have explicit expressions 
\begin{align}
\label{dressed-forms-2}
\omega^\u\defeq  \u\- \omega\, \u + \u\-d\u \quad \text{ and } \quad \alpha^\u \defeq \rho(\u)\-\alpha.
\end{align}
The dressing of  cocyclic connections and cocyclic tensorial forms is
\begin{align}
\label{twisted-dressed-forms}
\varpi^\u\defeq  C(\u)\- \varpi\, C(\u) + C(\u)\-dC(\u) \quad \text{ and } \quad \alpha^\u \defeq C(\u)\-\alpha,
\end{align}
where, from the 1-cocycle $C:P\times H \rarrow \mathsf G$ (where we distinguish the target group $\mathsf G$ of $C$ from that $G\subseteq H$ of the dressing field), one  defines the $K$-\emph{cocyclic dressing field} 
\begin{equation}
\begin{aligned}
\label{twisted-dressing}
C(\u)\defeq C(\ \,, \u): P &\rarrow \mathsf G, 
\hspace{3.5cm} \text{s.t.} 
\quad R^*_k C(\u)=C(\, \ , k)\-C(\u).
\\
p& \mapsto C\big(\u(p)\big)\defeq C\big(p, \u(p)\big),
\end{aligned}
\end{equation}
whose equivariance is secured by the 1-cocycle property \eqref{1-cocycle} and \eqref{prop-cocycle}:
\begin{align}
\label{cocyclic-dressing}
    C\big( \u(pk)\big)=C\big(pk, \u(pk)\big)
=C\big(pk, k\-\u(p) \big)
=C\big(pk, k\- \big)\, C\big( pkk\-, \u(p) \big)
=C\big(p,k \big)\- C\big(p, \u(p)\big).
\end{align}
Similarly, we find the $C(\u)^\kappa =C(\kappa)\- C(\u)$, from which, together with \eqref{GT-tens-forms}-\eqref{GT-connections}, the $\K\simeq \Aut_v(P)^K$-invariance of \eqref{twisted-dressed-forms} is easily checked. 

From the above, comparing \eqref{GT-tens-forms}-\eqref{GT-connections} and \eqref{dressed-forms-2}-\eqref{twisted-dressed-forms}, we derive the ``\emph{rule of thumb}" of the DFM: 
To find the dressing of a form $\beta \in \Omega^\bullet(P)$, on account of the formal similarity of the action of the dressing map $f_\u$ and a vertical automorphism $\Aut_v(P)^K$, first compute its $\K$-gauge transformation $\beta^\kappa \defeq \Xi^*\beta$ in terms of $\kappa \in \K \sim \Xi \in \Aut_v(P)^K$, and then simply substitute  $\kappa \rarrow \u$ in this expression to get the dressed, $K$-basic, form $\beta^\u$.

Let us stress the following important fact: It should be clear from its definition that a dressing field \emph{is not} a gauge group element, $\u \notin \K$, and that a dressing map $f_\u$ is \emph{not} a vertical automorphism.
So that despite a superficial formal resemblance with \eqref{GT-tens-forms}-\eqref{GT-connections}, dressed fields \eqref{dressed-forms-2}-\eqref{twisted-dressed-forms} are \emph{not} gauge transformations: dressed fields $\beta^\u$ are not points in the $\K$-gauge orbits $\O_\K[\beta]$ of the form $\beta$ -- and in particular must not be conflated with a gauge-fixing of $\beta$. 
For example, a dressed connection is no more a connection, $\omega^u \notin \C$.

We observe that a dressing field induces a (partial, i.e. Lie$K$-valued) flat Ehresmann connection $\omega_0\defeq -d\u \u\-$. 

As we stressed above, since dressings fields \eqref{field-dep-dressing-field} depend on the dynamical d.o.f. at play, dressed forms \eqref{dressed-forms-1}-\eqref{dressed-forms-2}-\eqref{twisted-dressed-forms} have a natural interpretation as \emph{relational variables}:
They  encode the $\K$-gauge-invariant, physical, relational d.o.f. among the dynamical d.o.f. describing the system.

\paragraph{Residual transformations of the 1st kind}

Being $K$-basic, i.e. $\K$-invariant, the dressed forms $\beta^\u$ may be expected to display residual transformations under what remains of the structure and gauge groups. 
For these to be meaningful, it must be the case that $K$ is a normal subgroup of the structure group, $K \triangleleft H$, so that their quotient is itself a group $H/K=J$. 
Then $P'=P/K$ is a $J$-principal subbundle of $P$, on which dressed form $\beta^\u$ ``live", with isomorphic vertical automorphism and gauge groups $\Aut(P')=\Aut_v(P)^J\simeq \J$. 
The structure group of $P$ can then be written (assuming the trivial homomorphism $\phi \,:\, J \to H$, $\phi(j)=j \in H$) as $H=J \ltimes K$, correspondingly its vertical automorphism groups is
$\Aut_v(P)=\Aut_v(P)^J \ltimes \Aut_v(P)^K$, and  its gauge group is $\H=\J \ltimes \K$.

Since the $J$-equivariance and $J$-horizontality of $\beta$ are already known by assumption, those of $\beta^\u$  depend only on the $J$-equivariance of $\u$: $R^*_j\u$. 
Alternatively, one may say that the $\J$-transformation of $\beta$ being known, to get that of $\beta^\u$ one needs only to find the $\J$-transformation of $\u$: $\u^\eta$. 
Finding the residual $J$-equivariance and/or $\J$-transformation of $\u$ will depend on the specifics of the situation under consideration. 

As an illustrative example, consider the case where
$R^*_j \u =j\-\u j$ for $j \in J$, so that $\u^\eta
=\eta\-\u\,\eta$ for $\eta \in \J$.
The~residual $\J$-transformations of the dressed forms \eqref{dressed-forms-2} are then easily found to be: 
$(\omega^\u)^\eta=\eta\- \omega^\u\, \eta + \eta\-d\eta$ and $(\alpha^\u)^\eta=\rho(\eta)\-\alpha^\u$.
They behave as standard connections and tensorial forms on $P'\subset P$. 
See \cite{Berghofer-et-al2023, Attard-Francois2016_I} for other cases.

\paragraph{Residual transformations of the 2nd kind}

The dressed forms may  exhibit residual transformations resulting from a possible ambiguity in the choice of dressing field.
Two dressings $\u, \u' \in \D r[K, G]$ may a priori be related by $\u'=\u\xi$, where 
$\xi \in \G \defeq \left\{\,   \xi:P \rarrow G\, |\,  R^*_k \xi = \xi  \, \right\}$.
We may write the $\G$-action on a dressing by $\u^\xi=\u\xi$. 
By~definition, $\G$ has no action on $\Omega^\bullet(P)$, so  $\beta^\xi=\beta$. 
The $\G$-action  on  dressed forms is thus easily found via
$(\beta^\u)^\xi \defeq (\beta^\xi)^{\u^\xi}=\beta^{\u\xi}$. 
For example, the residual $\G$-transformations of \eqref{dressed-forms-2} are found to be:
$(\omega^\u)^\xi \defeq (\omega^\xi)^{\u^\xi}=\omega^{\u\xi}=\xi\- \omega^\u \xi + \xi\-d\xi$ and $(\alpha^\u)^\xi \defeq(\alpha^\xi)^{\u^\xi}=\alpha^{\u\xi}=\rho(\xi\-) \alpha^\u$. 

For cocyclic dressing fields we get $C(\u)^\xi  \defeq C\big(\u^\xi \big)$, which is $C\big(p, \u(p)\xi(p)\big)
 = C\big(p, \u(p)\big)\,  C\big(p \u(p), \xi(p)\big) 
 = C\big(p, \u(p)\big)\, \, C\big(f_\u(p), \xi(p)\big)$, where we may write $f_\u (p)=p' \in P' \subset P$. This result we may denote by
\begin{align}
 \label{Ambig-twisted-dress-field}
 C(\u)^\xi  = C\big(\u^\xi \big) = C(\u)\, C(\xi).
\end{align}
So,  the residual $\G$-transformations of cocyclic dressed forms \eqref{twisted-dressed-forms} are easily found to be:
\begin{align}
\label{resid-trsf-dressed-twisted-forms}
  (\varpi^\u)^\xi = C(\xi)\- \varpi^\u C(\xi) + C(\xi)\- dC(\xi)
  \quad \text{ and } \quad (\alpha^\u)^\xi = C(\xi)\- \alpha^\u.
\end{align}
Transformations like \eqref{resid-trsf-dressed-twisted-forms} are the basis of the frame change covariance of the relational version of our geometric formulation of QM, both of which we now describe.

\section{Bundle geometric non-relativistic Mechanics and Quantum Mechanics}
\label{Bundle geometric non-relativistic Mechanics and Quantum Mechanics}

As said in the introduction, until now, the main applications of the DFM have been in classical general-relativistic Gauge Field Theory \cite{JTF-Ravera2024gRGFT},  the natural relational interpretation 
of the former
fitting the conceptual foundation of the latter \cite{JTF-Ravera2024c}.
In the following, the DFM will be applied for the first time in a quantum context, allowing a manifestly relational 
formulation of non-relativistic many-particle quantum mechanics.
This is made possible by first providing a bundle geometric description of both non-relativistic Mechanics and then Quantum Mechanics. 
 
\subsection{Bundle geometry of configuration space-time $\mC$}
\label{Bundle geometry of configuration space-time}

Our  geometric formulation of non-relativistic QM is based on the bundle geometry of the configuration space-time of $N$  classical structureless, point-like, particles, which  generalises the description of Galilean space-time.

The~configuration space-time  is a principal bundle $\mC \xrightarrow{\pi} \T$ over the $1$-dimensional non-compact manifold $\T$, representing the Newtonian  universal time line. 
The fiber over an instant $t\in\T$ is $\mC_{|t}\defeq \pi\-(t)= \EE^{3N}$, i.e. the configuration space of $N$ classical structureless point particles. 
A point $p=(p_1, \ldots, p_N) \in \mC_{|t} \subset \mC$, $\pi(p)=t \in \T$, is the instantaneous configuration of the $N$ particles. 
The structure group of $\mC$ is $H=\RR^{3N} \defeq \underbrace{\RR^3 \times \ldots \times \RR^3}_{N \text{ times}}$. 
\vspace{-3.5mm}

\noindent Its~right~action is  
\begin{equation}
\begin{aligned}
\label{right-action}
\mC \times H &\rarrow \mC,  \\
(p, X) &\mapsto R_Xp = p+X,  \qquad \text{ s.t. } \quad \pi \circ R_X = \pi.
\end{aligned}
\end{equation}
We observe that this action reflects the fact that $\EE^{3N}$ is an affine space over $\RR^{3N}$, because the Euclidean plane $\EE^3$ is affine over $\RR^3$.
Since $H$ is (additive) Abelian, we have 
$R_{X'}R_Xp=p+ X+ X' = R_X R_{X'}p$, 
i.e. $R_{X+X'}=R_{X'+X}$. 

As usual, the linearisation of the right action of $H$ defines the canonical vertical subbundle $V\mC \subset T\mC$, and a fundamental vertical vector  is
\begin{align}
\label{fund-vectors}
\chi^v_{|p} \defeq \tfrac{d}{d\tau}\, R_{X_\tau}p\ \big|_{\tau=0} = p+\chi \ \in V_p\mC , \quad 
\text{with } \quad 
\tfrac{d}{d\tau}\,X_\tau \, \big|_{\tau=0} \defeq \chi \in \LieH,
\end{align}
so that a fundamental vector field is $\chi^v \in \Gamma(V\mC)$. 
Because $H$ is Abelian, these are right invariant:
$R_{X*}\, \chi^v_{|p}=\chi^v_{|p+X}$.
Eq. \eqref{fund-vectors} defines the ``verticality map" $|^v : \LieH \rarrow \Gamma(V\mC) $, applying to any $\LieH$-valued (0-)form on $\mC$. 
As we shall see shortly, contrary to the case of Gauge Field Theory, vertical ``motion" along the fibers of $\mC$, with ``velocity" in $\Gamma(V\mC)$, is not always unphysical, or ``pure gauge": it  also contains genuine physical changes within the set of particles.
Of~course, the true \emph{physical} dynamics of the latter unfolds along a curve in $\mC$ that projects onto a curve  in $\T$. 

Over an open set $\U \subset \T$, i.e. a time interval, a local trivialisation is a map
\begin{equation}
\begin{aligned}
 \label{loc_triv_config}   
 \phi: \mC_{|\U} &\rarrow \U \times H, \\
 p &\mapsto \phi(p)\defeq \big( \pi(p), h(p) \big) = (t,x) = \big(t, (x_1, \ldots, x_N) \big).
\end{aligned}
\end{equation}
The fiber coordinate $h(p)=x$ may also be written 
$x=\{x_i\}_{i\in \{1, \ldots, N\}}=\{{x_i}^a e_a\}$ where $\{e_a\}$ is a basis of $\RR^3$, meaning that $x=\{x_i\}$ is an intrinsic vector of $\RR^{3N}$, independent of a choice of basis.
The trivialisation is s.t. $\phi \circ R_X = R_X \circ \phi$, i.e. $\phi(p+X)=\phi(p)+X$, meaning that the fiber coordinate satisfies
$ R^*_X h = h+X$, 
i.e. $h(p+X) = h(p) + X$. 
A local (trivialising) section is
$\s : \U \subset \T \rarrow \mC_{|\U} \simeq \U \times H$, 
$t \mapsto \s(t)= \big(t, x(t) \big)$, i.e. it gives the graph of the trajectories of the particles described by $p=\s(t)$. 
Two local sections are related by $\s'=R_g\, \s=\s+g$, where $g: \U \rarrow H$, $t \mapsto g(t)$ is a transition function of $\mC$. 
These are involved in \emph{gluings} of local representatives on $\T$ of forms on $\mC$, i.e. in \emph{passive gauge transformations}. 
Again, beware that here the latter may relate two physically distinct particle configurations. 
Since $\T$ is contractible, $\mC$ is \emph{globally trivial}, $\mC \simeq\T \times H$.

Under the action $R_X : \mC_{|t} \rarrow \mC_{|t}$, $p=(p_1, \ldots, p_N)\mapsto p+X=(p_1 +X_1, \ldots, p_N+X_N)$, of the structure group $H=\RR^{3N}$, each particle position is shifted by a different vector $X_i \in \RR^3$. 
It is clear that any  configuration of particles is left invariant by the subgroup
\vspace{-4mm}
\begin{align}
H_\Delta =\RR^{3N}_\Delta \defeq \text{diag}\RR^{3N} \simeq \RR^3, \quad \text{  whose elements are } \quad X=(\overbrace{X, \ldots , X}^{N \text{ times}}). 
\end{align}
Now, since any 
subgroup of an Abelian group is \emph{normal}, we have $H_\Delta \triangleleft H$, so that $H/H_\Delta$ is an Abelian group. 
The~structure group  thus decomposes as
\begin{align}
\label{semi-dir-H}
    H = H/H_\Delta \ltimes H_\Delta
\rdefeq H_\text{int} \ltimes H_\text{ext},
\end{align}
where the nomenclature is chosen to reflect the fact that $H_\text{int}$ implements physical changes in the configuration of particles as seen from any particle within it, while $H_\text{ext}$ only reflects a global change of the whole configuration as seen from a viewpoint external to the configuration.\footnote{So, $H_\text{ext}=H_\Delta$ can be seen as inducing unphysical transformation of the configuration. Yet, it is not so for the corresponding gauge subgroup, $\H_{\text{ext}}$. See below.
}

\begin{figure}[H]
\begin{center}
\includegraphics[width=0.6\textwidth]{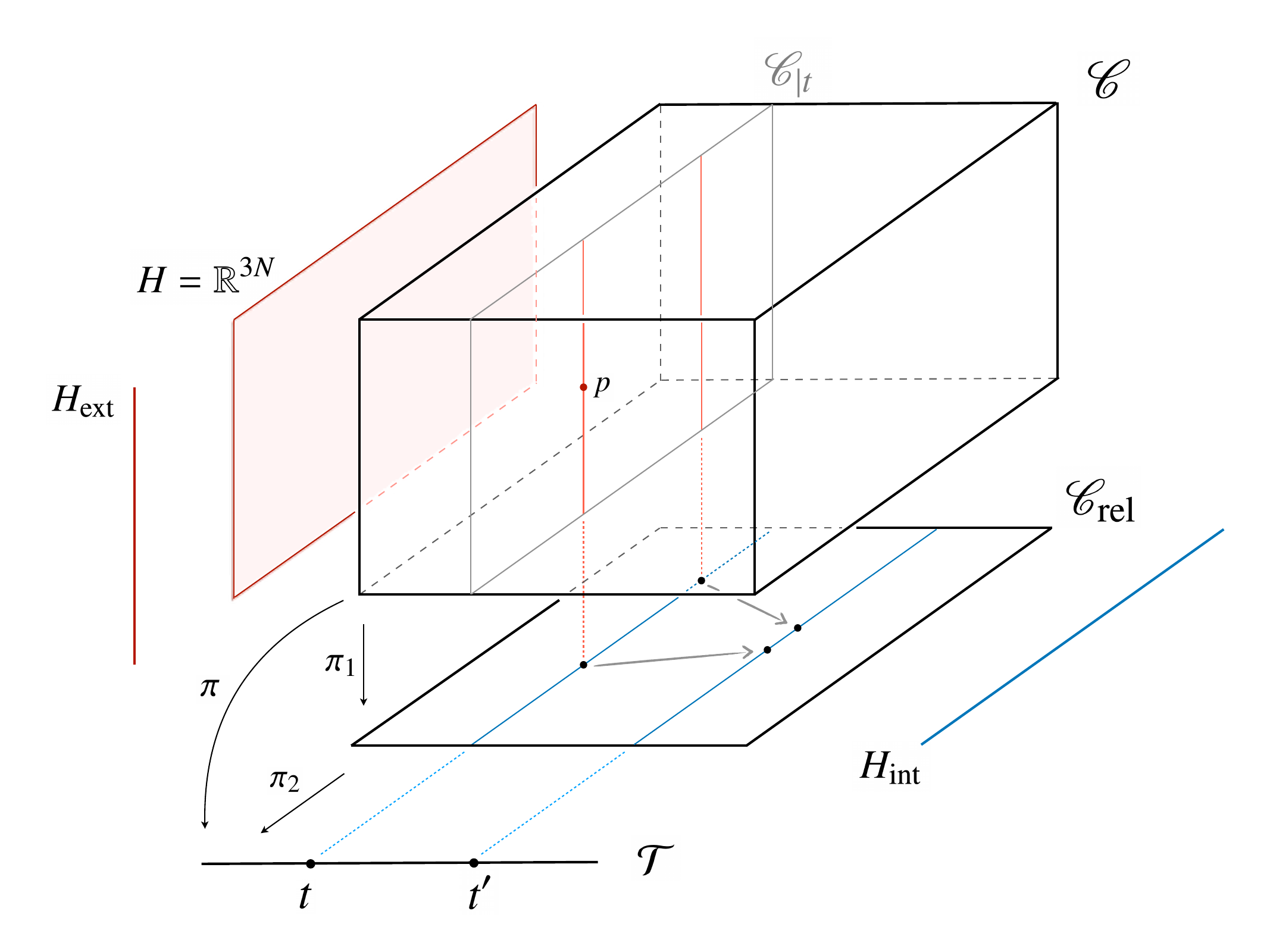}
\caption{The $2$-fibration structure of the configuration space-time  bundle $\mC$ of $N$ classical particles.}
\label{2-bundle-fig}
\end{center}
\end{figure}

The action of $H_\text{int}$ maps a  configuration to a physically distinct configuration, where the relative positions of the particles are different. 
The decomposition \eqref{semi-dir-H} implies that the configuration space-time bundle  is a $2$-fibration: 
\begin{equation}
\label{2-fibration}
\begin{aligned}
&\hspace{3cm }\mC \xrightarrow{H_\text{ext}}
\mC_\text{rel}\xrightarrow{H_\text{int}} \T, \\[2mm]
&\text{or} \hspace{-10mm}
\begin{tikzcd}[column sep=large, ampersand replacement=\&]
\ \&  \mC   
\arrow[rr, bend right=15,  start anchor={[yshift=-.5ex]}, end anchor={[yshift=-.5ex]}, "\pi"'] 
\arrow[r, "\pi_1"  ]         
\&  \mC_{\text{rel}} \defeq\mC /H_\text{ext}  
\arrow[r, "\pi_2"]  
\& \T,
\end{tikzcd}\qquad  \text{with} \quad \pi_2 \circ \pi_1 = \pi,
\end{aligned}
\end{equation}
where the quotient subbundle $\mC_\text{rel} \defeq\mC /H_\text{ext}  \xrightarrow{H_\text{int}} \T$ is the \emph{relational configuration space-time bundle} -- here realised as a quotient -- encoding the internal relative dynamics of the $N$ particles system. 
Fig. \ref{2-bundle-fig} sketches the $2$-fibration structure of $\mC$.

The group of automorphism of $\mC$ is 
$\Aut(\mC)\defeq  \big\{ \Xi \in \Diff(\mC)\, |\, \Xi \circ R_X = R_X \circ \Xi \big\}$, and projects onto $\Diff(\T)$, i.e. induces diffeomorphisms of the  time line. 
Its maximal normal subgroup is the group of vertical automorphisms 
$\Aut_v(\mC)\defeq \big\{\Xi \in \Aut(\mC)\, |\, \pi\circ \Xi=\pi \big\}$,  inducing the identity transformation on $\T$.

The latter is isomorphic to the gauge group
$\H \defeq \big\{\, \bs X: \mC \rarrow H\, |\, R^*_X  \bs X = \text{Conj}(-X) \bs X = \bs X \, \big\}$. 
The isomorphism being  $\Xi(p)=R_{\bs X(p)} p = p+ \bs X(p)$. 
It is noteworthy that the elements of the gauge group are \emph{basic}, being horizontal as usual (as 0-forms on $\mC$) and \emph{invariant} due to $H$ being Abelian. 
So, we have that $\H\simeq \Omega^0_\text{basic}(\mC, H)\simeq \Omega^0(\T,  H)$, i.e.  any $\bs X \in \H$ arises from some $\t{\bs X} : \T \rarrow H$ s.t. $\bs X =\pi^* \t{\bs X}$. 
Elements of the gauge group $\H$ thus depend only on $t\in \T$. 
The restriction of $\H$ to a trivial patch $\mC_{|\U} \subset \mC$ thus carries the same information as the local gauge group $\H_\text{loc}\defeq\big\{\t{\bs X} \defeq \s^*\bs X: \U\rarrow H\, |\,  
\t{\bs X}^{\t{\bs X}'}=\t{\bs X} \big\} \simeq \Omega^0(\U,  H)\rdefeq\Omega^0(\T,  H)_{|\U}$. 
The definition of $\H_\text{loc}$ is actually  independent of the choice of local section $\s$,  its elements having trivial gluings. 
The use of the same notation $\t{\bs X}$ for elements of $\H_\text{loc}$ and $\Omega^0(\T,  H)$ is meant to highlight that they are the same objects.

We observe that $\H$ includes \emph{Galilean boosts} via elements of the form 
$\bs X(t)=v_\text{ext} t$, where $v_\text{ext} \in H_\text{ext}$ is constant and interpretable as the relative velocity between the configuration $p$ of $N$ particles and an external observer $\O_\text{ext}$.  
The gauge group also encodes constant relative accelerations $a_\text{ext} \in H_\text{ext}$ between $p$ and $\O_\text{ext}$ via  $\bs X(t)=\tfrac{1}{2}a_\text{ext} t^2$, whereby inertial forces manifest. 
More generally, the gauge group 
$\H\simeq \Omega^0_\text{basic}(\mC, H)\simeq \Omega^0(\T,  H)$ encodes precisely what is known as ``\emph{extended Galilean transformations}": 
Indeed, the action $\Xi: \mC \rarrow \mC$, $p \mapsto \Xi(p)=p+\bs X(p)$, of a vertical automorphism $\Xi \in \Aut_v(\mC)\simeq \bs X\in \H$ is, in bundle coordinates, 
$(t, x)  \mapsto (t', x')=\big(t, x + \t {\bs X}(t) \big)$. 
The latter is precisely the usual definition of extended Galilean transformations as described e.g. in \cite{Greenberger1979}. 

The~characteristic 
SES of the  bundle $\mC$ is then
\begin{align}
\label{SES-mC}
\id_\mC\rarrow \Aut_v(\mC) \simeq \H \xrightarrow{\triangleleft} \Aut(\mC)  \xrightarrow{}  \Diff(\T) \rarrow  \id_\T.
\end{align}
Given that the structure group $H$ decomposes as \eqref{semi-dir-H}, the group of vertical automorphisms and the gauge group decompose correspondingly: $\Aut_v(\mC)=\Aut_v(\mC)^{H_\text{int}} \ltimes \Aut_v(\mC)^{H_\text{ext}}  \simeq \H = \H_\text{int} \ltimes \H_\text{ext}$,
with $\Aut_v(\mC)^{H_\text{int}}=\Aut_v(\mC_\text{rel})$. 
So we have the SES characterizing the relational configuration space-time bundle,
\begin{align}
\label{SES-mC-rel}
\id_{\mC_\text{rel}}\rarrow \Aut_v(\mC_\text{rel}) \simeq \H_\text{int} \xrightarrow{\triangleleft} \Aut(\mC_\text{rel})  \xrightarrow{}  \Diff(\T) \rarrow  \id_\T.
\end{align}


The action of $\Aut_v(\mC)\simeq \H$  on $\beta\in \Omega^\bullet(\mC)$ defines the \emph{active gauge transformations}  
$\beta^{\bs X} \defeq \Xi^* \beta$, geometrically found via $ \Xi^* \beta (\mathfrak X) = \beta( \Xi_* \mathfrak X)$, where $\mathfrak X \in \Gamma(T\mC)$ is a generic vector field on $\mC$ and, as special case of \eqref{pushforward_X},
\begin{align}
 \label{pushforward-mC}   
 \Xi_* \mathfrak X_{|p} = R_{\bs X(p)*} \mathfrak{X}_{|p} + \big\{d\bs X_{|p}(\mathfrak{X}_{|p}) \big\}^v_{|\Xi(p)}
 =R_{\bs X(p)*} \mathfrak{X}_{|p} + \big\{d\bs X_{|p}(\mathfrak{X}_{|p}) \big\}^v_{|p},
\end{align}
where the second equality arises because of the right invariance of fundamental vector fields. 
For example, for  an Ehresmann connection $\omega \in \Omega^1_\text{eq}(\mC, \Ad)$, s.t. $R^*_X\, \omega =\omega$ and $\omega(\chi^v)=\chi \in \LieH$, we have 
\begin{align}
\label{GT-standard-mC}
\omega^{\bs X} \defeq \Xi^* \omega = \omega + d\bs X,
\end{align}
which is an Abelian version of the first line of \eqref{GT-connections}.\footnote{
 We may observe that, as is standard in bundle geometry, the kernel of a connection defines an horizontal subbundle $\ker \omega \defeq H\mC$ s.t. $T_p\mC = V_p\mC \oplus H_p\mC$ at any $p\in \mC$: 
Given a curve $c(\tau) \in \T$ s.t. $c(0)=t$ and $c(1)=t'$,
it allows to define the horizontal lift $c^h(\tau) \in \mC$ s.t. $\tfrac{d}{d\tau}\, c^h(\tau)\, \big|_{\tau=0} \in \Gamma(H\mC)$ is a horizontal vector field.
In turn, this allows to define
the parallel transport map $c^h_\tau: \mC_t \rarrow \mC_{t'}$, $p \mapsto c^h_1(p)=p'$, which prescribes a standard of identification across distinct fibers, i.e. determines (for each choice of connection $\omega$) what it means for a configuration of particles $p$ to remains  ``the same" across time.}

Cocyclic geometry on $\mC$ is  defined via a 1-cocycle 
\begin{equation}
\label{1-cocycle-Abelian-0}
\begin{aligned}
C: \mC \times H &\rarrow G,  \\
(p, X) &\mapsto C(p, X), \quad \text{s.t. } \quad 
C(p, X+X')= C(p, X)\cdot C(p+X, X'),
\end{aligned}
\end{equation}
with corresponding linear 1-cocyle   $a(\chi, p) \defeq \tfrac{d}{d\tau} \, C(p, X_\tau) \,\big|_{\tau=0}$, 
\begin{equation}
\begin{aligned}
a: \mC \times \LieH &\rarrow \LieG, \\
    (p, \chi)    &\mapsto a(\chi, p), 
    \quad \text{s.t. }\ \  
    \chi^v \cdot a(\chi', p) - {\chi'}^v \cdot a(\chi, p) + [a(\chi, p),\  a(\chi', p)]_\LieG= 0. 
\end{aligned}
\end{equation}
So that cocyclic tensorial forms $\alpha \in \Omega^\bullet_\text{tens}(\mC, C)$ and cocyclic connections $\varpi \in \Omega^\bullet_\text{eq}(\mC, C)$, s.t. 
\begin{align}
    R^*_X\, \varpi_{|p+X}= C(p, X)\- \varpi_{|p} C(p, X) + C(p, X)\-dC(\ \, , X)_{|p} \quad \text{ and } \quad \varpi_{|p}(\chi^v_{|p}) =a(\chi, p),
\end{align}
have gauge transformations given by
\begin{align}
 \label{GT-twisted-mC}
 \alpha^{\bs X} = C(\bs X)\- \alpha \quad 
 \text{ and } \quad 
 \varpi^{\bs X} = C(\bs X)\-\varpi\, C(\bs X) + C(\bs X)\-dC(\bs X),
\end{align}
with the map $C(\bs X): \mC \rarrow G$ defined by $C\big(\bs X(p)\big)=C\big(p, \bs X(p)\big)$, being special cases of \eqref{GT-tens-forms}-\eqref{GT-connections}.

\subsection{Classical mechanics on $\mC$: The action as a cocyclic  connection}
\label{Classical mechanics on C}

The dynamics of the configuration of $N$ classical particles is given by the Lagrangian, which we take to be in general a \emph{cocyclic tensorial} 1-form $L \in \Omega^1_\text{tens}(\mC, \mathrm c)$ where the 1-cocycle $\mathrm c$ takes values in $\big(\Omega^1(\mC), +\big)$ seen as an additive Abelian group: It thus satisfies the defining property
\begin{align}
\label{1-cocycle-Abelian-1}
\mathrm c(p, X+X') = \mathrm c(p, X) + \mathrm c(p+X, X'),
\end{align}
the Abelian version of \eqref{1-cocycle}. 
A Lagrangian thus satisfies 
\begin{align}
\label{Lagrangian-1-form}
\iota_{\chi^v} L=0 \quad \text{and}  \quad
R^*_X L_{|p+x}=L_{|p} + \mathrm c(p, X).
\end{align}
The infinitesimal equivariance is then given by the Lie derivative,
\begin{align}
\label{L-class-anomaly}
\mathfrak L_{X^v} L
=\tfrac{d}{d\epsilon} R^*_{X_\epsilon} L\, \big|_{\epsilon=0}
=\tfrac{d}{d\epsilon} \mathrm c(\,\_\,, X_\epsilon) \, \big|_{\epsilon=0}
\rdefeq \mathrm a(\chi,\_\,), 
\end{align}
where $\tfrac{d}{d\epsilon} X_\epsilon\, \big|_{\epsilon=0} =\chi \in \LieH$, and  $\mathrm a(\chi;\_\,)$ is the $\LieH$-$1$-cocycle (linear in $\chi$) associated to  $L$. 
The \mbox{$\H\!\!\sim\!\Aut_v(\mC)$-transformation} of $L$ is then easily deduced to be
\begin{align}
\label{GT-L}
L^{\bs X}\defeq \Xi^* L = L + \mathrm c(\,\_\,, \bs X) \rdefeq  L+ \mathrm c(\bs X).
\end{align}
The Lie$\H\!\!\sim\!\aut_v(\mC)$-transformation of $L$ is $\mathfrak L_{\bs \chi^v}L = \mathrm a(\bs \chi,\,\_\,)\rdefeq \mathrm a(\bs \chi)$, with $\bs \chi\in$ Lie$\H$. The cocyle $\mathrm a(\bs\chi)$ will be shown to encode the Euler-Lagrange equations. 

In~a trivialisation $\phi$, i.e. in a bundle coordinate chart, we may therefore  write 
(using the same notation above) 
$L_{|(t, x)} =(\phi\-)^*L_{|p}=\mathcal L(t, x)\, dt$. 
Given a local section $\s: \U\subset \T \rarrow \mC$, $t\mapsto \s(t)\simeq \big(t, x(t) \big)$, where $\simeq$  is meant to indicate the expression in a bundle chart, the local representative of $L$ on $\T$ is the familiar expression:
\begin{align}
\label{Lagrangian-section}
 \s^*L_{|t} = L_{|\s(t)} \simeq L_{|(t,\, x(t))} =\mathcal L\big(t, x(t) \big)\, dt \quad  \in \Omega^1(\T, \mathrm c).
\end{align}
Given a curve $\gamma: I \subset \RR \rarrow \mC$, $\tau \mapsto \gamma(\tau) \simeq \big(t(\tau), x(\tau) \big)$, with tangent vector field 
$\dot \gamma 
= \tfrac{dx}{d\tau}\tfrac{\d}{\d x} + \tfrac{dt}{d\tau}\tfrac{\d}{\d t} 
= \dot x \tfrac{\d}{\d x} + \dot t \tfrac{\d}{\d t} \in \Gamma(T\mC)$,
the restriction of $L$ ``on" $\gamma$, actually its pullback to $I$, is:
\begin{align}
\label{Lagrangian-curve}
\ell_{\gamma|\tau}\defeq
\gamma^*L_{|\tau}= L_{|\gamma_\tau} \simeq L_{|(t(\tau),\, x(\tau))} = \mathcal L\big( t(\tau),x(\tau)\big)\,\gamma^* dt_{|\tau} 
=
\mathcal L\big( t(\tau),x(\tau)\big)\, \dot{t} d\tau  \quad\in \Omega^1(I, \mathrm c).
\end{align}
Observe that  \eqref{Lagrangian-curve} is the so-called ``parametrized" formulation of non-relativistic mechanics (as exposed in e.g. \cite{Rovelli2004}), while the standard \eqref{Lagrangian-section} is the ``deparametrized" or ``unparametrized" version. 
We obtain the latter from the former via the  choice $\dot t=1$, which identifies the parameter $\tau$ with the time variable $t$, getting again the familiar expression $\ell_{\gamma|\tau}= \mathcal L\big(\tau, x(\tau) \big)\, d\tau$. 
Remark also that $\dot x=\tfrac{dx}{dt} \dot t \rdefeq x' \dot t$, where $x'$ is the physical (measurable) velocity of $x$ -- i.e. the set of velocities $\{x'_i\}_{i=\{1,\ldots, N\}}$ of the $N$ particles under consideration. So, the condition $\dot t =1$ thus identifies $\dot x$ with the velocity $x'$.\footnote{We stress that $\dot t=1$ is not a ``gauge choice" (as many would be inclined to understand it), but actually reflects a dressing operation. We shall fully elaborate on this remark in a forthcoming paper \cite{JTF-Ravera2025RQ}.}
This dual picture is unsurprising: the image of a local section $\s$ is a curve in $\mC$ parametrized by $t \in \T\!$.

Even though there is no  action of $H$ and $\H$ on $I$, one may define indirectly their ``action" on $\ell \in \Omega^1(I)$ -- in the similar way that one may define their action on local representatives $\s^* L \in \Omega^1(\T)$. 
The equivariance of $\ell_\gamma$ may be defined via the action of $H$ on $\gamma$: since $R_X \gamma \rdefeq \gamma'$,
\begin{align}
\label{equiv-ell}
R^*_X \ell_\gamma \defeq {\gamma'}^*L=(R_X\circ \gamma)^*L = \gamma^* R^*_X L
=\gamma^*\big(L + \mathrm c(\, \_ \,, X) \big)
=\ell_\gamma+ \mathrm c_\gamma(\, \_ \,, X),
\end{align}
where $\mathrm c_\gamma(\, \_ \,, X)\defeq \mathrm c(\, \_\,, X) \circ \gamma:I \rarrow \Omega^1(\mC)$. Hence the notation $\ell_\gamma\in \Omega^1(\mC, \mathrm c)$ and the legitimacy of considering $\ell_\gamma$ as a cocyclic $1$-form. The same may be done for local representatives, using $\s'=R_X\s$. 
The infinitesimal equivariance is obtained by linearising the above, which may be written as 
\begin{align}
 \label{infinitesimal-equiv-ell}
 \delta_\chi \ell_\gamma = \mathrm a_\gamma(\chi, \,\_\,)\defeq \mathrm a(\chi, \,\_\,) \circ \gamma.
\end{align}
The action of the gauge group $\H\simeq \Aut_v(\mC)$ is likewise obtained: 
considering $\gamma^{\bs X}\defeq \Xi\circ \gamma$, one defines
\begin{align}
\label{GT-ell}
\ell^{\bs X}_\gamma\defeq \big(\gamma^{\bs X}\big)^*L
=(\Xi\circ \gamma)^*L = \gamma^* \Xi^* L
=\gamma^*\big(L + \mathrm c(\, \_ \,, \bs X) \big)
=\ell_\gamma+ \mathrm c_\gamma(\bs X),
\end{align}
where $\mathrm c_\gamma(\bs X)\defeq \gamma^*\mathrm c(\, \_\,, \bs X)  :I \rarrow \Omega^1(\mC)$.
This is analogous to the way one obtains the local version the active gauge transformation of $L$, and the way one defines passive gauge transformations of local representatives, i.e. gluings.  
The infinitesimal gauge transformation, i.e. the action of Lie$\H\sim \aut_v(\mC)$, is obtained by linearising the above, 
\begin{align}
 \label{infinitesimal-gauge-ell}
 \delta_{\bs\chi} \ell_\gamma = \mathrm a_\gamma(\bs\chi)\defeq \gamma^* \mathrm a(\bs\chi, \_\,),
\end{align}
which is also just $\gamma^* \big(\mathfrak L_{\bs\chi^v}L\big)$, as expected from consistency of the definitions. 
Observe that $\delta_{\bs\chi} \gamma =\bs\chi(\gamma)$. 

Let us observe 
that the tangent vector field $\dot \gamma \in \Gamma(T\mC)$ of a curve $\gamma$ is right-invariant:
This is clear from the fact that $\dot \gamma =\tfrac{d}{d\tau} \gamma(\tau)\, \big|_{\tau=0} \simeq \dot x \tfrac{\d}{\d x}+ \dot t \tfrac{\d}{\d t}$ and 
$R_{X*} \dot \gamma \defeq \tfrac{d}{d\tau} R_X \circ \gamma(\tau)\, \big|_{\tau=0} = \tfrac{d}{d\tau} \gamma(\tau) + X\, \big|_{\tau=0} = \dot \gamma$. 
Therefore, its pushforward by a vertical automorphism is, by \eqref{pushforward-mC},
$\Xi_* \dot \gamma = \dot\gamma + \big\{ 
d\bs X(\dot \gamma)\big\}^v$. We have  $d\bs X(\dot \gamma) = \tfrac{\d \bs X}{\d t} \dot t + \tfrac{\d \bs X}{\d x} \dot x \rdefeq \tfrac{d \bs X}{d \tau}\rdefeq \dot {\bs X}$,
and since $ \tfrac{\d \bs X}{\d x} =0$ as $\bs X \in \Omega^0_\text{inv}(\mC)$, we get $\dot {\bs X} =\tfrac{\d \bs X}{\d t} \dot t$.
So, 
\begin{align}
\label{pushforward-curve}
\Xi_* \dot \gamma = \dot\gamma + \big\{ 
 \dot{\bs X} \big\}^v \simeq (\dot x + \dot{\bs X} )\, \tfrac{\d}{\d x} + \dot t \tfrac{\d}{\d t}. 
\end{align}
The choice $\dot t=1$ gives $\dot{\bs X} =\tfrac{\d \bs X}{\d t} =\bs X'$, with $\bs X'$ the physical velocity shift induced by the extended Galilean transformation $\bs X \in \H$. 
In particular, the above specialises to vertical tangent vectors of vertical curves, s.t. $\dot t=0$.
 
\medskip
The action functional is defined by 
\begin{align}
\label{action-functional}
\S[\gamma] \defeq \int_\gamma L = \int_I \gamma^*L \rdefeq \int_I \ell_\gamma \simeq \int_I \mathcal L \big(t(\tau), x(\tau) \big)\, \dot t\, d{\tau}.
\end{align}
Here again, the standard ``unparametrised" version of the action is obtained by setting $\dot t=1$.\footnote{Or by defining it as $\mathcal S[\s]\defeq \int_{\U\subset \T} \s^* L$, for a local section $\s:\U \subset \T \rarrow \mC_{|\U}$, $t \mapsto \s(t)\simeq \big(t, x(t) \big)$ representing a trajectory/history. }
The equivariance of $\S[\gamma]$ is derived from  that of $\ell_\gamma$:
$R^*_X \S[\gamma] =\int_I R^*_X \ell_\gamma
=\int_I \ell_\gamma + \mathrm c_\gamma(\,\_\,, X) = \S[\gamma] + \int_I c_\gamma(\,\_\,, X)$. Which, for consistency, is cross-checked by
\begin{equation}
\label{equiv-S}
\begin{aligned}
 R^*_X \S[\gamma]\defeq&\, \S[\gamma'] 
 = \int_{\gamma'} L 
 = \int_I {\gamma'}^*L 
 = \int_I (R_X \circ \gamma)^*L
 = \int_I \gamma^* R_X^*L\\
 =\,& \int_\gamma R^*_X L 
 = \int_\gamma  L + \mathrm c(\,\_\,, X) 
 \rdefeq
 \mathcal S[\gamma] + \int_\gamma \mathrm c(\,\_\,, X). 
\end{aligned}
\end{equation}
Its infinitesimal equivariance is then defined by linearisation of the above,
\begin{equation}
\begin{aligned}
\label{inf-equiv-S}
\delta_{ \chi} \S_{|\gamma} 
\defeq \tfrac{d}{d\epsilon }  R^*_{X_\epsilon} \S[\gamma] \,\big|_{\epsilon=0}
=  \int_\gamma \tfrac{d}{d\epsilon }  R^*_{X_\epsilon} L \,\big|_{\epsilon=0}
=\int_\gamma \mathfrak L_{\chi^v} L
= \int_\gamma \mathrm a(\chi,\_\,),
=\int_I \mathrm a_\gamma(\chi,\_\,), 
\end{aligned}
\end{equation}
which is indeed the expression  $\delta_{ \chi} \S_{|\gamma} =\int_I \delta_{ \chi}\ell_\gamma$ intuitively expected from the definition of $\S[\gamma]$.
%
The $\H\!\sim\!\Aut_v(\mC)$-transformation of the action is likewise defined, via $\gamma^{\bs X}\defeq \Xi \circ \gamma$, as 
\begin{equation}
\begin{aligned}
\label{GT-S}
\S[\gamma]^{\bs X}\defeq \S[\gamma^{\bs X}]
= \int_\gamma \Xi^* L
= \int_{\gamma^{\bs X}} L
= \int_I (\gamma^{\bs X})^* L
= \int_I \ell_\gamma^{\bs X}
= \int_I \ell_\gamma + \mathrm c_\gamma(\bs X)
= \S[\gamma] + \int_\gamma \mathrm c(\bs X),
\end{aligned}
\end{equation}
with $\mathrm c(\bs X)\defeq \mathrm c(\,\_\,,\bs X)$.
Which is consistent with the intuitive formula $\S[\gamma]^{\bs X} = \int_I \ell_\gamma^{\bs X}$.
By linearisation, the Lie$\H\!\sim\!\aut_v(\mC)$-transformation of $\S$ is easily found: 
\begin{equation}
\begin{aligned}
\label{inf-GT-S}
\delta_{\bs \chi} \S_{|\gamma} 
\defeq \tfrac{d}{d\epsilon }\int_\gamma  \Xi_\epsilon^* L \,\big|_{\epsilon=0}
=\int_\gamma \mathfrak  L_{\bs \chi^v} L
= \int_\gamma \mathrm a(\bs \chi)
=\int_I \mathrm a_\gamma(\bs \chi) 
\rdefeq
\int_I \delta_{\bs X}\ell_\gamma.
\end{aligned}
\end{equation}
Equivalently, this is seen to be the Gateaux derivative along $\H\!\sim\!\Aut_v(\mC)$:
for $\gamma_\epsilon\defeq \Xi_\epsilon \circ \gamma = \gamma + \bs X_\epsilon (\gamma)$,  we have
\begin{align}
\label{inf-GT-S-bis}
   \delta_{\bs \chi} \S_{|\gamma} 
\defeq 
\tfrac{d}{d\epsilon}\, \S[\gamma_\epsilon] - \S[\gamma]\,\big|_{\epsilon=0} 
= \tfrac{d}{d\epsilon}\,
    \int_\gamma \mathrm c(\,\_\,, \bs X_\epsilon)\, \big|_{\epsilon=0}
= \int_\gamma \mathrm a(\bs \chi).
\end{align}
This allows to define the classical dynamics as the critical points (curves) of $\S$.

Considering the set $\P_{0,1}\defeq \big\{ \gamma: I \rarrow  \mC\, |\, \gamma(\d I)=\{p_0, p_1\}\big\}$ of paths with beginning and end points $p_0 \simeq (t_0, x_0)$ and  $p_1 \simeq (t_1, x_1)$, respectively. 
Consider also the gauge subgroup 
$\H_{0, 1} \defeq \big\{ \bs X \in \H\, |\, \bs X(t\leq t_0)=0=\bs X(t\geq t_1)  \text{ and } \bs X(t_0<t<t_1)\neq 0  \big\} \subset \H$. We have that $\H_{0,1}$ acts freely and transitively on $\P_{0,1}$, i.e. 
for any choice of initial reference curve $\gamma \in \P_{0,1}$, any other can be written as $\gamma^{\bs X}$ for some $\bs X \in \H_{0,1}$: so $\P_{0,1} \simeq \H_{0,1}$.
The classical physical trajectory $\gamma_c$, i.e. the classical history, between the configurations $p_0$ and $p_1$, is the critical point of $\S[\gamma]$ for $\gamma \in \P_{0,1}$. By \eqref{inf-GT-S}-\eqref{inf-GT-S-bis}, we see that this is given by the vanishing of the linearised $1$-cocycle. 
Said otherwise, the $1$-cocycle $\mathrm a(\bs\chi)$ encodes  the  Euler-Lagrange equations, i.e. the dynamics,  up to vanishing boundary terms:
\begin{align}
\label{critical-point}
\delta_{\bs \chi} \S_{|\gamma_c} 
= \int_{\gamma_c} \mathrm a(\bs \chi)
=\int_I \mathrm a_{\gamma_c}(\bs \chi)
=\int_I  E(\bs \chi, \gamma_c) \ 
\equiv 0, 
\quad \forall \ \bs \chi\in \text{Lie}\H_{0,1},
\end{align}
where $E(\bs \chi, \gamma)\defeq \bs\chi(\gamma)\cdot EL[\gamma]\,d\tau 
=\delta_{\bs\chi} \gamma \cdot  EL[\gamma]\,d\tau$ is the \emph{equations of motion 1-form}, with $EL[\gamma]$ the Euler-Lagrange equations for the curve $\gamma$.
The critical curve is s.t. $EL[\gamma_c]\equiv 0$.
Then, for any $\gamma\in \P_{0,1}$ there is a unique $\bs X \in \H_{0,1}$ s.t. $\gamma =\gamma_c^{\bs X}$, and  by \eqref{GT-S} one has
\begin{align}
 \label{Diff-S-S-critical}   
 \S[\gamma] = \S[\gamma_c] + \int_{\gamma_c} \mathrm c(\bs X).
\end{align}

One may define the function
$S_\gamma= \S[\gamma;\_\,,  p_0] \in \Omega_\text{tens}^0(\mC; c_\gamma)$ where 
$\gamma(\d I)=\{p_0, p\}$
and $c_\gamma(\_ \, ,X)\defeq \int_\gamma \mathrm c(\_ \, ,X)$ is a 1-cocycle inherited from \eqref{equiv-S}: i.e. we have $R^*_X S_\gamma = S_\gamma + c_\gamma(\_ \,, X)$. 
So its $\H$-transformation is immediately read to be $S_\gamma^{\bs X}\defeq \Xi^* S_\gamma = S_\gamma+ c_\gamma(\bs X)$, where $c_\gamma(\bs X)\defeq c_\gamma(\_, \bs X)$. 
For $\gamma =\gamma_c$, the cocyclic 0-form $S_{\gamma_c}$ is the \emph{Hamilton Principal Function} (HPF). It satisfies the Hamilton equation: 
$\tfrac{\d S_{\gamma_c}}{dt}  = -H\big(x, \tfrac{\d S_{\gamma_c}}{\d x}; t \big)$,
 where the canonical conjugate momenta is by definition $\mathrm \uppi\defeq \tfrac{\d S_{\gamma_c}}{\d x}$ and $H(x, \mathrm \uppi, t)$ is the Hamiltonian. 
 Defining the 1-cocycle $C_\gamma(\_ \, , X) \defeq \exp\{-\tfrac{i}{\hbar}c_\gamma(\_ \, ,X)\}$, we  find that the action induces 
 the \emph{flat cocyclic connection} $\varpi_{0,\gamma}\defeq -\tfrac{i}{\hbar} d S_{\gamma}$ on $\mC$, s.t. 
\begin{equation}
\begin{aligned}
\label{flat-cocyclic-conn}
R^*_X \varpi_{0,\gamma} &= \varpi_{0,\gamma} + C_\gamma(\_\,, X)\- dC_\gamma(\_\,, X). \\[1mm]
\iota_{\chi^v} \varpi_{0,\gamma} 
&= -\tfrac{i}{\hbar}\, \iota_{\chi^v} dS_{\gamma} = -\tfrac{i}{\hbar}\, L_{\chi^v} S_{\gamma}
= -\tfrac{i}{\hbar}\, \tfrac{d}{d\epsilon}\, R^*_{X_\epsilon} S_{\gamma} \, \big|_{\epsilon=0} 
= -\tfrac{i}{\hbar}\, \tfrac{d}{d\epsilon}\, c_\gamma(\_ \,,X_\epsilon) \, \big|_{\epsilon=0} 
= -\tfrac{i}{\hbar}\, \int_{\gamma} \mathrm a(\chi)\\[-1.5
mm]
&=\tfrac{d}{d\epsilon}\, C_\gamma(\_\, , X_\epsilon)\, \big|_{\epsilon=0}, 
\end{aligned}
\end{equation}
which is a special case of \eqref{twisted-connection}. 
From which follows that the $\H$-transformation of $\varpi_{0,\gamma}$ is 
\begin{equation}
\label{GT-twised-conn-dS}
\begin{aligned}
{\varpi_{0,\gamma}}^{\bs X} \defeq \Xi^* \varpi_{0,\gamma}= \varpi_{0,\gamma} +  C_\gamma(\bs X)\- dC_\gamma(\bs X),
\end{aligned}
\end{equation}
which is a special case of \eqref{GT-connections}. It is also $(dS_\gamma)^{\bs X}\defeq \Xi^*dS_\gamma= dS_\gamma + d\mathrm c_\gamma(\bs X)$,  consistent with the $\H$-transformation of $S_\gamma$ and $[d, \Xi^*]=0$.
Notice that $\iota_{\chi^v} \varpi_{0,\gamma} \neq 0$ since $\chi \in$ Lie$H$, contrary to what is done in \eqref{critical-point} to determine the critical point $\gamma_c$. Actually, since $E(\chi; \gamma_c)\equiv 0$ by definition of $\gamma_c$, the verticality property of $\varpi_{0,\gamma_c}$ gives the conserved Noether charge: 
In a bundle  chart, $\chi^v \simeq \chi \tfrac{\d}{\d x}$ and $dS_{\gamma_c} = \tfrac{\d S_{\gamma_c}}{\d x} dx + \tfrac{\d S_{\gamma_c}}{\d t} dt$, so $\iota_{\chi^v} \varpi_{0,\gamma_c} = -\tfrac{i}{\hbar}\, \chi\, \tfrac{\d S_{\gamma_c}}{\d x} \rdefeq -\tfrac{i}{\hbar}\, \chi\, \mathrm \uppi$.\footnote{This is also none other than the evaluation of $\chi^v$ on the canonical 1-form (also known as the presymplectic potential) $-\tfrac{i}{\hbar}\,\theta \defeq -\tfrac{i}{\hbar}\, \mathrm\uppi dx$, familiar in that form from Geometric Quantization. }
Finally, we may observe that \eqref{Diff-S-S-critical}  implies a relation analogous to \eqref{GT-twised-conn-dS}: For $\bs X \in \H_{0,1}$ we have
\begin{align}
 \label{Link-twist-crit-vs-non-crit}
 \varpi_{0,\gamma} = \varpi_{0,\gamma_c} +   C_{\gamma_c}(\bs X)\- dC_{\gamma_c}(\bs X).
\end{align}
We are now but a couple of steps from a formulation of  Quantum Mechanics.

\subsection{Quantum mechanics on $\mC$: Schrödinger equation as a cocyclic covariant derivative}
\label{Quantum mechanics on C} 

 One may  quite naturally consider the simplest possible cocyclic tensorial objects on $\mC$ to which  \eqref{flat-cocyclic-conn} is associated:  
\begin{equation}
\begin{aligned}
    \psi_\gamma : \mC &\rarrow \CC \qquad \text{s.t.} \quad \psi_{{\gamma}} \in \Omega^0_\text{tens}(\mC, C_\gamma) \\
    p &\mapsto \psi_\gamma(p) \\
    p+X & \mapsto \psi_\gamma(p+X)= C_\gamma(p, X)\- \psi_\gamma(p),
\end{aligned}    
\end{equation}
i.e. $R^*_X \psi_\gamma = C_\gamma(\_\,, X)\- \psi_\gamma$ with the 1-cocycle $C_\gamma: \mC \times H \rarrow U(1)$, $(p, X) \mapsto C_\gamma(p , X) \defeq \exp\{-\tfrac{i}{\hbar}c_\gamma(p ,X)\}$ defined via  $\S[\gamma]$.
Their gauge transformation is thus $\psi_\gamma^{\bs X} = C_\gamma(\bs X)\-\psi_\gamma=:\psi_{\gamma^{\bs X}}$, for $\bs X \in \H$.
 These can also be seen  as sections $\t \psi_\gamma$ of the cocyclic associated bundle $E^{C_\gamma}\defeq \mC \times_{C_\gamma} \CC$ over $\T$, so that $\Omega^0_\text{tens}(\mC, C_\gamma) \simeq \Gamma(E^{C_\gamma})$. 
 Let us name the latter the space of ``pre-wave functions".
One then has the cocyclic covariant derivative preserving cocyclic tensoriality,
\begin{equation}
\label{cocyclic-cov-der-QM}
\begin{aligned}
\b D_{0, \gamma} : \Omega^0_\text{tens}(\mC, C_\gamma) &\rarrow \Omega^1_\text{tens}(\mC, C_\gamma), \\
\psi_\gamma &\mapsto \b D_{0, \gamma}\, \psi_\gamma \defeq d\psi_\gamma+ \varpi_{0,\gamma}\,\psi_\gamma.
\end{aligned}
\end{equation}
Notice that from the horizontality property of $\b D_{0, \gamma}\, \psi_\gamma$ follows, in bundle coordinates, that
$-i \hbar\tfrac{\d}{\d x} =\tfrac{\d S_\gamma}{\d x}$.
Then, in view of \eqref{Link-twist-crit-vs-non-crit} it must be the case that 
\begin{equation}
\label{QM_-1}
 \begin{aligned}
 \psi_\gamma =  C_{\gamma_c}(\bs X)\- \psi_{\gamma_c},
 \quad
\text{and}
\quad
\b D_{0, \gamma}\, \psi_\gamma 
= C_{\gamma_c}(\bs X)\-\b D_{0, \gamma_c}\, \psi_{\gamma_c}
\quad
\text{for } \bs X \in \H_{0,1}.
 \end{aligned}   
\end{equation}
Now, from \eqref{cocyclic-cov-der-QM} one defines the linear space of ``wave functions" as  $\mK \defeq \ker \b D_{0, \gamma} \subset \Omega^0_\text{tens}(\mC, C_\gamma)$, i.e. the covariantly constant cocyclic pre-wave functions.
The name is justified by the observation that QM is encoded in 
 the statement 
 \begin{align}
 \label{QM-0}
\psi_{\gamma_c} \in \mK, 
\quad \text{\ i.e. } \quad
\b D_{0, \gamma_c}\, \psi_{\gamma_c}=0.
 \end{align}
 Indeed, remember from the previous section that $\varpi_{0, \gamma_c} = -\tfrac{i}{\hbar} dS_{\gamma_c}$ where $S_{\gamma_c}=\S[\gamma_c; \_\,, p_0]$ is the HPF, satisfying the Hamilton-Jacobi equation. So, in bundle coordinate 
 we have
 \begin{equation}
 \label{QM-1}
\begin{aligned}
\b D_{0, \gamma_c}\, \psi_{\gamma_c}
= d \psi_{\gamma_c} + \varpi_{0, \gamma_c}\psi_{\gamma_c}=0
\quad &\Rightarrow \quad
dt\,\tfrac{\d}{\d t} \psi_{\gamma_c} + dx\,\tfrac{\d}{\d x}\psi_{\gamma_c} 
- \tfrac{i}{\hbar} \left( 
dt\,\tfrac{\d S_{\gamma_c}}{\d t} 
+
dx\,\tfrac{\d S_{\gamma_c}}{\d x} 
\right)\psi_{\gamma_c} =0 \\[1mm]
&\hookrightarrow \quad 
 dt\,\left(\tfrac{\d}{\d t}\psi_{\gamma_c} + \tfrac{i}{\hbar} H(x, \mathrm \uppi, t)\,\psi_{\gamma_c}\right)
 +
 dx\,\left( \tfrac{\d}{\d x} -\tfrac{i}{\hbar} \mathrm \uppi \right)\,\psi_{\gamma_c} =0 .
\end{aligned}
\end{equation}
The horizontality property of {$\b D_{0, \gamma_c}\, \psi_{\gamma_c}$} means the identical vanishing of the coefficient of $dx$: which is  an \emph{eigenvalue problem} implying the quantum mechanical prescription for the momentum operator $\hat {\mathrm \uppi} \defeq -i\hbar\tfrac{\d}{\d x}$. 
It is then immediate that, upon defining (still in bundle coordinates) the position operator $\hat{\mathrm x} \psi = \mathrm x \psi$, one gets the canonical commutation relation: $[\hat{\mathrm x}, \hat{\mathrm \uppi}]= i\hbar \delta(\mathrm x - x)$.
These facts are ``kinematical", so to speak, arising from the geometric nature (the tensoriality) of the objects considered.
Cocyclic covariant constancy is a further dynamical constraint: It specifically implies the vanishing of the coefficient of $dt$ in \eqref{QM-1}, which is just the eigenvalue problem for the Schrödinger equation. 
In summary, we have 
\begin{equation}
\label{QM-3}
\begin{aligned}
 \b D_{0, \gamma_c}\, \psi_{\gamma_c} {=} 0 
\qquad \Rightarrow \qquad 
\hat {\mathrm \uppi} \defeq -i\hbar\tfrac{\d}{\d x} 
\quad \text{ and } \quad 
i\hbar \tfrac{\d}{\d t} \psi_{\gamma_c} = H(\hat{\mathrm x}, \hat{\mathrm \uppi}, t)\, \psi_{\gamma_c}.
\end{aligned}
\end{equation}
The cocyclic covariant constancy condition \eqref{QM-0} naturally encodes both the essential kinematical and  dynamical axioms of Quantum Mechanics: the momentum operator and the Schrödinger equation. 
It extends to any $\psi_\gamma \in \mK$ by  \eqref{QM_-1}.
We remark that, per section \ref{Cocyclic bundle geometry in a nutshell}, QM is equivalently encoded in the statement $\t\psi_\gamma \in \t \mK$, where $\t\mK:=\ker \b\nabla_0$ with 
$\b \nabla_0: \Gamma(E^{C_\gamma}) \rarrow \Omega^1(\T)\otimes E^{C_\gamma}$ the cocyclic covariant derivative induced on  associated bundles $E^{C_\gamma} \rarrow \T$. 

One defines the standard (Hermitian) inner product, seen  as a fiber-wise 
integration on $\mC$,
\begin{equation}
\begin{aligned}
\langle \_,\_ \rangle : 
\mK\times \mK
&\rarrow C^\infty(\T, \CC), \\
(\psi_\gamma, \psi_\gamma') &\mapsto \langle \psi_\gamma, \psi_\gamma'\rangle\defeq \int_{\mC} \psi_\gamma^* \psi_\gamma'\, \vol(\mC), 
\end{aligned}    
\end{equation}
with $\psi_\gamma^*$ the $\CC$-conjugate of $\psi_\gamma$, and $\vol(\mC)$ the $H_\Delta$-invariant volume form on fibers $\mC_{|t}=\EE^{3N}$, $R^*_X \vol(\mC)=\vol(\mC)$. 
Then, 
$\mH:=\Big(\mK; \langle \_,\_ \rangle \Big)$ is a Hilbert space. 

We observe that one may obtain \eqref{QM-0}-\eqref{QM-3} as the field equation of a ``meta-action" $\mathfrak S (\psi_\gamma, \varpi_{0,\gamma}) \defeq \langle \psi_\gamma, \iota_{\mathfrak X}\b D_{0,\gamma}\, \psi_\gamma \rangle$ -- yet, we are aware, a not very apt name since this functional is dimensionless --  where $\mathfrak X \in \Gamma(T\mC)$ is any non-vertical vector field ($\pi_* \mathfrak X \neq 0$), 
via the associated variational principle: $\tfrac{\delta}{\delta \psi} \mathfrak S =0 \ \Rightarrow\ \b D_{0, \gamma}\, \psi_\gamma=0$.
In typical quantum field-theoretical  parlance, $\mathfrak S (\psi_\gamma, \varpi_{0,\gamma})=\langle \psi_\gamma| \iota_{\mathfrak X}\b D_{0,\gamma}| \psi_\gamma \rangle$ may be understood as the expectation value of the cocyclic covariant derivative operator $\b D_{0, \gamma}$. QM arises from requiring stationarity of this expectation value.

\paragraph{Path integral}

The \emph{a priori} dependence of the above considerations on curves $\gamma \in \P_{0,1}$ between initial and final points $\gamma(\d I)=\{p_0, p \}$ may seem unusual, but it naturally leads to the path integral description of QM.
One may  formally define the quantity depending only on the boundary by functional integration
\begin{align}
 \label{propagator-0}
 K(p, p_0) \defeq \int_{\,\P_{0,1}}\!\! \D \gamma  \ \ \psi_\gamma
 \ = 
 \int_{\,\H_{0,1}}\!\! \D \bs X  \ \ \psi_{\gamma_r^{\bs X}}
 \ =
 \int_{\,\H_{0,1}}\!\! \D \bs X  \ \ C_{\gamma_r}(\bs X)\-\psi_{\gamma_r} ,
\end{align}
where $\D \gamma$ and $\D\bs X$ are  formal measures on the space of curves $\P_{0,1}$ and on $\H_{0,1}$ respectively.
The first equality arises from the isomorphism $\P_{0,1}\simeq \H_{0,1}$ mentioned above when discussing classical dynamics, and specified by any reference curve $\gamma_r \in \P_{0,1}$ so that for any other $\gamma \in \P_{0,1}$, $\exists! \bs X \in \H_{0,1}$ s.t. $\gamma=\gamma_r^{\bs X}$. 
Stated otherwise, $\H_{0,1}$ acts freely and transitively on $\P_{0,1}$.
One may then define the pre-wave function depending only on the final point by fiber integration over all initial conditions:
\begin{align}
\label{pre-wave-Psi}
\psi(p):= \int_{\mC_{|t_0}} K(p, p_0)\,  \psi(p_0)\,  \vol(\mC_{|t_0}),
\end{align}
where $\vol(\mC_{|t_0})$ is the volume form of the fiber $\mC_{|t_0} \subset \mC$ over $\pi(p_0)=t_0 \in \T$. 
In bundle coordinates, $\psi(p)\simeq \psi(t, x)$.
The quantity \eqref{propagator-0} is usually seen as a propagator, describing the ``transition amplitude" between the initial and final states.
We may take the classical history as reference curve $\gamma_r=\gamma_c$, so that 
\begin{align}
 \label{propagator-1}
 K(p, p_0) \defeq \int_{\,\P_{0,1}}\!\! \D \gamma  \ \ \psi_\gamma
 \ =
 \int_{\,\H_{0,1}}\!\! \D \bs X  \ \ C_{\gamma_c}(\bs X)\-\psi_{\gamma_c} .
\end{align}
Restricting furthermore the attention to wave functions $\psi_\gamma \in \mK$,  solutions of $\b D_{0, \gamma} \psi_\gamma=0$ of the form
\begin{align}
\label{Dirac-Feynman-weight}
\psi_\gamma
=
\psi_0 \exp\int_\gamma \varpi_{0, \gamma}
=
\psi_0 \exp\tfrac{i}{\hbar}S_\gamma
=
\psi_0 \exp\tfrac{i}{\hbar}\S[\gamma],
\end{align}
then the quantity  \eqref{propagator-1} becomes the standard Dirac-Feynman propagator, or path integral (PI),
\begin{align}
 \label{propagator-2}
 K(p, p_0) &= \int_{\,\P_{0,1}}\!\! \D \gamma  \ \ \psi_0 \exp\tfrac{i}{\hbar}\S[\gamma] \\
  &=
 \int_{\,\H_{0,1}}\!\! \D \bs X  \ \ C_{\gamma_c}(\bs X)\-\psi_0 \exp\tfrac{i}{\hbar}\S[\gamma_c]
 \ =
 \psi_0 \exp\tfrac{i}{\hbar}\S[\gamma_c]
 \int_{\,\H_{0,1}}\!\! \D \bs X  \ \ 
 \exp\tfrac{i}{\hbar}c_{\gamma_c}(\bs X). \notag
\end{align}
Then, $\psi(p)$ in \eqref{pre-wave-Psi} is a wave function.  
Notice how geometric considerations automatically provide, in view of \eqref{QM-3}, a splitting of the PI in a contribution from the classical history, $\psi_{\gamma_c}=\psi_0 \exp\tfrac{i}{\hbar}\S[\gamma_c] \in \mK$, 
and a contribution from the functional integral of the 1-cocycle
$c_{\gamma_c}(\bs X)\defeq\int_{\gamma_c} \mathrm c(\_, \bs X) = \int_I \mathrm c(\_, \bs X) \circ \gamma_c$, inherited from the Lagrangian. 
In the semi-classical (e.g. WKB) analysis, the former contribution is known to contain the essential dynamical physical information, while the latter is a normalization factor.

\medskip
The flat cocyclic connection $\varpi_{0,\gamma}$ is at the center of our geometric formulation of both classical and quantum mechanics.  
As an intriguing exercise, one may entertain the possibility of allowing a non-flat cocyclic connection $\varpi_\gamma$ and see what kind of generalisation of classical and quantum mechanics this leads to. 

\section{Relational bundle geometric Classical  and Quantum Mechanics}
\label{Relational bundle geometric classical  and quantum Mechanics}

In this section, we apply the Dressing Field Method to the configuration space-time bundle $\mC$ to obtain relational reformulations of Classical and Quantum Mechanics.
First, we consider how to realise the relational configuration space-time subbundle $\mC_\text{rel} \defeq\mC /H_\text{ext}  \xrightarrow{H_\text{int}} \T$, encoding the physical kinematics.
In particular, we show how the notion of residual transformations of the 2nd kind in the DFM, parametrizing the choice of dressing, here parametrizes the multiple realisations of $\mC_\text{rel}$ within $\mC$. 
Then, we show how to obtain the dressed Lagrangian/action and wave function encoding the physical classical and quantum dynamics. In the context of relational QM, the group of residual transformations of the 2nd kind implements a notion of ``physical reference frame covariance".

\subsection{Relational configuration space-time subbundle via DFM}
\label{Relational configuration space-time subbundle via DFM}

Given that the configuration space-time bundle $\mC$ is a 2-fibration 
\begin{equation}
\begin{aligned}
\begin{tikzcd}[column sep=large, ampersand replacement=\&]
\ \&  \mC   
\arrow[r, "H_\text{ext}"  ]         
\&  \mC_{\text{rel}} \defeq\mC /H_\text{ext}  
\arrow[r, "H_\text{int}"]  
\& \T,
\end{tikzcd} 
\end{aligned}
\end{equation}
illustrated in Fig. \ref{2-bundle-fig}, with gauge group $\H=\H_\text{int} \ltimes \H_\text{ext}$,
we may use the DFM to realise the relational configuration space-time $\mC_{\text{rel}}$, thereby reducing the gauge group $\H_\text{ext}$.
Specializing \ref{The dressing field method: basics} to this context, a 
 dressing field fit for that purpose is defined as
\begin{align}
\label{Dressing-conf-ST}
{\rm u}: \mC \rarrow H_\text{ext}=H_\Delta \triangleleft H   
\quad \text{ s.t. } \quad 
R^*_X {\rm u} = {\rm u} - X, 
\quad \text{ i.e. } \quad 
{\rm u}(p+X)={\rm u}(p) - X.
\end{align}
This is typically the rule for an Abelian dressing field. 
It is a $H_\text{ext}$-tensorial $0$-form, thus it gauge transforms as ${\rm u}^{\bs X}\defeq\Xi^*{\rm u}= {\rm u} - \bs X$ for $\bs X \in \H_\text{ext}$. 
It allows to define the map $f_{\rm u} : \mC \rarrow \mC$,
\begin{equation}
        p \mapsto f_{\rm u}(p):=p+{\rm u}(p), 
\quad  \text{ s.t. } \quad f_{\rm u} \circ R_X = f_{\rm u} 
\quad  \text{ and } \quad f_{\rm u} \circ \Xi = f_{\rm u} 
\quad \text{for} \quad 
\Xi \in \Aut_v(\mC)^{H_\text{ext}} \sim \bs X \in \H_\text{ext}, 
\end{equation}
whose image is a subbundle of $\mC$ isomorphic to $\mC_\text{rel}$: we thus denote it $\mC_\text{rel}^{\rm{u}}$. 

\medskip
The question is: Can we produce such a dressing field? 
As a matter of fact we can: 
Take $p\in \mC$ as describing an instantaneous configuration of $N$ particles, and, given bundle coordinates s.t. $p\simeq (t, x) \in \U \times H$ with $\U \subset \T$, let us define
\vspace{-3mm}
\begin{align}
\label{dressing-field-ith}
  \u_i(p)=\u_i(p_1, \ldots,p_i, \ldots  p_N) \defeq - x_i =(\,\overbrace{-x_i, \ldots, -x_i}^{N \text{ times}}\,) \ \, \in H_\text{ext}=\RR^{3N}_\Delta ,
\end{align}
which clearly satisfies \eqref{Dressing-conf-ST}. 
So, the dressing field $\u_i$ outputs the vertical bundle coordinate (the spatial coordinate) of the $i^{th}$ particle of the configuration.\footnote{It induces the Lie$H_\text{ext}$-valued Ehresmann connection $\omega_0=-d\u_i$, giving the position of the $i^{th}$ particle as the standard of identification of fibers of $\mC$ (i.e. of space) at distinct moments of time $\T$ -- recall, without a connection, there is no canonical identification between fibers.}
We have the associated map
\begin{align}
\label{dressing-map-ith}   
f_{\u_i}(p) =p + \u_i(p) 
=\left( p_1-x_i, \ldots, p_i-x_i, \ldots, p_N-x_i  \right) 
\simeq
\left(t;  x_1-x_i, \ldots, 0, \ldots, x_N-x_i  \right).
\end{align}
That is, $\text{Im} f_{\u_i}\rdefeq \mC^{\,\u_i}_\text{rel}$ is the realisation of the relational configuration space-time bundle $\mC_\text{rel}$ from the viewpoint of the $i^{th}$ particle: it is a description of the $N$-configuration \emph{from within}, by a participating ``observer". Notice that \eqref{dressing-map-ith} is independent of the choice of bundle coordinates (i.e. of trivialisation).

\paragraph{Residual transformations of the 1st kind} 

Since $H_\text{ext}$ is reduced via dressing, and thus acts trivially on  $\mC^{\,\u_i}_\text{rel}$, one expects that there is still a residual action of the non-reduced subgroup $H_\text{int}\subset H$. 
It is indeed so, but with an interesting refinement that brings a better understanding of the isomorphism between $\mC^{\,\u_i}_\text{rel}$ and $\mC_\text{rel}$.

The group $H_\text{int}$ acts via $R_Y p = p+Y$. But it is intuitive that from the viewpoint of  any  particle in the $N$-configuration, it has not moved, the $N-1$ others have. So is it in particular for the $i^{th}$ particle. 
We indeed find that the dressing field \eqref{dressing-field-ith} satisfies
\begin{align}
 \label{dressing-ith-res1}   
 R^*_Y\u_i = \u_i - Y_i, 
 \quad \text{ i.e. } \quad 
 \u_i(p+Y) = \u_i(p) - Y_i
 \quad \text{for } Y \in H_\text{int}, 
\end{align}
so that, given $X= Y + X_0  \in H =  H_\text{int} \ltimes  H_\text{ext}$, we find that 
\begin{align}
 \label{map-ith-res1}  
 f_{\u_i}(p + X) 
 = f_{\u_i}(p + Y) 
 = p + Y +  \u_i(p+Y) 
 &= f_{\u_i}(p) + (Y-Y_i) ,\\
 &= \left(t;  x_1-x_i, \ldots, 0, \ldots, x_N-x_i   \right)
    +
   \left( Y_1-Y_i, \ldots, 0, \ldots, Y_N-Y_i   \right).  \notag
\end{align}
 Which is as expected. We thus find that  $\mC^{\,\u_i}_\text{rel}$ has structure group $H^{\,\u_i}_\text{int} \defeq\{Y -Y_i \}$,  isomorphic, but not identical,~to~$H_\text{int}$. 
The group of residual gauge transformations  is then $\H^{\,\u_i}_\text{int}\simeq \Aut_v(\mC^{\,\u_i}_\text{rel})$, isomorphic to $\H_\text{int}$ but with an altered action. 
Remark also that the  dimension $3(N-1)$ of $H_\text{int}$ is more immediately read from the concrete realisation $H_\text{int}^\u$ \eqref{map-ith-res1}. 

We seize the opportunity to observe that this phenomenon is exactly what happens in the DFM treatment of the electroweak model \cite{Attard_et_al2017, Francois2018, Berghofer-et-al2023}, which dispenses with the notion of spontaneous symmetry breaking (SSB): 
After reduction of the $\SU(2)$ subgroup of  $\U(1) \times \SU(2)$ via dressing, the  group of residual gauge transformations is $\U(1)^{\bs u}$, isomorphic to the original $\U(1)$ but with an action on the $\SU(2)$-invariant dressed fields different from the action of $\U(1)$ on the initial ``bare" fields.\footnote{ 
This is usually framed, heuristically and somewhat confusingly (from a conceptual perspective), as saying that the $\U(1)^{\u}=\U(1)_\text{EM}$ is the ``electromagnetic" gauge group after SSB, and distinct from the gauge group $\U(1)=\U(1)_{\rm Y}$ before SSB, generating the so-called ``weak hypercharge" $\rm Y$. The DFM analysis in terms of residual transformations of the 1st kind is more direct technically and clearer conceptually.}
The Gell-Mann–Nishijima formula flows trivially from the construction of 
gauge-invariants; its analogue here is $\b x^{\,i} \defeq x-x_i$ which is $\H_\text{ext}$-invariant.

\paragraph{Residual transformations of the 2nd kind: frame covariance} 

As explained in \ref{The dressing field method: basics}, the DFM framework accounts for the possible multiplicity of choices of dressing via what we call residual transformations of the 2nd kind.
In the present case, it is clear that the definition of the dressing field \eqref{dressing-field-ith} allows for alternative choices: actually $N$ such choices $\{\u_i\}_{i \in \{1, \ldots, N\}}$. 
Considering e.g. the dressing field given by the $j^{th}$ particle in the $N$-configuration
\begin{align}
\label{dressing-field-jth}
  \u_j(p)=\u_j(p_1, \ldots,p_j, \ldots  p_N) \defeq - x_j 
  \ \, \in H_\text{ext},
\end{align}
we find that it is related to $\u_i$ as
\begin{align}
  \label{Res-2nd-i-j}  
  \u_j = \u_i + Z_{ij}, 
\end{align}
where $Z_{ij}$ is the element of $\RR_\Delta$ s.t. $p_j =p_i - Z_{ij} \in \EE^3$; as in bundle coordinates $p\simeq (t, x)$, it is $x_i-x_j=Z_{ij} \in \RR_\Delta$.

\noindent At given $t\in \T$, within the fiber $\mC_t$, dressings are thus related by the finite (permutation) group 
\begin{align}
\label{G-res-2nd-1}
G\defeq \left\{\, Z_{ij} \in \RR_\Delta\, |\, p_j =p_i - Z_{ij} \text{ for any } p_j, p_i \text{ in the $N$-configuration } p \,\right\}.
\end{align}
So, in general, dressing fields are related by the action of the finite local group
\begin{align}
\label{G-res-2nd-2}
\G\defeq \left\{\, \bs Z_{ij}: \mC \rarrow G\, |\, R^*_X\bs Z_{ij} = \bs Z_{ij} \ \text{ for } X \in H_\text{ext}  \, \right\}.
\end{align}
The $H_\text{ext}$-invariance of  $\bs Z \in \G$, implying $\bs Z^{\bs X}=\bs Z$ for $\bs X \in \H_\text{ext}$, follows from the definition \eqref{G-res-2nd-1}.
This action is
\begin{align}
\label{2-dress-Z}
\u_j = \u_i + \bs Z_{ij},
\qquad \text{ which we may denote generically } \qquad
\u^{\bs Z} = \u + \bs Z.
\end{align}
For later convenience, let us also denote 
\begin{align}
 \label{Generic-f-u-p}
f_{\u}(p)= p + \u(p) \simeq 
\left(\,t; \b x_1,\, \ldots\,, \b x_{N-1}\,  \right)=(t, \b x)
\end{align}
for the generic bundle coordinate on $\mC^{\,\u}_\text{rel}$. It is an instantaneous relational description of  $N$ particles as seen from any one of them, naturally accounting for $N-1$ relations. 
In this notation, the structure group of $\mC^{\,\u}_\text{rel}$ is $H^{\,\u}_\text{int} \defeq\{\b Y \}$. 
The notation may specialise, writing e.g. \eqref{dressing-map-ith} for the $i^{th}$ particle as: 
$f_{\u_i}(p) = p + \u_i(p) 
\simeq \big(t; {\b x}_1^{\,i}, \ldots, {\b x}_{N-1}^{\,i} \big) = (t, {\b x}^{\,i})$.

\begin{figure}[H]
\begin{center}
\includegraphics[width=0.6\textwidth]{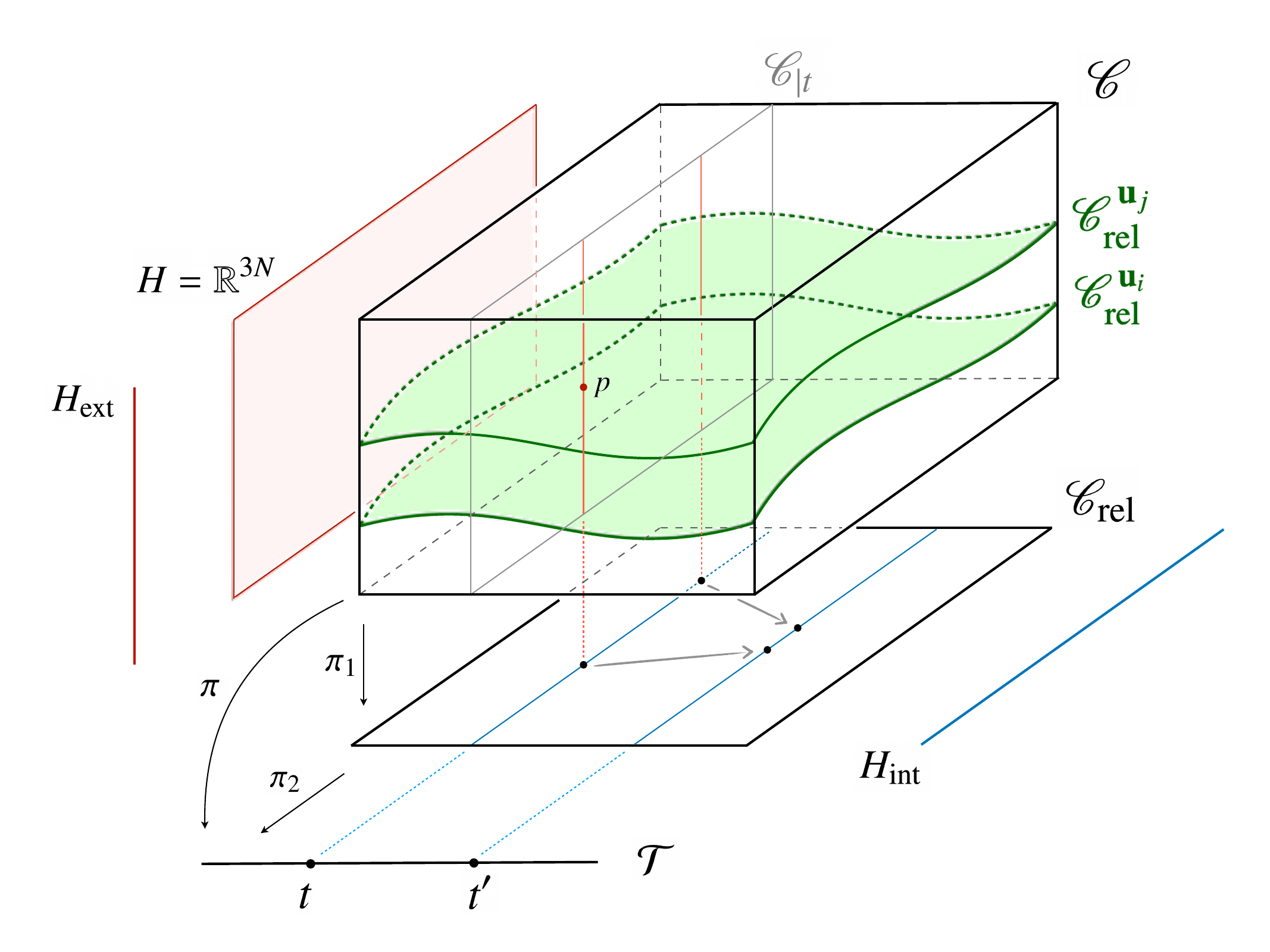}
\caption{Realisations of the relational space-time configuration bundle via dressing and frame covariance.}
\label{2-bundle-bis-fig}
\end{center}
\end{figure}

Another choice of dressing field $\u_j$ induces another realisation of the relational configuration spacetime bundle Im$f_{\u_j}\rdefeq\mC^{\,\u_j}_\text{rel}$, which is a $H^{\,\u_j}_\text{int} \defeq\{Y -Y_j \}$-bundle, isomorphic to $\mC_\text{rel}$. 
It describes  the $N$-configuration relationally, from the viewpoint of the $j^{th}$ particle. 
We have the isomorphisms
$\mC^{\,\u_j}_\text{rel} \simeq \mC^{\,\u_i}_\text{rel}\simeq \mC_\text{rel}$, as illustrated by Fig. \ref{2-bundle-bis-fig}.

The group $\G$ thus parametrizes the multiple realisations of $\mC_\text{rel}$, allowing to switch between internal relational viewpoints: Indeed, we have e.g. 
\vspace{-3mm}
\begin{equation}
\label{Ref-frame-change}
\begin{aligned}
f_{\u_i}(p) &\simeq \big(\,t;\, x_1 - x_i\, , \ldots,\,  \overbrace{x_i-x_i}^{0}\, , \ldots,\, \overbrace{x_j-x_i}^{-\bs Z_{ij}(p)}\, ,\,  \ldots x_N-x_i \, \big),  \\[2mm]
f_{\u_j}(p) &\simeq \big(\, t;\, x_1 - x_j\, , \ldots,\, \underbrace{x_i-x_j}_{\bs Z_{ij}(p)}\, , \ldots,\, \underbrace{x_j-x_j}_{0}\, , \ldots x_N-x_j \, \big) = f_{\u_i}(p) + \bs Z_{ij}(p), 
\end{aligned}
\end{equation}
which showcases the relational symmetry of the viewpoint permutation.
In the generic notation of \eqref{2-dress-Z}-\eqref{Generic-f-u-p},
\begin{align}
\label{f-u-Z}
f_{\u^{\bs Z}}(p)= f_{\u}(p) + \bs Z(p). 
\end{align}
In other words, it is clear that the group $\G$ of transformations of the 2nd kind  precisely encodes the notion of \emph{physical reference frame covariance}  of the relational description. 

As seen in section \ref{Classical mechanics on C}, a kinematical history is  a curve $\gamma: I\rarrow \mC$. Its ``dressing" is the projection on $\mC^\u_\text{rel}$, 
\begin{equation}
\label{relational-kin}
\begin{aligned}
\gamma^\u\defeq  f_\u \circ \gamma : I&\rarrow \mC^\u_\text{rel}, \\[1mm]
\tau &\mapsto \gamma^\u(\tau)= \gamma(\tau) + \u\big(\gamma(\tau)\big) \\
&\phantom{\mapsto \gamma^\u(\tau)}
\simeq \big(\,t(\tau)\,; \b x_1(\tau),\, \ldots\,, \b x_{N-1}(\tau)\,  \big)=\big(\, t(\tau), \b x(\tau)\,\big).
\end{aligned}
\end{equation}
It~is~a \emph{relational kinematical history}  of a $N$ particles configuration, as described from the internal viewpoint of any one of them, e.g. the $i^{th}$ if $\u=\u_i$. 
By definition it is $\H_\text{ext}$-invariant: we define $(\gamma^\u)^{\bs X} \defeq f_\u \circ \Xi\circ \gamma = f_\u \circ \gamma = \gamma^\u$, for $\bs X \in \H_\text{ext}$ generating $\Xi \in \Aut_v(\mC)^{H_\text{ext}}$.
On account of \eqref{map-ith-res1}, it supports the residual transformation of the 1st kind
\begin{align}
\label{kin-res-1st}
(\gamma^\u)^{{\bs Y}}&= \gamma^\u + \b{\bs Y} (\gamma) \rdefeq (\gamma^\u)^{\b{\bs Y}}, \quad \text{ for } {\bs{Y}} \in \H_\text{int} \simeq \b{\bs{Y}} \in \H_\text{int}^\u,\\
\phantom{(\gamma^\u)^{\b{\bs Y}}}
&\simeq \big(\,t(\tau)\,;  \b x_1(\tau),  \ldots, \b x_{N-1}(\tau)\,   \big)
    +
   \big( \b Y_1(\tau),  \ldots, \b Y_{N-1}(\tau)\,   \big),  \notag
\end{align}
which describes a \emph{physically distinct} relational kinematical history in $\mC^\u_\text{rel}$. 
The infinitesimal variation of a physical relational kinematics is then:
$\delta_{\bs\Upsilon}\, \gamma^\u \defeq \tfrac{d}{d\epsilon} \, (\gamma^\u)^{\bs Y_\epsilon} - \gamma^\u \, \big|_{\epsilon=0} =\b{\bs \Upsilon}(\gamma)$, with $\tfrac{d}{d\epsilon} {\bs Y_\tau} \big|_{\epsilon=0} = \bs\Upsilon \in$ Lie$\H_\text{int}$.

On account of \eqref{Ref-frame-change}-\eqref{f-u-Z}, $\gamma^\u$ support the action of $\G$ 
\begin{align}
\label{kin-res-2nd}
(\gamma^\u)^{\bs Z}= \gamma^\u + \bs Z (\gamma), \quad \text{ for } \bs Z \in \G,
\end{align}
where $(\gamma^\u)^{\bs Z}$ and $\gamma^\u $ describe the \emph{same} relational kinematical history: 
specialising the notation, this is indeed by~\eqref{Ref-frame-change}
\begin{equation}
\label{kin-res-2nd-bis}
\begin{aligned}
 \gamma^{\u}(\tau) &=   \gamma^{\u_i}(\tau)
\simeq \big(\,t(\tau)\,; {\b x}^{\, i}_1(\tau),\, \ldots\,, {\b x}^{\, i}_{N-1}(\tau)\,  \big)=\big(\, t(\tau),  {\b x}^{\, i}(\tau)\,\big),  \quad \text{and}\\[1mm]
(\gamma^\u)^{\bs Z}(\tau)&=\gamma^{\u_j}(\tau)= \gamma^{\u_i}(\tau) + \bs Z_{ij} \big(\gamma(\tau)\big)
\simeq \big(\,t(\tau)\,; {\b x}^{\, j}_1(\tau),\, \ldots\,, {\b x}^{\, j}_{N-1}(\tau)\,  \big)=\big(\, t(\tau),  {\b x}^{\, j}(\tau)\,\big).
\end{aligned}
\end{equation}
Equation \eqref{kin-res-2nd} demonstrates the physical reference frame covariance of the relational description of a kinematical history.
We can now move on to the relational description of classical and quantum mechanics.


\subsection{Relational Classical  Mechanics, and physical frame covariance}
\label{Relational Classical Mechanics}

As a special case of the general formulae
\eqref{dressed-forms-1}-\eqref{twisted-dressed-forms} of the DFM, the dressed Lagrangian 1-form is 
\begin{align}
 \label{dressed-L}
 L^{\u}\defeq f_\u^* L = L + \mathrm c( \u), 
\end{align}
where we used the notation $\mathrm c(\u)\defeq\mathrm c(\_\,, \u)$. It is $H_\text{ext}$-basic by construction, hence $\H_\text{ext}$-invariant.
This is checked explicitly using the $\H_\text{ext}$-transformation of $L$ \eqref{GT-L} and the fact that, by the Abelian version \eqref{1-cocycle-Abelian-1} of \eqref{prop-cocycle}, we get, 
\begin{align}
 \label{Abelian-cocyclic-dressing-X}
 \mathrm c(\u)^{\bs X}(p) \defeq&\, \mathrm c(\_\,, \u)^{\bs X}(p)
 = \big(\Xi^*  \mathrm c(\_\,, \u)\big) (p)
 = \mathrm c\big(\Xi(p), \Xi^*\u(p)\big)
 = \mathrm c\big(p+ \bs X(p), - \bs X(p)+ \,\u(p) \big) \notag\\
 =&\, \mathrm c\big(p+ \bs X(p), - \bs X(p) \big) 
 +
  \mathrm c\big(p+ \bs X(p) - \bs X(p),\,  \u(p)\big)
= -\mathrm c\big(p, \bs X(p)\big) 
+
 \mathrm c\big(p,\,  \u(p)\big), \notag\\[1mm]
 \mathrm c(\u)^{\bs X} =&\, \mathrm c(\u) - \mathrm c(\bs X), \qquad \text{for } \bs X \in \H_\text{ext}, 
\end{align}
i.e. an Abelian version of (the gauge transformation of) a cocyclic dressing field \eqref{cocyclic-dressing}.
Remark that we thus have 
\begin{align}
\label{Abelian-cocyclic-dressing-chi}
\mathfrak L_{\bs\chi^v}\mathrm c(\u) = -\mathrm a(\bs\chi) \quad \text{ for } \bs\chi \in \text{Lie}\H_\text{ext},
\end{align}
which is the ``classical anomaly" \eqref{L-class-anomaly}.
We may also observe that the notion of cocyclic dressing field $\mathrm c(\u)$, as it appears here in Mechanics, encompasses the notion of \emph{Wess-Zumino counterterms} introduced to ``restore" gauge-invariance and/or implement anomaly cancellation --
see e.g. section 12.3 in \cite{GockSchuck}, Chap.15 in \cite{Bonora2023}, or the end of Chap.4 in \cite{Bertlmann}. See \cite{JTF-Ravera2024gRGFT} for further discussion of this point in the context of general-relativistic Gauge Field Theory.

Notice that \eqref{dressed-L} can be obtained from \eqref{GT-L} substituting the gauge parameter by the dressing field, $\bs X \rarrow \u$.
This illustrates the DFM \emph{rule of thumb} described in section \ref{The dressing field method: basics}: To obtain the dressing of an object, compute first its gauge transformation, then substitute $\bs X \rarrow \u$ in the result. We shall often see instances of it in what follows.

On account of \eqref{map-ith-res1}, $L^\u$ exhibits  residual $\H_\text{int}\simeq \H^\u_\text{int}$-transformations (of the 1st kind), which by \eqref{dressed-L} may be found by: $(L^\u)^{Y} = L^{\bs Y} + \mathrm{c}(\u)^{\bs Y}$. The $\H_\text{int}$-transformation of $L$ is known as a special case of \eqref{GT-L}, so we need only to find  $\mathrm{c}(\u)^{\bs Y}$. For this let us first notice the very simple lemma 
\begin{align}
 \label{small-lemma1} 
& \mathrm{c}(p, \u + \bs{Y}) 
 =\mathrm{c}(p, \u) + \mathrm{c}(p + \u,  \bs{Y})
=\mathrm{c}(p, \bs{Y}) + \mathrm{c}(p +\bs{Y},  \u), \notag\\
&\Rightarrow \quad  
\mathrm{c}(p +\bs{Y},  \u) - \mathrm{c}(p + \u,  \bs{Y}) 
=\mathrm{c}(p, \u) - \mathrm{c}(p, \bs{Y}),
\end{align}
where we used the definition  of an Abelian 1-cocycle. So, for $\bs Y \in \H_\text{int}$ generating $\Xi \in \Aut_v(\mC)^{H_\text{int}}$ and $\u=\u_i$:
\begin{align}
 \label{Abelian-cocyclic-dressing-Y}   
 \mathrm{c}(\u_i)^{\bs Y}(p) \defeq&\,
\mathrm c(\_\,, \u_i)^{\bs Y}(p)
 = \big(\Xi^*  \mathrm c(\_\,, \u_i)\big) (p)
 = \mathrm c\big(\Xi(p), \Xi^*\u_i(p)\big)
 = \mathrm c\big(p+ \bs Y(p), \,\u_i(p) - \bs Y_i(p)  \big) \notag\\
 =&\, 
 \mathrm c\big(p+ \bs Y(p)+ \,\u_i(p), - \bs Y_i(p) \big) 
 +
\mathrm c\big(p+ \bs Y(p), \,\u_i(p) \big) \notag\\
=&\,
\mathrm c\big(p+ \,\u_i(p),\bs Y(p)  - \bs Y_i(p) \big) 
-
\mathrm c\big(p+ \,\u_i(p),\bs Y(p) \big)
+
\mathrm c\big(p+ \bs Y(p), \,\u_i(p) \big)\notag\\
=&\, 
\mathrm c\big(p+ \,\u_i(p),\b{\bs Y}^i(p) \big) 
+
\mathrm c\big(p, \,\u_i(p) \big)
- 
\mathrm c\big(p, \bs Y(p) \big)\notag\\
=&\, \mathrm c\big(p, \u_i(p) + \b{\bs Y}^i(p)\big) - \mathrm c\big(p, {\bs Y}\big),
\notag \\[1mm]
\hookrightarrow\quad \mathrm{c}(\u)^{\bs Y}
 =&\,  \mathrm{c}(\u)
    + 
    \mathrm c\big(f_\u(\_),\b{\bs Y} \big) 
    - 
    \mathrm{c}(\bs Y)\\
 = &\, \mathrm{c}(\u + \b{\bs Y}) - \mathrm{c}({\bs Y}), \notag
\end{align}
using again the defining property of $\mathrm c$. 
We thus compute the residual $\H_\text{int}\simeq \H^\u_\text{int}$-transformation of $L^\u$ to be
\begin{align}
 \label{dressed-L-res-1}
 (L^\u)^{\bs Y} 
 &= L^{\bs Y} + \mathrm{c}(\u)^{\bs Y}
= L + \mathrm{c}(\bs Y) 
    + \mathrm{c}(\u)\ 
    +\  \mathrm c\big(f_\u(\_),\b{\bs Y} \big) 
    - \mathrm{c}(\bs Y)
 = L +  \mathrm{c}(\u) +   \mathrm c\big(f_\u(\_),\b{\bs Y} \big)  \notag\\
 &= L^\u +  \mathrm c\big(f_\u(\_),\b{\bs Y} \big),
\end{align}
which is also $(L^\u)^{\bs Y}= L + \mathrm{c}(p, \u + \b{\bs Y})$. 
Eq. \eqref{dressed-L-res-1} is precisely what geometric reasoning immediately gives: 
Considering $L^\u$ as a form on the subbundle $\mC^\u_\text{rel}=$Im$f_\u$, its gauge group $\H^\u_\text{int}$ acts in exactly the same formal way as $\H_\text{int}$ acts on $L$, i.e. for $\b{\bs Y} \in \H^\u_\text{int}$ generating $\Xi \in \Aut_v(\mC^\u_\text{rel})$ we have
\begin{align}
\label{dressed-L-res-2}
(L^\u)^{\b{\bs Y}}\defeq \Xi^* L^\u 
= L^\u + \mathrm{c}(f_\u(\_), \b{\bs Y}). 
\end{align}
We also see that indeed $(L^\u)^{\bs Y}=(L^\u)^{\b{\bs Y}}$, as expected from $\H_\text{int}\simeq \H^\u_\text{int}$, but ensured by the peculiar $\H^\u_\text{int}$-transformation \eqref{Abelian-cocyclic-dressing-Y} of the cocyclic dressing field $\mathrm{c}(\u)$. 
For $\bs\Upsilon \in$ Lie$\H_\text{int}$ we read from \eqref{dressed-L-res-1}  the infinitesimal transformation 
\begin{align}
 \label{dressed-L-res-3}   
 \mathfrak L_{\,\bs\Upsilon^v} L^\u 
 =  \mathfrak L_{\,\b {\bs\Upsilon}^v} L^\u
 = \mathrm{a}\big(\b{\bs\Upsilon}, f_\u(\_) \big).
\end{align}
This cross-checked via \eqref{dressed-L} and from the transformation of the cocyclic dressing read from \eqref{Abelian-cocyclic-dressing-Y}
\begin{align}
\label{Abelian-cocyclic-dressing-Upsilon}
\mathfrak L_{\,\bs\Upsilon^v} \mathrm c(\u) 
= \mathrm{a}\big(\b{\bs\Upsilon}, f_\u(\_) \big) - \mathrm{a}(\bs\Upsilon),
\end{align}
by which we get,
\begin{equation}
\label{Link-bare-dressed-a}
\begin{aligned}
\mathfrak L_{\,\bs\Upsilon^v} L^\u 
=\mathfrak L_{\,\bs\Upsilon^v} L + 
  \mathfrak L_{\,\bs\Upsilon^v} \mathrm c(\u)
= \mathrm{a}(\bs\Upsilon) + \mathrm{a}\big(\b{\bs\Upsilon}, f_\u(\_) \big) - \mathrm{a}(\bs\Upsilon)
=\mathrm{a}\big(\b{\bs\Upsilon}, f_\u(\_) \big).
\end{aligned}
\end{equation}
This computation is instructive by itself for the following reason:
As we  observed in section \ref{Classical mechanics on C}, the infinitesimal 1-cocycle $\mathrm{a}(\bs\Upsilon)= \mathrm{a}(\bs\Upsilon,\_)$ essentially encodes the Euler-Lagrange equations,
and therefore the dressed  infinitesimal cocycle $\mathrm{a}\big(\b{\bs\Upsilon}, f_\u(\_) \big)$ -- which may be understood as a ``residual classical anomaly", in analogy with field theory -- will similarly be shown to encode the \emph{relational Euler-Lagrange equations}. So, eq. \eqref{Link-bare-dressed-a} establishes the link between the bare and relational equations of motions.

On account of \eqref{2-dress-Z}, $L^\u$ supports $\G$-transformations (of the 2nd kind), which by \eqref{dressed-L} may be found by writing: $(L^\u)^{\bs Z} = L^{\bs Z} + c(\u)^{\bs Z}= L + c(\u)^{\bs Z}$. 
So we just need to determine $c(\u)^{\bs Z}$, which is easily done:
\begin{align}
 \label{Abelian-cocyclic-dressing-Z}
\mathrm{c}(\u)^{\bs Z}(p)
\defeq \mathrm{c}\big(p, \u(p)\big)^{\bs Z}
= \mathrm{c}\big(p, \u^{\bs Z}(p)\big)
&= \mathrm{c}\big(p, \u(p)+{\bs Z}(p)\big) 
 \notag\\
&= \mathrm{c}\big(p, \u(p) \big) 
   + \mathrm{c}\big(p +\u(p), {\bs Z}(p)\big), \notag\\[1mm]
\hookrightarrow
\mathrm{c}(\u)^{\bs Z}  = \mathrm{c}(\u+{\bs Z}) 
= \mathrm{c}(\u) + \mathrm{c}\big( f_\u(\_)&,{\bs Z} \big). 
\end{align}
This is  the Abelian version of \eqref{Ambig-twisted-dress-field}.
We thus find the $\G$-transformation of $L^\u$ to be
\begin{align}
 \label{dressed-L-res-2.0}
 (L^\u)^{\bs Z} = L^\u + \mathrm{c}\big(f_\u(\_), \bs Z \big),
\end{align}
which is also $(L^\u)^{\bs Z}=L^{\u^{\bs Z}}= L + \mathrm{c}(\u+{\bs Z})$.
Eq. \eqref{dressed-L-res-2.0} is but a special case of the general formulae \eqref{resid-trsf-dressed-twisted-forms} of the DFM. 

Despite their formal similarity, equations \eqref{dressed-L-res-1} and \eqref{dressed-L-res-2.0} carry very different meanings: 
Eq. \eqref{dressed-L-res-1} specifies how the description of the Lagrangian form changes from the point of view of the same particle participating in a distinct physical configuration,
a change which is continuous and parametrized by the Lie group $\H_\text{int}\simeq \H^\u_\text{int}$.
So that both $(L^\u)^{\bs Y}$ and $(L^\u)$ ``live" on the same subbundle $\mC^{\u}_\text{rel}$.
While \eqref{dressed-L-res-2.0} specifies how the description of the Lagrangian changes when shifting to a distinct particle in the same physical configuration, 
a change which is discrete and parametrized by the finite group $\G$ \eqref{G-res-2nd-1}: 
Specialising the notation, \eqref{dressed-L-res-2.0} is 
$L^{\u_j} = L^{\u_i} + \mathrm{c}\big(f_{\u_i}(\_), \bs Z_{ij} \big)$, and indicates how to relate the relational descriptions of the Lagrangian on the bundles $\mC^{\u_i}_\text{rel}$ and $\mC^{\u_j}_\text{rel}$ in Fig. \ref{2-bundle-bis-fig}.

Since both $(L^\u)^{\bs Z}= L^{\u_j}$ and $(L^\u)= L^{\u_i}$ are the same object seen from two distinct perspectives, it is of interest to see explicitly how their residual $\H_\text{int}$-transformations are related. 
By \eqref{dressed-L-res-1} we know that
\begin{equation}
\begin{aligned}
(L^{\u_j})^{\bs Y}&= L^{\u_j} + \mathrm{c}\big(f_{\u_j}(\_), \b{\bs Y}^j \big), \quad \text{ with } \b{\bs Y}^j \defeq \bs Y - \bs Y_j, \\
(L^{\u_i})^{\bs Y}&= L^{\u_i} + \mathrm{c}\big(f_{\u_i}(\_), \b{\bs Y}^i \big), \quad \text{ with } \b{\bs Y}^i \defeq \bs Y - \bs Y_i.
\end{aligned}
\end{equation}
By \eqref{dressed-L-res-2.0}, we have $(L^{\u_j})^{\bs Y} = (L^{\u_i})^{\bs Y} + \mathrm{c}\big(f_{\u_i}(\_), \bs Z_{ij} \big)^{\bs Y}$. To compute the $\mathrm{c}\big(f_{\u_i}(\_), \bs Z_{ij} \big)^{\bs Y}$ the $\H_\text{int}$-transformation of $\bs Z=\bs Z_{ij}$ is needed. Since by definition $p_j=p_i -\bs Z_{ij}(p)$, under the action of $\H_\text{int}$ we have 
$p_j+\bs Y_j =p_i + \bs Y_i -\bs Z_{ij}^{\bs Y}(p)$, so that $p_i -\bs Z_{ij}(p)+\bs Y_j =p_i + \bs Y_i -\bs Z_{ij}^{\bs Y}(p)$, and we obtain finally
\begin{align}
\label{Y-trsf-Z}
\bs Z_{ij}^{\bs Y} 
= \bs Z_{ij} + \big(\bs Y_i - \bs Y_j\big)
=\bs Z_{ij} + \big(\b{\bs Y}^j - \b{\bs Y}^i\big).
\end{align}
So, we may compute
\begin{align}
\label{usefull-identity}
\mathrm{c}\big(f_{\u_i}(\_), \bs Z_{ij} \big)^{\bs Y}
&=
\mathrm{c}\big(f_{\u_i}(\_)^{\bs Y}, \bs Z_{ij}^{\bs Y} \big)  \notag\\
&=
\mathrm{c}\big(f_{\u_i}(\_) + \b{\bs Y}^i, \bs Z_{ij}+ (\b{\bs Y}^j - \b{\bs Y}^i )\big) \notag\\
&= 
\mathrm{c}\big(f_{\u_i}(\_) , \b{\bs Y}^i + (\b{\bs Y}^j - \b{\bs Y}^i ) + \bs Z_{ij}\big)
-
\mathrm{c}\big(f_{\u_i}(\_), \b{\bs Y}^i \big) \notag\\
&=
\mathrm{c}\big(f_{\u_i}(\_) , \b{\bs Y}^j + \bs Z_{ij}\big)
-
\mathrm{c}\big(f_{\u_i}(\_), \b{\bs Y}^i \big) \notag\\
&=
\mathrm{c}\big(f_{\u_i}(\_), \bs Z_{ij} \big)
+
\mathrm{c}\big(f_{\u_i}(\_)+ \bs Z_{ij},\b{\bs Y}^j  \big)
-
\mathrm{c}\big(f_{\u_i}(\_),\b{\bs Y}^i  \big) \\
&=
\mathrm{c}\big(f_{\u_i}(\_), \bs Z_{ij} \big)
+
\mathrm{c}\big(f_{\u_j}(\_),\b{\bs Y}^j  \big)
-
\mathrm{c}\big(f_{\u_i}(\_),\b{\bs Y}^i  \big). \notag
\end{align}    
Remark the similarity with \eqref{Abelian-cocyclic-dressing-Y}. 
We thus find 
\begin{align}
\label{dressed-Ls-Y}
(L^{\u_j})^{\bs Y} 
&= (L^{\u_i})^{\bs Y} + \mathrm{c}\big(f_{\u_i}(\_), \bs Z_{ij} \big)^{\bs Y} \notag\\
&= L^{\u_i} 
+
\cancel{\mathrm{c}\big(f_{\u_i}(\_), \b{\bs Y}^i \big)}
\ + \ 
\mathrm{c}\big(f_{\u_i}(\_), \bs Z_{ij} \big)
+
\mathrm{c}\big(f_{\u_i}(\_)+ \bs Z_{ij},\b{\bs Y}^j  \big)
-
\cancel{\mathrm{c}\big(f_{\u_i}(\_),\b{\bs Y}^i  \big)} \notag\\
&=
L^{\u_j} + \mathrm{c}\big(f_{\u_i}(\_)+ \bs Z_{ij},\b{\bs Y}^j  \big) 
= L^{\u_j} + \mathrm{c}\big(f_{\u_j}(\_),\b{\bs Y}^j  \big),
\end{align}
as expected. Again, the subtle cocyclic structure enforces consistency.
The linear version of the above is of special interest.
From  \eqref{dressed-L-res-3}  we have that
\begin{equation}
\begin{aligned}
 \label{dressed-L-res-3.0}   
 \mathfrak L_{\,\bs\Upsilon^v} L^{\u_j} 
 = \mathrm{a}\big(\b{\bs\Upsilon}^j, f_{\u_j}(\_) \big), 
 \quad \text{and}\quad
  \mathfrak L_{\,\bs\Upsilon^v} L^{\u_i} 
 = \mathrm{a}\big(\b{\bs\Upsilon}^i, f_{\u_i}(\_) \big).
\end{aligned}
\end{equation}
By \eqref{dressed-L-res-2.0} we must have 
$\mathfrak L_{\bs\Upsilon^v} L^{\bs u_j} =\mathfrak L_{\bs\Upsilon^v} L^{\bs u_i} 
+ \mathfrak L_{\bs\Upsilon^v} \mathrm{c}\big(f_{\u_i}(\_), \bs Z_{ij} \big)$,
and by \eqref{usefull-identity} we get 
\begin{align}
 \label{useful-identity-2}   
 \mathfrak L_{\bs\Upsilon^v} \mathrm{c}\big(f_{\u_i}(\_), \bs Z_{ij} \big)
 =
 \mathrm{a}\big(\b{\bs \Upsilon}^j, f_{\u_i}(\_)+ \bs Z_{ij}  \big)
-
\mathrm{a}\big( \b{\bs \Upsilon}^i, f_{\u_i}(\_)  \big)
=
 \mathrm{a}\big(\b{\bs \Upsilon}^j, f_{\u_j}(\_)  \big)
-
\mathrm{a}\big( \b{\bs \Upsilon}^i, f_{\u_i}(\_)  \big).
\end{align}
Remark the similarity with \eqref{Abelian-cocyclic-dressing-Upsilon}.
So, analogously to \eqref{Link-bare-dressed-a}, we have
\begin{align}
\label{Lie-deriv-dressed-Ls-Upsilon}
\mathfrak L_{\bs\Upsilon^v} L^{\bs u_j} =\mathfrak L_{\bs\Upsilon^v} L^{\bs u_i} 
+ \mathfrak L_{\bs\Upsilon^v} \mathrm{c}\big(f_{\u_i}(\_), \bs Z_{ij} \big) 
= \cancel{\mathrm{a}\big(\b{\bs\Upsilon}^i, f_{\u_i}(\_) \big) }
\  + \ 
\mathrm{a}\big(\b{\bs \Upsilon}^j, f_{\u_i}(\_)+ \bs Z_{ij}  \big)
-
\cancel{\mathrm{a}\big( \b{\bs \Upsilon}^i, f_{\u_i}(\_)  \big)}
= 
\mathrm{a}\big(\b{\bs\Upsilon}^j, f_{\u_j}(\_) \big),
\end{align}
as expected. 
As we show below, at the end of this section, this computation is key to establish the correspondence between  formulations of  the \emph{relational variational principle} as seen from distinct perspectives within the same kinematical history $\gamma$. 

The dressed Lagrangian restricted along a kinematical history $\gamma$ is,
\begin{equation}
\begin{aligned}
\label{dressed-ell-1}
\ell^\u_\gamma \defeq \gamma^* L^\u 
= (f_\u \circ \gamma)^* L \rdefeq (\gamma^\u)^*L  \defeq&\, \ell_{\gamma^\u} \\
=&\, ( \mathcal{L} \circ \gamma^\u)\  (\gamma^{\u})^*dt
=( \mathcal{L} \circ \gamma^\u)\ \dot{t}\,d\tau, \\[1mm]
\text{i.e. } \quad
\ell^\u_\gamma (\tau)= \ell_{\gamma^\u}(\tau)= 
\mathcal{L}\big(t(\tau), \b x(\tau)\,\!\big)\ \dot{t}\, d\tau \ \,   \in \ &\, \Omega^1(I, {\rm c}).
\end{aligned}
\end{equation}
This expression is conceptually clear: it is the  relational parametrized Lagrangian from the viewpoint of one of the $N$ particles in $\gamma$, which thus describes the kinematics as $\gamma^\u$. 
From \eqref{dressed-L} we get the  computationally useful expression
\begin{align}
\label{dressed-ell-2}
\ell_{\gamma^\u}=\ell^\u_\gamma \defeq \gamma^* L^\u 
= \gamma^*\left(L + \mathrm c(\u)\right)
\rdefeq \ell_\gamma + \mathrm c_\gamma(\u),
\end{align}
giving the relational Lagrangian in terms of the bare Lagrangian and its 1-cocycle.
It could have been obtained from the $\H_\text{ext}$-transformation of $\ell$ \eqref{GT-ell} via the DFM rule of thumb. 
The residual $\H_\text{int}\simeq \H^\u_\text{int}$-transformation is, by \eqref{dressed-L-res-1},
\begin{equation}
\label{dressed-ell-res-1}
\begin{aligned}
(\ell^\u_\gamma)^{\bs Y}
     = (\ell^\u_\gamma)^{\b{\bs Y}} 
     \defeq \gamma^*\left( L^\u + \mathrm{c}(f_\u(\_), \b{\bs Y} \big) \right)
    &= \ell^\u_\gamma + \mathrm{c}\big(f_\u(\gamma), \b{\bs Y}(\gamma)\big), \\[1mm]
\text{or equivalently} \quad 
(\ell_{\gamma^\u})^{\bs Y}
    = (\ell_{\gamma^\u})^{\b{\bs Y}}
    &=  \ell_{\gamma^\u} + \mathrm{c}_{\gamma^u}(\b{\bs Y}).
\end{aligned}
\end{equation}
It 
is   the relational parametrized Lagrangian from the internal viewpoint of the same particle in a \emph{physically distinct} kinematical history $(\gamma^\u)^{\b Y}$ given by \eqref{kin-res-1st}. %
This may be checked by 
$(\ell_{\gamma^\u})^{\bs Y}\defeq \gamma^* \Xi^* f_\u^* L = (\Xi\circ \gamma)^*f_\u^* L= (\gamma^{\bs Y})^* f_\u^* L= (f_\u \circ \gamma^{\bs Y})^*L$. 
Then observe that 
$f_{\u_i} \circ \gamma^{\bs Y}
=  \gamma^{\bs Y} + \u(\gamma^{\bs Y})
=\gamma+\bs Y(\gamma) + \u_i\big(\gamma+\bs Y(\gamma)  \big) 
=\gamma+\bs Y(\gamma) + \u_i(\gamma) -\bs Y_i(\gamma) 
=f_{u_i}(\gamma) + \b{\bs Y}^i(\gamma)
=\gamma^{\u_i} +  \b{\bs Y}^i(\gamma)
\rdefeq (\gamma^{\u_i})^{\b{\bs Y}}
$. 
So that $(\ell_{\gamma^\u})^{\bs Y}=\ell_{(\gamma^\u)^{\b{\bs Y}}}$.
The infinitesimal transformation is then
\begin{align}
 \label{dressed-ell-res-1-bis}   
 \delta_{\bs\Upsilon} \,\ell_\gamma^\u 
\defeq
\tfrac{d}{d\epsilon} \, (\ell_\gamma^\u)^{\bs Y_\epsilon} - \ell_\gamma^\u \, \big|_{\epsilon=0}
= \tfrac{d}{d\epsilon} \mathrm{c}_{\gamma^u}(\b{\bs Y_\epsilon}) \, \big|_{\epsilon=0} \rdefeq \mathrm{a}_{\gamma^\u}(\b{\bs\Upsilon}),
\end{align}
which is the variation of the relational Lagrangian under variation of the physical relational kinematical  history $\delta \gamma^\u =\b{\bs\Upsilon}(\gamma)$.
This is consistent with \eqref{dressed-L-res-3}, writing $\delta_{\bs\Upsilon} \,\ell_\gamma^\u = \gamma^* \big(\mathfrak{L}_{\,\bs\Upsilon^v} L^\u \big) = \gamma^* \mathrm{a}\big(\b{\bs\Upsilon}, f_\u(\_) \big)$.

The $\G$-transformation of $\ell^\u_\gamma$ is, by \eqref{dressed-L-res-2.0}, 
\begin{equation}
\label{dressed-ell-res-2}
\begin{aligned}
(\ell^\u_\gamma)^{\bs Z}
\defeq \gamma^*\big( L^\u + \mathrm{c}(f_\u(\_), {\bs Z}) \big)
= \ell^\u_\gamma + \mathrm{c}\big(f_\u(\gamma), {\bs Z}(\gamma)\big)
&=\ell^\u_\gamma + \mathrm{c}\big(\gamma^\u, {\bs Z}(\gamma)\big), \\[1mm]
\text{or equivalently} \quad 
(\ell_{\gamma^\u})^{\bs Z}
    &=  \ell_{\gamma^\u} + \mathrm{c}_{\gamma^\u}( {\bs Z}),
\end{aligned}
\end{equation}
which is the relational parametrized Lagrangian from the internal viewpoint of a distinct particle in the same physical kinematical history $(\gamma^\u)^{\bs Z}$ given by
\eqref{kin-res-2nd}.
Eq. \eqref{dressed-ell-res-2} showcases the physical reference frame covariance of the relational dynamics, controlled by the finite group $\G$: Taking a concrete case, \eqref{dressed-ell-res-2} is
$\ell^{\u_j}_\gamma = \ell^{\u_i}_\gamma + \mathrm{c}\big(\gamma^{\u_i}, \bs{Z}_{ij}(\gamma) \big)$. 
\medskip

Consequently, by \eqref{dressed-L} and \eqref{dressed-ell-1}-\eqref{dressed-ell-2} we find the dressed, relational action to be
\begin{equation}
\label{dressed-S}
\begin{aligned}
   \mathcal S[\gamma]^\u 
    \defeq \int_\gamma L^\u 
    = \int_I \ell^\u_\gamma 
    =&\, \int_I( \mathcal{L} \circ \gamma^\u)\ \dot{t}\,d\tau
    \simeq \int_I \mathcal{L}\big(t(\tau), \b x(\tau)\,\!\big)\ \dot{t}\, d\tau \\[2mm]
    \rotatebox[origin=c]{180}{\text{\LARGE$\Lsh$}} =&\, \int_I \ell_\gamma + \mathrm c_\gamma(\u) 
    \rdefeq \S[\gamma] + c_\gamma(\u),
\end{aligned}
\end{equation}
giving both the conceptually clear expression and the computationally useful form. The latter, as usual, could be directly read from the $\H_\text{ext}$-transformation \eqref{GT-S} of $\S[\gamma]$ by the DFM rule of thumb. 
Quite naturally, it is the action associated to the relational kinematics $\gamma^\u$, as it is clear from 
\begin{align}
\label{dressed-S-bis}
 \mathcal S[\gamma]^\u 
    \defeq \int_\gamma L^u 
    = \int_\gamma f^*_\u L
    = \int_{f_\u \circ \gamma} L 
    = \int_{\gamma^\u} L 
    \rdefeq S[\gamma^\u]. 
\end{align}
From \eqref{dressed-L-res-1} and \eqref{dressed-ell-res-1} we find that the $\H_\text{int}\simeq \H_\text{int}^\u$-transformation of the dressed action is
\begin{equation}
\label{dressed-S-res-1}
\begin{aligned}
(\S[\gamma]^\u)^{\bs Y}\defeq &\,
  \int_\gamma (L^\u)^{\bs Y}
 =\int_\gamma L^\u + \mathrm{c}\big(f_\u(\_), \b{\bs Y} \big)
 = \S[\gamma]^\u + c_{\gamma^u}(\b{\bs Y}), \\[1mm]
 \text{or equivalently } \quad
 (\S[\gamma^\u])^{\bs Y}\defeq&\,\int_I (\ell_{\gamma^\u})^{\bs Y}
 = \int_I \ell_{\gamma^\u} + \mathrm{c}_{\gamma^u}(\b{\bs Y}) 
 =\S[\gamma^\u] + c_{\gamma^u}(\b{\bs Y}).
\end{aligned}
\end{equation}
It is the dressed action  seen from the viewpoint of the same particle in a physically distinct kinematical history $(\gamma^\u)^{\b Y}$, given by \eqref{kin-res-1st}: indeed, one also writes 
$(\S[\gamma^\u])^{\bs Y}\defeq\S[(\gamma^\u)^{\bs Y}]=\S[(\gamma^\u)^{\b{\bs Y}}]$.
The infinitesimal residual transformation is then, for $\Upsilon \in$ Lie$\H_\text{int}$,
\begin{equation}
\begin{aligned}
\label{dressed-S-res1-bis}
\delta_{\bs \Upsilon} \S^\u_{|\gamma} 
\defeq \tfrac{d}{d\epsilon }\int_\gamma  \Xi_\epsilon^* L^\u \,\big|_{\epsilon=0}
=\int_\gamma \mathfrak L_{\bs \Upsilon^v} L^\u
= \int_\gamma \mathrm{a}\big(\b{\bs\Upsilon}, f_\u(\_) \big)
=\int_I \mathrm a_{\gamma^\u}(\b{\bs\Upsilon}) 
\rdefeq
\int_I \delta_{\bs \Upsilon}\ell_\gamma^\u.
\end{aligned}
\end{equation}
Equivalently, this is the Gateaux derivative of $\S[\gamma^\u]$ along $\H^\u_\text{int}\!\sim\!\Aut_v(\mC^\u_\text{rel})$: 
 for $\gamma^\u_\epsilon\defeq (\gamma^\u)^{{\bs Y}_\epsilon} = \gamma^\u + \b{\bs Y}_\epsilon (\gamma)$ and infinitesimal variation of the relational kinematical history $\delta_{\bs \Upsilon} \gamma^\u = \b{\bs\Upsilon}(\gamma)$, 
we have
\begin{align}
\label{dressed-S-gateau}
   \delta_{\bs \Upsilon} \S_{|\gamma^\u} 
\defeq 
\tfrac{d}{d\epsilon}\, \S[\gamma_\epsilon^\u] - \S[\gamma^\u]\,\big|_{\epsilon=0} 
= \tfrac{d}{d\epsilon}\,
    \int_\gamma \mathrm c\big(f_\u(\_), {\bs Y}_\epsilon \big)\, \big|_{\epsilon=0}
= \int_\gamma \mathrm a\big(\b{\bs\Upsilon} , f_\u(\_) \big).
\end{align}
This allows to define the critical points  of $\S^\u$, i.e.  the \emph{classical relational dynamics}.

We proceed as in section \ref{Classical mechanics on C}; Consider $\P_{0,1}^\u\defeq \big\{ \gamma^\u: I \rarrow  \mC^\u_\text{rel}\, |\, \gamma^u(\d I)=\{q_0, q_1\}\big\}$ the set of paths with beginning and end points $q_0=f_\u(p_0)\simeq (t_0, \b x_0)$ and  $q_1=f_\u(p_1)\simeq (t_1, \b x_1)$, 
i.e. $\P_{0,1}^\u =\big\{ \gamma^u\, |\, \gamma\in \P_{0,1} \big\}$.
Consider also the gauge subgroup 
$\H_\text{int\,$|\, 0, 1$} \defeq \big\{ \bs Y \in \H_\text{int}\, |\, \bs Y(t\leq t_0)=0=\bs Y(t\geq t_1)  \text{ and } \bs Y(t_0<t<t_1)\neq 0  \big\} \subset \H_\text{int}$, isomorphic to the similarly defined (\emph{mutatis mutandis})
$\H^\u_\text{int\,$|\, 0, 1$} 
\subset \H^\u_\text{int}$. 
We have that $\H_\text{int\,$|\, 0, 1$}\simeq \H^\u_\text{int\,$|\, 0, 1$}$ acts freely and transitively on $\P^\u_{0,1}$, i.e. 
for any choice of initial reference curve $\gamma^\u \in \P^\u_{0,1}$, any other is written as $(\gamma^\u)^{\bs Y}= (\gamma^\u)^{\b{\bs Y}}$ for some $\bs Y \in \H_\text{int\,$|\, 0, 1$} \simeq \b{\bs Y} \in \H^\u_\text{int\,$|\, 0, 1$}$: so $\P^u_{0,1} \simeq \H_\text{int\,$|\, 0, 1$} \simeq \H^\u_\text{int\,$|\, 0, 1$}$.

The \emph{classical relational history} $\gamma^\u_c$ between  $q_0$ and $q_1$ is the critical point of $\S[\gamma^\u]=\S[\gamma]^\u$ for $\gamma^\u \in \P^\u_{0,1}$. By \eqref{dressed-S-res1-bis}-\eqref{dressed-S-gateau}, it is given by the vanishing of the  $1$-cocycle $\mathrm{a}(\b{\bs\Upsilon})$ on $\gamma^\u_c$. 
It encodes  the dressed  Euler-Lagrange equations, i.e. the relational dynamics (up to vanishing boundary terms) as seen from within any particle in $\gamma$:
\begin{align}
\label{dressed-critical-point}
\delta_{\bs \Upsilon} \S_{|\gamma^\u_c} 
= \int_{\gamma^u_c} \mathrm a(\b{\bs \Upsilon})
=\int_I \mathrm a_{\gamma^\u_c}(\b{\bs \Upsilon})
=\int_I  E(\b{\bs\Upsilon}, \gamma^\u_c) \ 
\equiv 0, 
\quad \forall \ \bs \Upsilon \simeq \b{\bs \Upsilon}\in \text{Lie}\H_\text{int\,$|\, 0, 1$} \simeq \text{Lie}\H^\u_\text{int\,$|\, 0, 1$},
\end{align}
where $E(\b{\bs\Upsilon}, \gamma^\u)\defeq \b{\bs\Upsilon}(\gamma) \cdot EL[\gamma^\u]\,d\tau
=\delta_{\bs \Upsilon} \gamma^\u \cdot EL[\gamma^\u]\,d\tau$ is the equations of motion 1-form, with $EL[\gamma^\u]$ the Euler-Lagrange equations for the curve $\gamma^\u$.
The critical curve is s.t. $EL[\gamma^\u_c]\equiv 0$.
Then, for any $\gamma^\u\in \P^\u_{0,1}$ there is a unique $\bs Y \in \H^\u_\text{int\,$|\, 0, 1$}$ s.t. $\gamma^\u =(\gamma^\u_c)^{\bs Y}$, and  by \eqref{dressed-S-res-1} one has
\begin{align}
 \label{Diff-dressed-S-S-critical}   
 \S[\gamma^\u] = \S[\gamma^\u_c] + \int_{\gamma^\u_c} \mathrm c(\b{\bs Y}).
\end{align}
It is clear that $\gamma^\u_c=f_\u(\gamma_c)$, i.e. that the variational principle is well-behaved under dressing, meaning the \emph{relational variational principle} \eqref{dressed-critical-point} indeed produces the relational version of the dynamics obtained via the ``bare" variational principle \eqref{critical-point}. 
This can be seen from a direct computation, similar to \eqref{Link-bare-dressed-a}: 
using \eqref{Abelian-cocyclic-dressing-Upsilon} we have indeed
\begin{align}
\label{link-bare-dressed-var-princ}
 \delta_{\bs\Upsilon} S^\u_{|\gamma} 
 = \int_\gamma \mathfrak L_{\bs \Upsilon^v} L^\u
 =  \int_\gamma \mathfrak L_{\bs \Upsilon^v} L\  + \int_\gamma \mathfrak L_{\bs \Upsilon^v} \mathrm{c}(\u)
 &=  \delta_{\bs\Upsilon} S_{|\gamma} 
   + \int_\gamma \mathfrak L_{\bs \Upsilon^v} \mathrm{c}(\u),
   \qquad  \forall \bs\Upsilon \in \text{Lie}\H_\text{int},\\
 &=\int_\gamma \mathrm{a}(\bs\Upsilon)\ + \int_\gamma  \mathrm{a}\big(\b{\bs\Upsilon}, f_\u(\_) \big) - \mathrm{a}(\bs\Upsilon)  
 =\int_\gamma \mathrm{a}\big(\b{\bs\Upsilon}, f_\u(\_) \big), \notag
\end{align}    
which of course reproduces \eqref{dressed-critical-point}.
But for $\bs\Upsilon \in \text{Lie}\H_\text{int\,$|\, 0, 1$}$, by \eqref{critical-point}-\eqref{dressed-critical-point}, eq. \eqref{link-bare-dressed-var-princ} establishes the correspondence between the bare and the relational formulations of the variational principle.
Notice the remarkable fact that from the above we see that \emph{both} the bare and relational Euler-Lagrange equations can be obtained from the  Lie$\H_\text{int\,$|\, 0, 1$}$-transformation \eqref{Abelian-cocyclic-dressing-Upsilon} of the cocyclic dressing field $\mathrm{c}(\u)$. 

The $\G$-transformation of $\S[\gamma]^\u=\S[\gamma^\u]$ is, by \eqref{dressed-L-res-2.0} and \eqref{dressed-ell-res-2},
\begin{equation}
\label{dressed-S-res2}
\begin{aligned}
(\S[\gamma]^\u)^{\bs Z} \defeq&\,
\int_\gamma (L^\u)^{\bs Z}
= \int_\gamma L^\u + \mathrm{c}\big(f_\u(\_), \bs Z \big)
= \S[\gamma]^\u + c_{\gamma^\u}(\bs Z),\\[1mm]
 \text{or equivalently } \quad 
 \S[\gamma^\u]^{\bs Z} \defeq&\, 
 \int_I (\ell_{\gamma^\u})^{\bs Z}
 = \int_I \ell_{\gamma^\u} + \mathrm{c}_{\gamma^\u}( {\bs Z})
 = \S[\gamma^\u] + c_{\gamma^\u}( {\bs Z}).
\end{aligned}
\end{equation}
It the relational action from the internal viewpoint of a distinct particle in the same physical kinematical history $(\gamma^\u)^{\bs Z}$ given by
\eqref{kin-res-2nd}:
Specializing the notation in \eqref{dressed-S-res2}, it is
$\S[\gamma^{\u_j}] = \S[\gamma^{\u_i}] + {c}\big(\gamma^{\u_i}, \bs{Z}_{ij}(\gamma) \big)$.
It shows the physical reference frame covariance of the relational dynamics.

We can go a step further, establishing the correspondence between the relational formulation of the variational principle as written, e.g., from the viewpoints of the $i^{th}$ and the $j^{th}$ particles. It can be obtained by linearising a computation all but similar to \eqref{dressed-Ls-Y}, using \eqref{usefull-identity}, relating $\S[\gamma^{\u_j}]^{\bs Y}$ and $\S[\gamma^{\u_i}]^{\bs Y}$. 
Or it is obtained directly from \eqref{Lie-deriv-dressed-Ls-Upsilon}, or using \eqref{useful-identity-2}, 
\begin{align}
\label{link-dressed-var-princ-i-j}
 \delta_{\bs\Upsilon} S^{\u_j}_{|\gamma} 
 = \int_\gamma \mathfrak L_{\bs \Upsilon^v} L^{\u_j}
 &=  \int_\gamma \mathfrak L_{\bs \Upsilon^v} L^{\u_i} + \int_\gamma \mathfrak L_{\bs \Upsilon^v} \mathrm{c}\big(f_{\u_i}(\_), \bs Z_{ij} \big)  \notag \\
 &=  \delta_{\bs\Upsilon} S^{\u_i}_{|\gamma} 
   + \int_\gamma \mathfrak L_{\bs \Upsilon^v}\mathrm{c}\big(f_{\u_i}(\_), \bs Z_{ij} \big),
   \qquad  \forall \bs\Upsilon \in \text{Lie}\H_\text{int},\\
 &= \int_\gamma \cancel{\mathrm a\big(\b{\bs\Upsilon}^i , f_{\u_i}(\_) \big) }
 +
 \int_\gamma  \mathrm{a}\big(\b{\bs \Upsilon}^j, f_{\u_j}(\_)  \big)
-
\cancel{\mathrm{a}\big( \b{\bs \Upsilon}^i, f_{\u_i}(\_)  \big)}
 =
 \int_\gamma \mathrm{a}\big(\b{\bs \Upsilon}^j, f_{\u_j}(\_)  \big), \notag
\end{align}    
which reproduces \eqref{dressed-S-res1-bis}, as expected. 
Remark the similarity with \eqref{link-bare-dressed-var-princ}, using \eqref{Link-bare-dressed-a}-\eqref{Abelian-cocyclic-dressing-Upsilon}.
And, similarly, for $\bs\Upsilon \in \text{Lie}\H_\text{int\,$|\, 0, 1$}$, by \eqref{dressed-critical-point}, eq. \eqref{link-dressed-var-princ-i-j} establishes the equivalence between the relational formulations of the variational principle written by $i^{th}$ and the $j^{th}$ particles.
We also notice the remarkable fact that from the above we see that \emph{both} the relational Euler-Lagrange equations written from the viewpoints of the $i^{th}$ and the $j^{th}$ particles can be obtained from the  Lie$\H_\text{int\,$|\, 0, 1$}$-transformation \eqref{useful-identity-2} of the cocyclic quantity $\mathrm{c}\big(f_{\u_i}(\_), \bs Z_{ij} \big)$. 
\medskip

The dressed action induces the function 
$S^\u_\gamma=S_{\!\gamma^\u}= \S[\gamma^\u;\_\,,  q_0]$ with  
$\gamma^\u(\d I)=\{q_0, q\}$: 
It is $\H_\text{ext}$-invariant by construction and cocyclic $\H_\text{int}\simeq \H^\u_\text{int}$-tensorial, $S_{\!\gamma^\u} \in
\Omega_\text{tens}^0(\mC^\u_\text{rel}; c_{\gamma^\u})$.
It is then immediate that its $\H_\text{int}\simeq \H^\u_\text{int}$-transformation is given by $(S_{\gamma^\u})^{\bs Y}=(S_{\gamma^\u})^{\b{\bs Y}}= S_{\gamma^\u}+ c_{\gamma^\u}(\b{\bs Y})$. 
For $\gamma^\u =\gamma^\u_c$, the cocyclic 0-form $S_{\gamma^\u_c}$ is the \emph{dressed} Hamilton Principal Function, and  satisfies the dressed Hamilton equation: 
$\tfrac{\d S^\u_{\gamma_c}}{dt} 
=
-H\big(\b{x}, \tfrac{\d S^\u_{\gamma_c}}{\d \b{x}}; t \big)
$,
 where the dressed canonical conjugate momenta is by definition $\b{\mathrm  \uppi}\defeq \tfrac{\d S^\u_{\gamma_c}}{\d \b x}$ and $H(\b x, \b{\mathrm \uppi}, t)$ is the dressed Hamiltonian. 

Defining the $U(1)$-valued $H^\u_\text{int}$-1-cocycle
 $C_{\gamma^\u}(\_ \,, \b Y) 
 = C_{\gamma}(f_\u(\_)\,, \b Y) 
 \defeq \exp\{-\tfrac{i}{\hbar}c_{\gamma^\u}(\_ \ , \b Y)\}
 = \exp\{-\tfrac{i}{\hbar} \int_\gamma \mathrm{c}(f_\u(\_)\,, \b Y)\}$,
 the dressed action induces 
 the \emph{ dressed flat cocyclic connection} $\varpi^\u_{0,\gamma}\defeq -\tfrac{i}{\hbar} d S^\u_\gamma$ on $\mC^\u_\text{rel}$, s.t. 
\begin{equation}
\begin{aligned}
\label{dressed-flat-cocyclic-conn}
R^*_{\b Y} \varpi^\u_{0,\gamma} &= \varpi^\u_{0,\gamma} + C_{\gamma^\u}(\_\,, \b Y)\- dC_{\gamma^\u}(\_\,, \b Y), \\[1mm]
\iota_{{\b\Upsilon}^v} \varpi^\u_{0,\gamma} 
&= -\tfrac{i}{\hbar}\, \iota_{{\b\Upsilon}^v} dS^\u_\gamma = -\tfrac{i}{\hbar}\, L_{\chi^v} S^\u_\gamma
= -\tfrac{i}{\hbar}\, \tfrac{d}{d\epsilon}\, R^*_{\b Y_\epsilon} S^\u_\gamma \, \big|_{\epsilon=0} 
= -\tfrac{i}{\hbar}\, \tfrac{d}{d\epsilon}\, c_{\gamma^\u}(\_ \,,\b Y_\epsilon) \, \big|_{\epsilon=0} 
= -\tfrac{i}{\hbar}\, \int_{\gamma} \mathrm a\big({\b\Upsilon}, f_\u(\_) \big)\\[-1.5
mm]
&=\tfrac{d}{d\epsilon}\, C_{\gamma^\u}(\ \, , \b Y_\epsilon)\, \big|_{\epsilon=0}, 
\end{aligned}
\end{equation}
which is the dressing of \eqref{flat-cocyclic-conn}: By \eqref{dressed-S}, indeed,
\begin{align}
\label{dressed-varpi-ds}
\varpi^\u_{0, \gamma}
=
\varpi_{0, \gamma }   +  C_{\gamma}(\u)\- dC_\gamma(\u),
\end{align}
a special case of \eqref{twisted-dressed-forms}.
It follows that the $\H_\text{int}\simeq\H_\text{int}^\u$-transformation of $\varpi^\u_{0,\gamma}$ is 
\begin{equation}
\label{GT-dressed-twised-conn-dS}
\begin{aligned}
({\varpi^\u_{0,\gamma}})^{\b{\bs Y}} \defeq \Xi^* \varpi^\u_{0,\gamma}= \varpi_{0,\gamma}^\u +  C_{\gamma^\u}(\b{\bs Y})\- dC_{\gamma^
\u}(\b{\bs Y}),
\end{aligned}
\end{equation}
which is  $(dS^\u_\gamma)^{\b{\bs Y}}= dS^\u_\gamma + d\mathrm c_{\gamma^\u}(\b{\bs Y})$,  as expected from $\H_\text{int}\simeq\H_\text{int}^\u$-transformation of $S^\u_\gamma$ and $[d, \Xi^*]=0$ for $\Xi\in \H_\text{int}^\u$.
Since $E(\b{\bs Y}; \gamma^\u_c)\equiv 0$ by definition of $\gamma^\u_c$, the verticality property of $\varpi^\u_{0,\gamma_c}=\varpi_{0,{\gamma^\u_c}}$ gives the dressed conserved Noether charge: 
In a bundle  chart, $\b \Upsilon^v \simeq \chi \tfrac{\d}{\d \b x}$ and $dS^\u_{\gamma_c} = \tfrac{\d S^\u_{\gamma_c}}{\d \b x} d\b x + \tfrac{\d S^\u_{\gamma_c}}{\d t} dt$, so $\iota_{\b\Upsilon^v} \varpi^\u_{0,\gamma_c} = -\tfrac{i}{\hbar}\, \chi\, \tfrac{\d S^\u_{\gamma_c}}{\d \b x} \rdefeq -\tfrac{i}{\hbar}\, \b \Upsilon\, \b{\mathrm \uppi}$.
Finally, notice that \eqref{Diff-dressed-S-S-critical}  implies a relation, analogous to \eqref{GT-dressed-twised-conn-dS}: For $\b{\bs Y} \in \H_{\text{int}\,|\,0,1}$ s.t. $\gamma^\u=(\gamma^\u_c)^{\b{\bs Y}}$, we have
\begin{align}
 \label{Link-dressed-twist-crit-vs-non-crit}
 \varpi^\u_{0,\gamma} = \varpi^\u_{0,\gamma_c} +   C_{\gamma^\u_c}(\b{\bs Y})\- dC_{\gamma^\u_c}(\b{\bs Y}).
\end{align}
\medskip
Under the action of $\G$, by \eqref{dressed-S-res2}, we have 
$(S^\u_\gamma)^{\bs Z}=S_{\!(\gamma^\u)^{\bs Z}} = S^\u_\gamma + c_{\gamma^\u}(\bs Z)$. 
Specializing, $S^\u_\gamma=S^{\u_i}_\gamma$ and $(S^\u_{\gamma})^{\bs Z}=S^{\u_j}_{\gamma}$ are the same object seen from distinct viewpoint within $\gamma$.
In particular, $(S^\u_{\gamma_c})^{\bs Z}=S^{\u_j}_{\gamma_c}$ is the dressed HPF described from the $j^{th}$ particle.
This induces the cocyclic connection 
 $(\varpi^\u_{0,\gamma})^{\bs Z}\defeq -\tfrac{i}{\hbar} d (S^\u_\gamma)^{\bs Z}$ on $\mC^{\u^{\bs Z}}_\text{rel} = \mC^{\u_j}_\text{rel}$, explicitly,
\begin{equation}
\label{dressed-twised-conn-res2}
({\varpi^\u_{0,\gamma}})^{\bs Z} 
=
\varpi_{0,\gamma}^\u +  C_{\gamma^\u}(\bs Z)\- dC_{\gamma^\u}(\bs Z), 
\quad
\text{ i.e. }
\quad 
({\varpi^{\u_j}_{0,\gamma}}) 
=
\varpi_{0,\gamma}^{\u_i} +  C_{\gamma^{\u_i}}(\bs Z_{ij})\- dC_{\gamma^{\u_i}}(\bs Z_{ij}).
\end{equation}
This is another key instance showing how the discrete group $\G$  relates the relational descriptions on $\mC^{\u_i}_\text{rel}$ and $\mC^{\u_j}_\text{rel}$, and implement the physical reference frame covariance of this description.
\medskip

After this thorough presentation of relational classical mechanics, we can now turn to the formulation of relational quantum mechanics.


\subsection{Relational Quantum Mechanics, and physical frame covariance}
\label{Relational Quantum Mechanics}

 We define the dressed pre-wave function, using 
\eqref{dressed-forms-1}-\eqref{twisted-dressed-forms} or the DFM rule of thumb, 
\begin{align}
 \label{dressed-psi-gamma}
\psi_\gamma^{\u}\defeq f^*_\u \psi_\gamma = C_\gamma(\u)\-\psi_\gamma, 
\end{align}
where $ C_\gamma(\u)\defeq C_\gamma(\_\,, \u)=\exp\{-\tfrac{i}{\hbar}c_\gamma\big(\_\, ,\u(\_)\big)\}$. It is $H_\text{ext}$-basic on $\mC$ by construction, and $\H_\text{ext}$-invariant as is checked  using the $\H_\text{ext}$-transformation of $\psi_\gamma$, $\psi_\gamma^{\bs X} = C_\gamma(\bs X)\-\psi_\gamma$, the fact that by \eqref{Abelian-cocyclic-dressing-X} one has $C_\gamma(\u)^{\bs X}=C_\gamma(\bs X)\-C_\gamma(\u)$. 
It is thus a 0-form on $\mC^{\,\u}_\text{rel}$, as it is  clear from \eqref{Generic-f-u-p}, and in bundle coordinates
\begin{align}
 \label{dressed-psi-bis}
\psi_\gamma^\u(p)
= \psi_\gamma\circ f_{\u}(p)
= \psi_\gamma \big(p + \u(p)\big) 
\simeq 
\psi_\gamma \left(\,t; \b x_1,\, \ldots\,, \b x_{N-1}\,  \right)
= \psi_\gamma (t, \b x),
\end{align}
which also shows that we may write $\psi^\u_\gamma=\psi_{\gamma^\u}$.
Specializing this to e.g. $\u=\u_i$, it is $\psi^{\u_i}_\gamma(p)\simeq \psi_\gamma(t, \b x^i)
=\psi_\gamma(t, \b x^i_1, \dots, \b x^i_{N-1})
=\psi_\gamma(t; x_1 - x_i, \dots, 0, \dots, x_N-x_i )$, i.e. the pre-wave function as seen from the $i^{th}$ particle. 
Again, residual $\H_\text{int}\simeq \H^\u_\text{int}$-transformations are naturally expected and easily written, since, from \eqref{Abelian-cocyclic-dressing-Y}, we have 
\begin{align}
\label{C(u)-Y}    
C_\gamma(\u)^{\bs Y} = C_\gamma(\bs Y)\-C_\gamma(\u)\, C_{\gamma^\u}(\b{\bs Y}),
\end{align}
so that we find 
\begin{align}
\label{dressed-psi-Y}
(\psi_\gamma^\u)^{\bs Y}
=
(\psi_{\gamma^\u})^{\bs Y}
&=
\big( C_\gamma(\u)^{\bs Y}\big)\-\psi_\gamma^{\bs Y},  \notag\\
&=
 C_{\gamma^\u}(\b{\bs Y})\-  C_\gamma(\u)\-C_\gamma(\bs Y)
 \cdot
 C_\gamma(\bs Y)\- \psi_\gamma, \notag \\
 &=
 C_{\gamma^\u}(\b{\bs Y})\- \psi_{\gamma^\u} 
 \rdefeq
 (\psi_{\gamma^\u})^{\b{\bs Y}}.
\end{align}
The dressed pre-wave function is thus a cocyclic tensorial 0-form on $\mC^{\,\u}_\text{rel}$,  $\psi^\u_\gamma=\psi_{\gamma^\u} \in \Omega^0_\text{tens}\big(\mC^{\,\u}_\text{rel}, C_{\gamma^\u}\big)$.
So it can also be seen as a section of the cocyclic associated bundle 
$E^{C_{\gamma^\u}}\defeq \mC^{\,\u}_\text{rel} \times_{C_{\gamma^\u}} \CC$.
So that $\Omega^0_\text{tens}\big(\mC^{\,\u}_\text{rel}, C_{\gamma^\u}\big) \simeq \Gamma\big(E^{C_{\gamma^\u}}\big)$ is the space of ``dressed pre-wave functions". 

The dressed cocyclic connection \eqref{dressed-varpi-ds}-\eqref{GT-dressed-twised-conn-dS} induces the  cocyclic covariant derivative preserving this space
\begin{equation}
\label{dressed-cocyclic-cov-der-QM}
\begin{aligned}
\b D_{0, {\gamma^\u}} : \Omega^0_\text{tens}\big(\mC^{\,\u}_\text{rel}, C_{\gamma^\u} \big) &\rarrow \Omega^1_\text{tens}\big(\mC^{\,\u}_\text{rel}, C_{\gamma^\u} \big), \\
\psi_{\gamma^\u} &\mapsto \b D_{0, {\gamma^\u}}\, \psi_{\gamma^\u} \defeq d\psi_{\gamma^\u}+ \varpi_{0,{\gamma^\u}}\,\psi_{\gamma^\u}.
\end{aligned}
\end{equation}
It is the dressing of the bare cocyclic covariant derivative; by 
\eqref{dressed-varpi-ds}-\eqref{dressed-psi-gamma} we have
$\b D_{0, {\gamma^\u}}\, \psi_{\gamma^\u} = C_\gamma(\u)\- \b D_{0, {\gamma}}\, \psi_{\gamma}$.
From the horizontality  of $\b D_{0, \gamma^\u}\, \psi_{\gamma^\u}$ follows that in bundle coordinates 
$-i \hbar\tfrac{\d}{\d \b x} =\tfrac{\d S_{\gamma^\u}}{\d \b x}$.
By \eqref{Link-dressed-twist-crit-vs-non-crit}, we have that
\begin{equation}
\label{critic-vs-non-critic}
 \begin{aligned}
 \psi_{\gamma^\u} =  C_{\gamma^\u_c}(\b{\bs Y})\- \psi_{\gamma^\u_c},
 \quad
\text{and}
\quad
\b D_{0, \gamma^\u}\, \psi_{\gamma^\u} 
= C_{\gamma^\u_c}(\b{\bs Y})\-\b D_{0, \gamma^\u_c}\, \psi_{\gamma^\u_c}
\quad
\text{for } \b{\bs Y} \in \H^\u_{\text{int}\,|\, 0,1}.
 \end{aligned}   
\end{equation}
The space of dressed wave function is 
${\mK}^\u \defeq \ker \b D_{0, {\gamma^\u}} \subset \Omega^0_\text{tens}\big(\mC^\u_\text{rel}, C_{\gamma^\u}\big)$, i.e. the covariantly constant  dressed pre-wave functions.
A relational version of QM is encoded in the statement
\begin{align}
 \label{dressed-QM-0}
\psi_{\gamma^\u_c} \in \mK^\u, 
\quad \text{\ i.e. } \quad
\b D_{0, \gamma^\u_c}\, \psi_{\gamma^\u_c}=0.
 \end{align}
Since $S_{\gamma^\u_c}=\S[\gamma^\u_c; \_\,, q_0]$ is the dressed HPF, satisfying the relational Hamilton-Jacobi equation, in bundle coordinates we have
\begin{equation}
 \label{dressed-QM-1}
\begin{aligned}
\b D_{0, \gamma^\u_c}\, \psi_{\gamma^\u_c}
= d \psi_{\gamma^\u_c} + \varpi_{0, \gamma^\u_c}\psi_{\gamma^\u_c}=0
\quad &\Rightarrow \quad
dt\,\tfrac{\d}{\d t} \psi_{\gamma^\u_c} + d\b x\,\tfrac{\d}{\d \b x}\psi_{\gamma^\u_c} 
- \tfrac{i}{\hbar} \left( 
dt\,\tfrac{\d S_{\gamma^\u_c}}{\d t} 
+
d\b x\,\tfrac{\d S_{\gamma^\u_c}}{\d \b x} 
\right)\psi_{\gamma^\u_c} =0 \\[1mm]
&\hookrightarrow \quad 
 dt\,\left(\tfrac{\d}{\d t}\psi_{\gamma^\u_c} + \tfrac{i}{\hbar} H(\b x, \b{\mathrm \uppi}, t)\,\psi_{\gamma^\u_c}\right)
 +
 d\b x\,\left( \tfrac{\d}{\d \b x} -\tfrac{i}{\hbar} \b{\mathrm \uppi }\right)\,\psi_{\gamma^\u_c} =0 .
\end{aligned}
\end{equation}
This gives the relational version of the quantum mechanical prescription for the momentum operator $\hat {\b{\mathrm \uppi}} \defeq -i\hbar\tfrac{\d}{\d \b{x}}$, from which  we get the relational canonical commutation relation: $[\hat{\b{\mathrm x}}, \hat{\b{\mathrm \uppi}}]= i\hbar \delta(\b{\mathrm x} - \b x)$.
So, altogether we have
\begin{equation}
\label{dressed-QM-3}
\begin{aligned}
 \b D_{0, \gamma^\u_c}\, \psi_{\gamma^\u_c}\equiv 0 
\qquad \Rightarrow \qquad 
\hat {\b{\mathrm \uppi}} \defeq -i\hbar\tfrac{\d}{\d \b x} 
\quad \text{ and } \quad 
i\hbar \tfrac{\d}{\d t} \psi_{\gamma^\u_c} = H(\hat{\b{\mathrm x}}, \hat{\b{\mathrm \uppi}}, t)\, \psi_{\gamma^\u_c}.
\end{aligned}
\end{equation}
It extends to any $\psi_{\gamma^\u} \in \mK^\u$ by  \eqref{critic-vs-non-critic}.
Relational QM is equivalently encoded in 
$\t\psi_{\gamma^\u} \in \t \mK^\u$, where $\t\mK^\u:=\ker \b\nabla^\u_0$ with 
$\b \nabla^\u_0: \Gamma(E^{C_{\gamma^\u}}) \rarrow \Omega^1(\T)\otimes E^{C_{\gamma^\u}}$ the cocyclic covariant derivative induced on  associated bundles $E^{C_{\gamma^\u}} \rarrow \T$. 

\paragraph{Dressed path integral}

The above allows to define the dressed, or relational, version of the PI, 
defined on $\P_{0,1}^\u\defeq \big\{ \gamma^\u: I \rarrow  \mC^\u_\text{rel}\, |\, \gamma^u(\d I)=\{q_0, q_1\}\big\}=\big\{ \gamma^u\, |\, \gamma\in \P_{0,1} \big\}$ the space of paths with beginning and end points $q_0=f_\u(p_0)\simeq (t_0, \b x_0)$ and  $q_1=f_\u(p_1)\simeq (t_1, \b x_1)$.
We define the dressed propagator,
\begin{align}
 \label{dressed-propagator-0}
  K^\u(p, p_0)
  \defeq
  f^*_\u K (p, p_0)
  =
 K(q, q_0) \defeq \int_{\,\P^\u_{0,1}}\!\! \D \gamma^\u   \ \psi_{\gamma^\u}
 \ = 
 \int_{\,\H^\u_{\text{int}\,|\, 0,1}}\!\!\!\!\!\!\!\! \D \b{\bs Y}  \ \ \psi_{(\gamma^\u_r)^{\b{\bs Y}}}
 \ =
 \int_{\,\H^\u_{\text{int}\,|\, 0,1}}\!\!\!\!\!\!\!\! \D \b{\bs Y}   \ C_{\gamma^\u_r}(\b{\bs Y})\-\psi_{\gamma^\u_r} ,
\end{align}
where $\D \gamma^\u$ and $\D\b{\bs Y}$ are  formal measures on  $\P^\u_{0,1}$ and  $\H^\u_{\text{int}\,|\, 0,1}$ respectively.
The dressed pre-wave function depending only on the final point is
\begin{align}
\label{dressed-pre-wave-Psi}
\psi^\u(p)
\defeq f^*_\u \psi(p)
= \psi(q)
=\int_{\mC^{\,\u}_{\text{rel}\,|t_0}}\!\!\!\! K(q, q_0)\,  \psi(q_0)\,  \vol(\mC^{\,\u}_{\text{rel}\,|t_0}),
\end{align}
with $\vol(\mC^{\,\u}_{\text{rel}\,|t_0})$  the volume form of the fiber $\mC^{\,\u}_{\text{rel}\,|t_0} \subset \mC^{\,\u}_\text{rel}$ over $\pi(q_0)=t_0 \in \T$. 
In bundle coordinates, $\psi(q)\simeq \psi(t, \b x)$.
Taking the relational classical history $\gamma^\u_r=\gamma^\u_c$ as reference curve, we write 
\begin{align}
 \label{dressed-propagator-1}
 K(q, q_0) \defeq \int_{\,\P^\u_{0,1}}\!\! \D \gamma^\u  \ \psi_{\gamma^\u}
 \ =
 \int_{\,\H^\u_{\text{int}\,|\, 0,1}}\!\!\!\!\!\!\!\!  \D \b{\bs Y}  \ \ C_{\gamma^\u_c}(\b{\bs Y})\-\psi_{\gamma^\u_c} .
\end{align}
Further restricting attention to dressed wave functions $\psi_{\gamma^\u} \in \mK^\u$,  which are solutions of $\b D_{0, \gamma^\u} \psi_{\gamma^\u}=0$ of the form
\begin{align}
\label{dressed-Dirac-Feynman-weight}
\psi_{\gamma^\u}
=
\psi_0 \exp\int_{\gamma^\u} \varpi_{0, \gamma^\u}
=
\psi_0 \exp\tfrac{i}{\hbar}S_{\gamma^\u}
=
\psi_0 \exp\tfrac{i}{\hbar}\S[\gamma^\u],
\end{align}
then the propagator  \eqref{dressed-propagator-1} becomes the \emph{relational path integral}
\begin{align}
 \label{dressed-propagator-2}
 K(q, q_0) &= \int_{\,\P^\u_{0,1}}\!\! \D \gamma^\u  \ \ \psi_0 \exp\tfrac{i}{\hbar}\S[\gamma^\u] \\
  &=
 \int_{\,\H^\u_{\text{int}\,|\,0,1}}\!\!\!\!\!\!\!\! \D \b{\bs Y}  \ \ C_{\gamma^\u_c}(\b{\bs Y})\-\psi_0 \exp\tfrac{i}{\hbar}\S[\gamma^\u_c]
 \ =
 \psi_0 \exp\tfrac{i}{\hbar}\S[\gamma^\u_c]
 \int_{\,\H^\u_{\text{int}\,|\,0,1}}\!\!\!\!\!\!\!\!  \D \b{\bs Y}  \ \ 
 \exp\tfrac{i}{\hbar}c_{\gamma^\u_c}(\b{\bs Y}). \notag
\end{align}
Then $\psi^\u(p)=\psi(q)$ in \eqref{dressed-pre-wave-Psi} is a relational wave function, e.g. as seen from the $i^{th}$ particle if $\u=\u_i$.
As in the bare case, 
we obtain the 
splitting of the  relational PI in a contribution from the  classical history, $\psi_{\gamma^\u_c}=\psi_0 \exp\tfrac{i}{\hbar}\S[\gamma^\u_c]$, containing the essential dynamical physical information,  and a normalisation factor given by  functional integration of the 1-cocycle $c_{\gamma^\u_c}(\b{\bs Y})\defeq\int_{\gamma^\u_c} \mathrm c(\_, \b{\bs Y}) = \int_I \mathrm c(\_, \b{\bs Y}) \circ \gamma^\u_c$.

\paragraph{Reference frame covariance} 

The dressed pre-wave functions support $\G$-tranformations. 
The most direct way to write it is to notice that from \eqref{Abelian-cocyclic-dressing-Z} follows that the cocyclic dressing $C_\gamma(\u)$ transforms as
\begin{align}
 \label{C(u)-Z}
 C_\gamma(\u)^{\bs Z} = C_\gamma(\u)\,C_{\gamma^\u}(\bs Z),
\end{align}
as a special case of \eqref{Ambig-twisted-dress-field},
while $\psi_\gamma$ is $\G$-invariant by construction. 
A dressed pre-wave function \eqref{dressed-pre-wave-Psi} $\G$-transforms~as
\begin{align}
 \label{G-trsf-dressed-psi}
 \big(\psi_\gamma^{\u} \big)^{\bs Z}
 &= 
 \big(C_\gamma(\u)^{\bs Z}\big)\-(\psi_\gamma)^{\bs Z}
 =C_{\gamma^\u}(\bs Z)\- C_\gamma(\u)\- \psi_\gamma \notag\\
 &=
 C_{\gamma^\u}(\bs Z)\- \psi_\gamma^{\u}.
\end{align}
This is a case of \eqref{resid-trsf-dressed-twisted-forms}, and can also be written  as $(\psi_{\gamma^\u})^{\bs Z}= C_{\gamma^\u}(\bs Z)\- \psi_{\gamma^\u}$.
This result is also obtained from defining 
$\big(\psi_\gamma^{\u} \big)^{\bs Z} 
=\psi_\gamma^{(\u^{\bs Z})}  
\defeq 
f^*_{\u^{\bs Z}} \psi_\gamma$ via \eqref{f-u-Z}, 
which is thus the same pre-wave function seen from a different viewpoint. 
The~result \eqref{G-trsf-dressed-psi}  showcases the covariance of the relational description under change of physical reference frame. 
Specializing e.g. to $\u=\u_i$ and $\u^{\bs Z}=\u_j=\u_i + \bs Z_{ij}$, by \eqref{Ref-frame-change} we have, in bundle coordinates,
\begin{align}
\label{dressed-psi-gamma-Z}
\psi^{\u_j}_\gamma(p) 
&\simeq 
\psi_\gamma\big(t, \b x^j \big)
=
\psi_\gamma\big(t; x_1 -x_j, \ldots, x_i-x_j, \ldots, 0, \ldots, x_N-x_j \ \big), \notag\\
\psi^{\u_i}_\gamma(p) 
&\simeq 
\psi_\gamma\big(t, \b x^i \big)
=
\psi_\gamma\big(t; x_1 -x_i, \ldots, 0, \ldots, x_j-x_i, \ldots, x_N-x_i \ \big), \notag\\[1mm]
\text{and}& \quad 
\psi_\gamma\big(t, \b x^j \big) 
=
 C\big((t, \b x^i) , \bs Z_{ij}\big)\- \psi_\gamma\big(t, \b x^i \big).
\end{align}
Correspondingly, by
\eqref{dressed-twised-conn-res2} we have that
\begin{equation}
\label{D-psi-Z}
\begin{aligned}
 \big(\b D_{0, {\gamma^\u}}\, \psi_{\gamma^\u}\big)^{\bs Z}  
 &=C_{\gamma^\u}(\bs Z)\- \b D_{0, {\gamma^\u}}\, \psi_{\gamma^\u}, \\
 \hookrightarrow
 \b D_{0, {\gamma^{\u_j}}}\, \psi_{\gamma^{\u_j}} 
 &=C_{\gamma^{\u_i}}(\bs Z_{ij})\- \b D_{0, {\gamma^{\u_i}}}\, \psi_{\gamma^{\u_i}}.
\end{aligned}
\end{equation}
This shows that $\mK^{\u_j}\defeq \ker\b D_{0, {\gamma^{\u_j}}}  \simeq \mK^{\u_i}\defeq \b D_{0, {\gamma^{\u_i}}}$, which in turn implies the physical reference frame covariance of the relational description of QM \eqref{dressed-QM-3} -- here displayed on $\mC^{\u_i}_\text{rel}$ and $\mC^{\u_j}_\text{rel}$ -- as one should expect, a covariance explicitly controlled by the cocyclic structure. 
It is clear it similarly controls the frame covariance of the relational PI, via $K^{(\u^{\bs Z})}(p, p_0)$  and $\psi^{(\u^{\bs Z})}(p)$, 
so that e.g. relational wave functions on $\mC^{\u_i}_\text{rel}$ and $\mC^{\u_j}_\text{rel}$ 
are, in bundle coordinates,
\begin{align}
\label{dressed-wave-functions-Z}
\psi^{\u_j}(p) 
&\simeq 
\psi\big(t, \b x^j \big)
=
\psi\big(t; x_1 -x_j, \ldots, x_i-x_j, \ldots, 0, \ldots, x_N-x_j \ \big), \notag\\
\psi^{\u_i}(p) 
&\simeq 
\psi\big(t, \b x^i \big)
=
\psi \big(t; x_1 -x_i, \ldots, 0, \ldots, x_j-x_i, \ldots, x_N-x_i \ \big), \notag\\[1mm]
\text{and}& \quad 
\psi\big(t, \b x^j \big) 
=
 C\big((t, \b x^i) , \bs Z_{ij}\big)\- \psi\big(t, \b x^i \big).
\end{align}
 Observe that the change of physical reference frame is a unitary transformation of the relational quantum state $\psi^\u$, as it is implemented by the $U(1)$-valued $1$-cocyle $C(\bs Z)$. 

\section{Conclusion} 
\label{Conclusion}  

It is interesting to interpret eqs. \eqref{G-trsf-dressed-psi}-\eqref{dressed-wave-functions-Z}, key  results of the relational formulation of QM via DFM:
A first reading would be to say that a reference subsystem $s$, here a particle $s_i=p_i$, ``sees itself" as classical, and only the other subsystems (here the other $N-1$ particles in the $N$ particles configuration) are described  quantum mechanically from its perspective. 
Yet, the $\G$-transformations -- the transformations of the 2nd kind in the DFM framework -- implementing physical frame covariance ensure a ``quantum democracy" in that any subsystem is as legitimate as any to witness (write) the quantum dynamics. 

A more refined and accurate understanding would be to say that there is no meaningful way for a given reference subsystem $s_i=p_i$ to assign \emph{itself} a \emph{relational quantum state}: it can only do so for other subsystems, which are assigned a wave function with respect to $s_i=p_i$:  the dressed/relational wave function $\psi^{\u_i}$ \eqref{dressed-psi-gamma}/\eqref{dressed-pre-wave-Psi}. 
Said otherwise still, for a given subsystem  $s_i=p_i$, it only makes sense to assign a quantum state to the set of \emph{relations} between itself and the rest of the world, which is what $\psi^{\u_i}$ represents.
But the physical reference frame covariance implemented by $\G$-transformations ensures the equivalence (interchangeability) of any subsystem viewpoint to describe the collective relational quantum dynamics.

The reader will have certainly noticed that the dressed version of QM is very much alike the bare one: 
The dressed wave function $\psi^{\u_i}$ for $N-1$ particles assigned by the $i^{th}$ particle from within the $N$ particles configuration is not unlike the bare wave function $\psi$ for the $N$ particles configuration assigned by an idealized ``external observer", i.e. an ``external" physical system $s_\text{ext}$.
A benefit of the relational formulation via DFM is to clearly indicate that
the initial bare $\psi$ can indeed be understood as the dressed wave function  assigned to the ``internal" system $s_\text{int}=\{N  \text{ particles configuration}\}$ by $s_\text{ext}$ from within the total system $s_\text{tot}=s_\text{int} \cup s_\text{ext}$ ($=\{ (N+1) \text{ particles configuration}\}$). 

 A further benefit of the DFM approach to relational QM is its built-in treatment of physical reference frame covariance by $\G$-transformations: 
It dispels the possible a priori misconception that QM relies on  ``classical observers/systems" with a special status, 
showing instead how it naturally instantiate the above mentioned ``quantum democracy".
This is
broadly in line with heuristic ideas behind the literature on ``quantum reference frames" (QRFs), as discussed e.g. in \cite{Giacomini-et-al2019}.\footnote{We avoid the terminology ``QRF" as in our estimation it is potentially misleading,  suggesting a ``quantum fuzziness" of the reference subsystem $s_i$ \emph{in its own frame}.}

The relational formulation of QM via DFM thus shows explicitly something that might be overlooked on first reading:
that QM is a description of physics \emph{from within}
 -- without the need for an all-encompassing external viewpoint (``god's eye view"), nor of a partition between classical measuring devices and quantum systems (``Heisenberg cut").
This is broadly in line with Rovelli's  ``relational interpretation" of QM \cite{Rovelli1996, Rovelli2004, Rovelli-Vidotto2014}.

Remark also that the relational structure of QM we focus on here rests on that of classical mechanics, the DFM treating both in essentially the same way. 
From which follows that QM is plainly seen to be about quantization of relational d.o.f.: 
in the simple setup tackled here,
 the network of relative  positions of classical particles. 
\medskip

There are a number of developments that we shall pursue. 
First, our framework may be naturally generalised to the mechanics of extended bodies, undergoing rotations, or of particles with spin. 
The latter case would be treated elegantly via bundle supergeometry, thereby going back to Berezin original motivation for  the introduction of supergeometry in (established) physics \cite{Berezin-Marinov1977}.
Another future natural development will be to extend our framework to write a  bundle geometric formulation of relativistic QM --  which  may e.g. reproduce and generalise \cite{Chiou2013, Seidewitz2006, Redmount-Suen1993} -- whose  relational version would be obtained via the DFM.
But the immediate follow-up to the present work will be the companion \cite{JTF-Ravera2025RQ}, where we shall recast classical mechanics as a 1-dimensional ``general-relativistic" gauge field theory, to which the DFM as developed in \cite{JTF-Ravera2024gRGFT} applies. 
We will thereby highlight another key facet of its relational structure and, in echo to the observation of the previous paragraph, we will establish the notion of \emph{relational quantization}. 
A notion we shall then develop for standard general-relativistic Gauge Field Theory.


\section*{Acknowledgment}  

J.F. is supported by the Austrian Science Fund (FWF), grant \mbox{[P 36542]}, 
and by the Czech Science Foundation (GAČR), grant GA24-10887S.
L.R. is supported by the research grant GrIFOS, funding scheme PNRR Young Researchers, MSCA Seal of Excellence (SoE), CUP E13C24003600006, ID SOE2024$\_$0000103, of which this paper is part.
She would also like to thank the University of Graz (Graz, Austria), the Department of Mathematics and Statistics and the Department of Physics at the Faculty of Science of MUNI (Brno, Czechia) for the hospitality during her visits in which part of this work was developed. 

\section*{Data Availability Statement}

Data sharing not applicable to this article as no datasets were generated or analysed during the current study.

\appendix
\section{The case of free particles}\label{Free particles}

We consider a configuration of $N$ free particles represented by a point $p=(p_1, \ldots, p_N)\in \mC_{|t}$ at time $t\in \T$, 
and a vector $\mathfrak X_{|p}\in T_p\mC$ which in a bundle chart reads $\mathfrak X=\mathfrak{X}^x \d_x + \mathfrak{X}^t \d_t$, with $\mathfrak{X}^x \d_x \in T_p(\mC_{|t})=V_p\mC\simeq \RR^{3N}$.
At another configuration $p'=R_X p=p+X \in \mC_{|t}$, with $X \in H$, 
we have that $R_{X*} \mathfrak X$ and $\mathfrak X$ have the same $t$-component, i.e. the same projection as $\pi_* R_{X*} \mathfrak X = \pi_* \mathfrak X$. Furthermore, using the canonical parallel transport on $\mC_{|t}\simeq \EE^{3N}$, allowing to identify $V_p\mC \simeq V_{p'}\mC\simeq \RR^{3N}$, we may identify $R_{X*}(\mathfrak{X}^x \d_x)\simeq \mathfrak{X}^x \d_x$,
so that $R_{X*}\mathfrak{X}= \mathfrak{X}$. 
Correspondingly, for $\Xi \in \Aut_v(\mC)$ generated by $\bs X \in \H$ via $\Xi(p)= p + \bs X(p)$, we have by\eqref{pushforward-mC}
$\Xi_* \mathfrak{X} = \mathfrak{X} + \{d\bs X(\mathfrak{X})\}^v$. 
We~have $d\bs X(\mathfrak{X}) = \mathfrak{X}^x \d_x \bs X + \mathfrak{X}^t \d_t \bs X = \mathfrak{X}^t \bs X'$, since $\d_x \bs X = 0$ for $\bs X \in \H$ are basic on $\mC$ so $x$-independent, so that
$\Xi_* \mathfrak{X}\simeq \big(\mathfrak{X}^x +  \mathfrak{X}^t \bs X'\big) \d_x + \mathfrak{X}^t \d_t $.

The fiber metric is here the standard flat metric on $\RR^{3N}\simeq V\mC$, $\langle\_ \,,\_ \rangle: \RR^{3N} \times \RR^{3N} \rarrow \RR$, $(w, w') \mapsto \langle w, w'\rangle$. 
The particles have masses $m=(m_1, \dots, m_N)$, and  
we define the free Lagrangian on $\mC$ as 
$L_{|p}= \tfrac{m}{2} \langle v(\mathfrak X), v(\mathfrak X)\rangle_{|p}\, dt_{|p} = \sum_{k=1}^ N\tfrac{m_k}{2} \langle v_k(\mathfrak X), v_k(\mathfrak X)\rangle_{|p}\, dt_{|p}$, where  $dt \in \Omega^1_\text{basic}(\mC)$ 
and we defined $v\defeq \tfrac{dx}{dt}$, so that $v(\mathfrak X)=\tfrac{\mathfrak{X}^x}{\mathfrak{X}^t}$. 
It gauge transforms as
\begin{align}
 \label{equiv-free-L}
 L^{\bs X}\defeq \Xi^* L 
 &=
 \tfrac{m}{2} \langle v(\Xi_* \mathfrak X) , v(\Xi_* \mathfrak X)\rangle\, \Xi^* dt
 =
 \tfrac{m}{2} \langle v(\mathfrak X) +  \bs X' , v(\mathfrak X)+ \bs X' \rangle\, dt \notag\\
  &=
  \tfrac{m}{2} \langle v(\mathfrak X), v(\mathfrak X)\rangle \, dt \ +\ 
  m \left( 
  \langle v(\mathfrak X), \bs X'\rangle
  +
  \tfrac{1}{2} \langle \bs X', \bs X'\rangle
  \right) \, dt \notag\\
  &= L + \mathrm c(p, \bs X).
\end{align}
 One  checks that $\mathrm c(p, X)$ is a $1$-cocycle for the action of the structure group $H$ of $\mC$:
 \begin{align}
 \label{1-cocycle-Abelian-free}
\mathrm c(p, \bs X+ \bs Y)
&=
m \left( 
  \langle v(\mathfrak X), \bs X' + \bs Y'\rangle
  +
  \tfrac{1}{2} \langle \bs X'+ \bs Y' , \bs X'+ \bs Y'\rangle
  \right) \, dt \notag \\
&=
m \left( 
  \underline{\langle v(\mathfrak X), \bs X' \rangle }
  + \langle v(\mathfrak X), \bs Y'\rangle
  +
  \underline{\tfrac{1}{2} \langle \bs X', \bs X'\rangle} +\langle \bs X' , \bs Y'\rangle
  +\tfrac{1}{2} \langle \bs Y', \bs Y'\rangle
  \right) \, dt  \notag\\
 & = \underline{\mathrm c(p, \bs X)} 
   +  \ \left( \langle v(\mathfrak X) + \bs X', \bs Y'\rangle  
   + \tfrac{1}{2} \langle \bs Y', \bs Y' \rangle \right)\, dt  \notag\\
 &=
\mathrm c(p, \bs X) + \mathrm c(p+\bs X, \bs Y). 
 \end{align}
The infinitesimal gauge transformation 
of $L$ is thus, for $\b\chi = \tfrac{d}{d\epsilon} \bs X_\epsilon \big|_{\epsilon=0} \in $ Lie$\H$, 
\begin{align}
\mathfrak L_{\bs\chi^v} L 
=
\tfrac{d}{d\epsilon} \mathrm c(p, \bs X_\epsilon) \big|_{\epsilon=0}
=
m \langle v(\mathfrak X), \bs\chi '\rangle\, dt \rdefeq \mathrm a(\bs\chi, p).  
\end{align}

Given a kinematical history $\gamma: I \rarrow \mC$, $\tau \mapsto \gamma(\tau)\simeq \big(t(\tau), x(\tau) \big)$, we have $\dot\gamma \simeq \dot x \d_x + \dot t \d_t $. 
We have $v(\dot \gamma)=\tfrac{\dot x}{\dot t}=x'$, the physical velocity (and $\dot x=x'$ for the choice $\dot t=1$).
By \eqref{pushforward-curve} we have 
$\Xi_* \dot\gamma 
= \dot\gamma + \{\dot{\bs X}\}^v
\simeq \big(\dot x +  \dot t \bs X'\big) \d_x + \dot t\d_t $, 
so $v(\Xi_* \dot \gamma) = x'+\bs X'$. 
The~restriction of $L$ on $\gamma$ is 
$\ell_\gamma \defeq \gamma^* L
\simeq \tfrac{m}{2} \langle  x',  x' \rangle\, \gamma^* dt
=\tfrac{m}{2} \langle \dot x, \dot x \rangle\, {\dot t}\- d\tau$, which is the so-called ``parametrized" Lagrangian -- reducing for $\dot t=1$ to the standard unparametrized $\ell_\gamma = \tfrac{m}{2} \langle  x',  x' \rangle\,  dt$. 
The $\H$-transformation of $\ell_\gamma$ is, by \eqref{GT-ell} and above, 
\begin{equation}
\label{GT-ell-free-1}
\begin{aligned}
 \ell_\gamma^{\bs X} 
  = \ell_\gamma + \mathrm c_\gamma(\bs X)
  &=\ell_\gamma + \mathrm c\big(\gamma, \bs X(\gamma)\big) \\
 &\simeq \ell_\gamma +  m \left( 
  \langle \dot x, \dot{\bs X}\rangle
  +
  \tfrac{1}{2} \langle \dot{\bs X}, \dot{\bs X}\rangle
  \right) \, {\dot t}\-  d\tau.
\end{aligned}
\end{equation}
The linearisation of which gives 
\begin{equation}
\begin{aligned}
\delta_{\bs\chi} \ell_\gamma 
= \mathrm a_\gamma(\bs \chi) = \mathrm a \big(\chi(\gamma), \gamma \big)
&= m \langle \dot x, \dot {\bs\chi} \rangle\,  {\dot t}\- d\tau \\
&= -m \langle x'' , \bs \chi \rangle \,  {\dot t} d\tau  
+ 
\left( \tfrac{d}{d\tau} m\langle x', \bs \chi \rangle \right) d\tau ,
\end{aligned}
\end{equation}
where we identify the Euler-Lagrange 1-form $E(\bs\chi, \gamma)= \bs\chi(\gamma)\cdot EL[\gamma] d\tau = - \langle \bs \chi , m x'' \rangle \, \dot td\tau$, and the presymplectic potential giving the conserved Noether charge $ \langle m x', \bs \chi \rangle 
$.
The gauge transformation in the unparametrized version for $\dot t=1$ reads
\begin{equation}
\label{GT-ell-free-2}
\begin{aligned}
 \ell_\gamma^{\bs X} 
 &= \ell_\gamma +  m \left( 
  \langle  x', {\bs X}'\rangle
  +
  \tfrac{1}{2}\langle {\bs X}', {\bs X}'\rangle
  \right) \, dt.
\end{aligned}
\end{equation}
So, $\delta_{\bs\chi} \ell_\gamma 
= \mathrm a_\gamma(\bs \chi) = - \langle \bs \chi, mx''\rangle dt + \tfrac{d}{dt} \langle \bs \chi, m x' \rangle dt$, where we recognise the standard expression of the canonical momentum $\mathrm \uppi=mx'$ and of the Euler-Lagrange 1-form 
$E(\bs\chi, \gamma)=\bs\chi(\gamma)\cdot EL[\gamma] dt = - \langle \bs\chi, \mathrm \uppi' \rangle dt$.

The equations \eqref{GT-ell-free-1}-\eqref{GT-ell-free-2} give the transformation of the free Lagrangian under the so-called ``extended Galilean transformations" (or boosts) \cite{Greenberger1979}, 
as discussed in section \ref{Bundle geometry of configuration space-time}. 
Standard Galilean boosts are reproduced for $\bs X(t)=\mathrm v t$ with $\mathrm v \in H$, in which case the Lagrangian is quasi-invariant -- i.e. invariant up to a boundary term -- as
\begin{equation}
\label{GT-ell-free-3}
\begin{aligned}
 \ell_\gamma^{\bs X} 
&= \ell_\gamma +  m \left( 
  \langle  x', \mathrm v \rangle
  +
  \tfrac{1}{2}\langle \mathrm v, \mathrm v \rangle
  \right) \, dt \\
&= \ell_\gamma + \frac{d}{dt}\left(\, m \,\left( 
\langle  x, \mathrm v \rangle
  +
\tfrac{1}{2} \langle\mathrm v, \mathrm v \rangle \, t
\right)
\right)\, dt \\
&\rdefeq 
\ell_\gamma + \frac{d}{dt} \Delta\big((t, x), \mathrm v\big ) \, dt.
\end{aligned}
\end{equation}
This reproduces eq. (3.1.5) in \cite{DeAzc-Izq}, where $\Delta\big((t, x), \mathrm v\big )$ is called a 1-cocycle for the Galilean group, satisfying a special case of the defining property \eqref{1-cocycle-Abelian-1}-\eqref{1-cocycle-Abelian-free} of $\mathrm c(p, X)$ and $\mathrm c_\gamma(\bs X)$.
The rigid limit $\bs X \rarrow X$ of \eqref{GT-ell-free-1}/\eqref{GT-ell-free-2} gives the trivial $H$-equivariance of $\ell_\gamma$, $R^*_X \ell_\gamma = \ell_\gamma$, i.e. the well-known invariance of the free Lagrangian under translations. 

The action functional is $\S[\gamma]=\int_\gamma L = \int_I \ell_\gamma = \int_I \tfrac{m}{2} \langle \dot x, \dot x \rangle\,  {\dot t}\-  d\tau$, or, in the unparametrised form, 
$\S[\gamma]= \int \tfrac{m}{2} \langle  x', x' \rangle\, dt$.
Its $\H$-transformation is
\begin{equation}
\label{H-tr-action-free}
\begin{aligned}
 \S[\gamma]^{\bs X} 
  = \S[\gamma ]+\int_I \mathrm c_\gamma(\bs X) 
 &\simeq \S[\gamma] + \int_I m \left( 
  \langle \dot x, \dot{\bs X}\rangle
  +
  \tfrac{1}{2} \langle \dot{\bs X}, \dot{\bs X}\rangle
  \right) \, {\dot t}\-  d\tau \\[1mm]
  & \simeq \S[\gamma] + \int m \left( 
  \langle  x', {\bs X}'\rangle
  +
  \tfrac{1}{2} \langle {\bs X}', {\bs X}'\rangle
  \right) \, dt, 
  \quad \text{for } \dot t=1.
\end{aligned}
\end{equation}
 Which linearises as 
\begin{equation}
\begin{aligned}
\delta_{\bs\chi} \S[\gamma] 
= \int_I \mathrm a_\gamma(\bs \chi) 
&\simeq \int_I -m \langle x'', \bs \chi \rangle \,  {\dot t} d\tau  
+ 
\left( \tfrac{d}{d\tau} m\langle x', \bs \chi \rangle \right) d\tau \\[1mm]
&\simeq  \int -\langle \bs \chi, mx''\rangle dt 
 +
\tfrac{d}{dt} \langle \bs \chi, m x' \rangle \,dt, 
\quad
\text{for } \dot t =1, \\
&=
\int \bs\chi(\gamma)\cdot EL[\gamma]\, dt + \langle  \bs \chi (\gamma), \mathrm \uppi \rangle\big|^{t_1}_{t_0}, 
\quad
\text{with } \bs\chi(\gamma) =\delta_\chi \gamma.
\end{aligned}
\end{equation}
From which we obtain  the variational principle: $\delta_{\bs\chi} \S[\gamma_c]\equiv 0$ for any $\bs\chi \in$ Lie$\H_{0, 1}$, giving  the equation of motion $EL[\gamma_c]=m x''= \mathrm \uppi'=0$, and the conserved Noether charge $\mathrm \uppi$.

The wave function for $N$ free particles is  $\psi_\gamma(p) \simeq \psi\big(t, x\big)=\psi \big(t; x_1, \dots, x_N \big)$ and its $\H$-transformation is 
\begin{equation}
\label{GT-dressed-psi-free}
\begin{aligned}
  &\psi^{\bs X}_\gamma = C_\gamma(\bs X)\- \psi_\gamma
  = \exp \left\{\tfrac{i}{\hbar} \int_I \mathrm c_\gamma(\bs X) \right\}\  \psi_\gamma, \\[1mm]
 & \hookrightarrow \quad 
  \psi\big(t, x+ \bs X \big)= \exp\left\{ \tfrac{i}{\hbar} \int_I m \left( 
  \langle \dot x, \dot{\bs X}\rangle
  +
  \tfrac{1}{2} \langle \dot{\bs X}, \dot{\bs X}\rangle
  \right) \, {\dot t}\-  d\tau \right\} \ \psi\big(t, x \big) \\
  & \phantom{ \hookrightarrow \quad 
  \psi\big(t, x+ \bs X \big)\ \, }
  = \exp\left\{ \tfrac{i}{\hbar}\int m \left( 
  \langle  x', {\bs X}'\rangle
  +
  \tfrac{1}{2} \langle {\bs X}', {\bs X}'\rangle
  \right) \, dt \right\} \ \psi\big(t, x \big), \quad \text{for } \dot t=1.
\end{aligned}
\end{equation}
For the special case $\bs X(t)= \mathrm v t$, by \eqref{GT-ell-free-3} this reduces to
\begin{equation}
\begin{aligned}
\psi\big(t, x+ \bs X \big)
  &= \exp\left\{ \tfrac{i}{\hbar}\int 
  \frac{d}{dt} \Delta\big((t, x), \mathrm v\big ) \, dt
  \right\} \ \psi\big(t, x \big) 
  = \exp\left\{ \tfrac{i}{\hbar} 
  \Delta\big((t, x), \mathrm v\big ) 
  \right\} \ \psi\big(t, x \big) \\
  &= \exp\left\{ \tfrac{i}{\hbar}\, m \,\left( 
\langle  x, \mathrm v \rangle
  +
\tfrac{1}{2} \langle\mathrm v, \mathrm v \rangle \, t
\right)
\right\} \ \psi\big(t, x \big),
\end{aligned}
\end{equation}
 reproducing eq. (3.1.9) in \cite{DeAzc-Izq}, the well-known transformation of the wave function under Galilean boosts.

\paragraph{Relational description via the DFM}

Here we aim only to illustrate in a simple case how the relational, dressed, Lagrangian $\ell_\gamma^\u$  and  physical frame covariance may be computed using the cocycle. 
Taking, for concreteness, first the viewpoint of the $i^{th}$ particle, i.e. the dressing $\u(p)=\u_i(p)=-x_i$, 
we have, 
\begin{align}
\mathrm c_\gamma(\u_i) 
 \simeq 
 \mathop{\textstyle \sum}_{k=1}^N\, m_k \left(
 \langle \dot x_k, \dot \u_i\rangle +
 \tfrac{1}{2}\langle \dot \u_i, \dot \u_i \rangle
\right) {\dot t}\-  d\tau.
\end{align}
Using \eqref{dressed-ell-2} we thus have the dressed Lagrangian
\begin{align}
\label{dressed-ell-free}
\ell_\gamma^{\u_i} = \ell_\gamma + \mathrm c_\gamma(\u_i)
&\simeq 
\mathop{\textstyle \sum}_{k=1}^N\, 
\tfrac{m_k}{2} \big( 
\langle \dot x_k, \dot x_k\rangle 
+
2 \langle \dot x_k, \dot \u_i\rangle
+
\langle \dot \u_i, \dot \u_i\rangle
\big)\, {\dot t}\-  d\tau  \notag\\
&=
\mathop{\textstyle \sum}_{k=1}^N\,  
\tfrac{m_k}{2} \big( 
\langle \dot x_k, \dot x_k\rangle 
-
2 \langle \dot x_k, \dot x_i\rangle
+
\langle \dot  x_i, \dot x_i\rangle
\big)\, {\dot t}\-  d\tau  
= 
\mathop{\textstyle \sum}_{k=1}^N\,  
\tfrac{m_k}{2} \langle \dot x_k - \dot x_i, \dot x_k - \dot x_i\rangle \, {\dot t}\-  d\tau  \notag\\
&=
\mathop{\textstyle \sum}_{k=1}^{N}\,  
\tfrac{m_k}{2} \langle \dot{\b x}^{\,i}_k , \dot{\b x}^{\,i}_k \rangle \, {\dot t}\-  d\tau
= \tfrac{m}{2} \langle \dot{\b x}^{\,i} , \dot{\b x}^{\,i} \rangle \, {\dot t}\-  d\tau \notag\\
&=
\mathop{\textstyle \sum}\limits_{\substack{k=1 \\ k\neq i}}^{N}
\, \tfrac{m_{k}}{2} \langle \dot{\b x}^{\,i}_{k} , \dot{\b x}^{\,i}_{k} \rangle \, {\dot t}\-  d\tau
= \ell_{\gamma^{\u_i}}.
\end{align}
This Lagrangian describes the relational dynamics of $N-1$ particles with respect to the $i^{th}$ particle. It is manifestly $\H_\text{ext}$-invariant, and its residual transformation (1st kind) under $\H^{\u_i}_\text{int}\simeq\H_\text{int}$ is, by \eqref{dressed-ell-res-1}, analogous to \eqref{GT-ell-free-1}:
\begin{align}
\label{res1-dressed-ell-free}
(\ell_\gamma^{\u_i})^{\b{\bs Y}} 
  = \ell_\gamma^{\u_i} + \mathrm c_{\gamma^{\u_i}}(\b{\bs Y})
  &=\ell_\gamma^{\u_i} + \mathrm c\big(\gamma^{\u_i}, \b{\bs Y}(\gamma)\big)  \notag\\
 &\simeq 
 \ell_\gamma^{\u_i} +  m \left( 
  \langle \dot{\b x}^{\,i}, \dot{\b{\bs Y}}^{\,i}\rangle
  +
  \tfrac{1}{2} \langle \dot{\b{\bs Y}}^{\,i}, \dot{\b{\bs Y}}^{\,i}\rangle
  \right) \, {\dot t}\-  d\tau, 
\end{align}
with $\b{\bs Y}^{\,i}=\bs Y - \bs Y_i=\big(\bs Y_1 - \bs Y_i, \ldots, 0, \ldots, \bs Y_N - \bs Y_i \big)$ by \eqref{map-ith-res1} and \eqref{kin-res-1st}.
It is easily checked directly by plugging  ${\b x}^{\b{\bs Y}}=\b x + \b{\bs Y}$ in the expression \eqref{dressed-ell-free} for $\ell_\gamma^{\u_i}$.

Similarly, the dressed wave function for free particles is, by \eqref{dressed-psi-gamma}-\eqref{dressed-psi-bis}, 
\begin{align}
\psi^{\u_i}_\gamma = C_\gamma(\u_i)\- \psi_\gamma
  \simeq \psi\big(t, \b x^{\,i} \big)
    = \psi\big(t, \b x^{\,i}_1, \ldots, 0, \ldots, x^{\,i}_{N} \big),
\end{align}
and describes the relational quantum state between the $i^{th}$ particle and the $N-1$ others.
It is by construction $\H_\text{ext}$-invariant, and its $\H^{\u_i}_\text{int}\simeq\H_\text{int}$-transformation is,
by \eqref{dressed-psi-Y}, similar to \eqref{GT-dressed-psi-free}:
\begin{align}
\label{dressed-psi-free-Y}
  &\big(\psi^{\u_i}_\gamma\big)^{\b{\bs Y}} = C_\gamma(\b{\bs Y})\- \psi_\gamma
  = \exp \left\{\tfrac{i}{\hbar} \int_I \mathrm c_\gamma(\b{\bs Y}) \right\}\  \psi_\gamma, \notag\\[1mm]
 & \hookrightarrow \quad 
  \psi\big(t, {\b x}^{\,i} + \b{\bs Y}^{\,i}\big)= \exp\left\{ \tfrac{i}{\hbar} \int_I m \left( 
  \langle \dot{{\b x}}^{\,i}, \dot{\b{\bs Y}}^{\,i}\rangle
  +
  \tfrac{1}{2} \langle \dot{\b{\bs Y}}^{\,i}, \dot{\b{\bs Y}}^{\,i}\rangle
  \right) \, {\dot t}\-  d\tau \right\} \  \psi\big(t, \b x^{\,i} \big).
\end{align}

We may now look at how physical frame covariance is implemented by $\G$-transformations. 
By  \eqref{kin-res-2nd}-\eqref{kin-res-2nd-bis}, the change from the reference frame of the $i^{th}$ particle to that of the $j^{th}$ particle is given by
$\gamma^{\u_j}=(\gamma^{\u_i})^{\bs Z_{ij}}= \gamma^{\u_i} + \bs Z_{ij}(\gamma)$  for $\bs Z_{ij}=x_i - x_j \in \G$. 
Then, by \eqref{dressed-ell-res-2}, we have
\begin{align}
\ell^{\u_j}_\gamma 
&= 
\ell^{\u_i}_\gamma + \mathrm{c}\big(\gamma^{\u_i}, \bs{Z}_{ij}(\gamma) \big) \notag\\
&=
\ell^{\u_i}_\gamma +
 m \left( 
  \langle \dot{\b x}^{\, i}, \dot{\bs Z}_{ij}\rangle
  +
  \tfrac{1}{2} \langle \dot{\bs Z}_{ij}, \dot{\bs Z}_{ij}\rangle
  \right) \, {\dot t}\-  d\tau \notag \\
&= 
\mathop{\textstyle \sum}_{k=1}^N\,  
\tfrac{m_k}{2} \big( 
\langle \dot x_k, \dot x_k\rangle 
-
2 \langle \dot x_k, \dot x_i\rangle
+
\langle \dot  x_i, \dot x_i\rangle
\big)\, {\dot t}\-  d\tau \notag\\[-1mm]
&\phantom{blabla}\quad +\quad 
m_k
\left(
 \langle \dot x_k- \dot x_i, \dot x_i - \dot x_j \rangle +
 \tfrac{1}{2}\langle \dot x_i - \dot x_j, \dot x_i - \dot x_j\rangle
\right) {\dot t}\-  d\tau \notag\\
&=
\mathop{\textstyle \sum}_{k=1}^N\,  
\tfrac{m_k}{2} \Big( 
\langle \dot x_k, \dot x_k\rangle 
-
\xcancel{2 \langle \dot x_k, \dot x_i\rangle}
+
\bcancel{\langle \dot  x_i, \dot x_i\rangle }
 \notag\\[-1mm]
&\phantom{blabla}\quad +\quad
\xcancel{2\langle \dot x_k, \dot x_i \rangle }
-
2\langle \dot x_k, \dot x_j \rangle
-
\bcancel{2\langle \dot x_i, \dot x_i \rangle}
+
\cancel{2\langle \dot x_i, \dot x_j \rangle}
\ \ + \ \ 
\bcancel{\langle \dot x_i, \dot x_i \rangle}
- 
\cancel{2\langle \dot x_i, \dot x_j \rangle}
+ 
\langle \dot x_j, \dot x_j \rangle
\Big)\, {\dot t}\-  d\tau \notag\\
&= 
\mathop{\textstyle \sum}_{k=1}^N\,  
\tfrac{m_k}{2}
\langle \dot x_k - \dot x_j , \dot x_k - \dot x_j  \rangle \, {\dot t}\-  d\tau \notag\\
&= 
\mathop{\textstyle \sum}_{k=1}^N\,  
\tfrac{m_k}{2}
\langle\dot{\b x}_k^{\,j}  , \dot{\b x}_k^{\,j}  \rangle \, {\dot t}\-  d\tau 
= 
\tfrac{m}{2} \langle \dot{\b x}^{\,j} , \dot{\b x}^{\,j} \rangle \, {\dot t}\-  d\tau ,
\end{align}
showing explicitly that the cocyclic structure controls the physical frame covariance.
The frame covariance of the wave function is, by 
\eqref{dressed-psi-gamma-Z}-\eqref{dressed-wave-functions-Z},
\begin{align}
  &\psi^{\u_j}_\gamma=\big(\psi^{\u_i}_\gamma\big)^{\bs Z_{ij}} = C_{\gamma^{\u_i}}(\bs Z_{ij})\- \psi_\gamma^{\u_i}
  = \exp \left\{\tfrac{i}{\hbar} \int_I \mathrm c_{\gamma^{\u_i}}(\bs Z_{ij}) \right\}\  \psi_\gamma^{\u_i}, \notag\\[1mm]
 & \hookrightarrow \quad 
  \psi\big(t, {\b x}^{\,j} \big)= \exp\left\{ \tfrac{i}{\hbar} \int_I 
  m \left( 
  \langle \dot{\b x}^{\, i}, \dot{\bs Z}_{ij}\rangle
  +
  \tfrac{1}{2} \langle \dot{\bs Z}_{ij}, \dot{\bs Z}_{ij}\rangle
  \right) \, {\dot t}\-  d\tau 
  \right\} \  \psi\big(t, \b x^{\,i} \big).
\end{align}
It describes the relational quantum state between the $j^{th}$ particle and the $N-1$ others.
It is by construction $\H_\text{ext}$-invariant, and its $\H^{\u_j}_\text{int}\simeq\H_\text{int}$ is, \emph{mutatis mutandis}, the same as \eqref{dressed-psi-free-Y}.
We thus see the ``relational quantum democracy"  at play, whereby any subsystem can be used to describe the collective quantum state, and change of subsystem is implemented by 
$\G$-transformations via the cocyclic structure.
\medskip

All the above applies identically for a configuration of $N$ harmonic oscillators, whose dynamics is given by the Lagrangian 
$L= \tfrac{m}{2} \langle v, v\rangle\, dt 
- \tfrac{\mathrm k}{2} \langle x, x\rangle\, dt 
= \sum_{l=1}^ N\tfrac{m_l}{2} \langle v_l, v_l\rangle\, dt 
- \tfrac{\mathrm k_l}{2} \langle x_l, x_l\rangle\, dt$, 
with spring constants $\mathrm k = \big(\mathrm k_1, \ldots, \mathrm k_N\big)$, whose $\H$-transformation is given by the 1-cocycle 
$ \mathrm c \big(p, \bs X\big)
\defeq m \left( 
  \langle v, \bs X' \rangle
  +
  \tfrac{1}{2} \langle \bs X', \bs X' \rangle
  \right) \, dt
  -
  \mathrm k \left( 
  \langle x, \bs X \rangle 
  +
  \tfrac{1}{2} \langle \bs X, \bs X\rangle
  \right) \,
  dt$.

{
\normalsize 
 \bibliography{Biblio11}
}

\end{document}